\documentstyle[12pt,twoside,psfig]{report}
\newcommand{\bqa}{\begin{eqnarray}}
\newcommand{\eqa}{\end{eqnarray}}
\newcommand{\beq}{\begin{equation}}
\newcommand{\eeq}{\end{equation}}
\newlength{\head}
\newcommand{\heading}[2]{\markboth{\settowidth{\head}{#1}
     \addtolength{\head}{1.22cm}
     \protect\hspace{-10.5mm}	
     \protect\rule[-0.15cm]{\textwidth}{0.2mm}
     \protect\hspace{-\head}
     \bf{\hfill #1} }
     {\protect\rule[-0.15cm]
     {\textwidth}{0.2mm} \protect\hspace{-\textwidth}
     \protect\hspace{-0.5cm}
     \protect\bf{ \arabic{chapter}.\arabic{section} \hspace{2mm} #2} }} 
 
\newcommand{\headingx}[2]{\markboth{\settowidth{\head}{#1}
     \addtolength{\head}{1cm}
     \protect\hspace{-10.5mm}	
     \protect\rule[-0.15cm]{\textwidth}{0.2mm}
     \protect\hspace{-\head}
     \bf{\hfill #1} }
     {\protect\rule[-0.15cm]
     {\textwidth}{0.2mm} \protect\hspace{-\textwidth}
     \protect\hspace{-0.5cm}
     \protect\bf{ #2} }} 

\begin{document}

\pagestyle{empty}


\begin{titlepage}
\begin{center}
\vspace*{2cm}
{\Huge {\bf On Effective Field Theories}} \\
\vspace{0.4cm}
{\Huge {\bf at Finite Temperature}} \\
\vspace{3cm}
{\Large {\bf Jens O. Andersen}} \\
\vspace{2cm}
\centerline{\psfig{figure=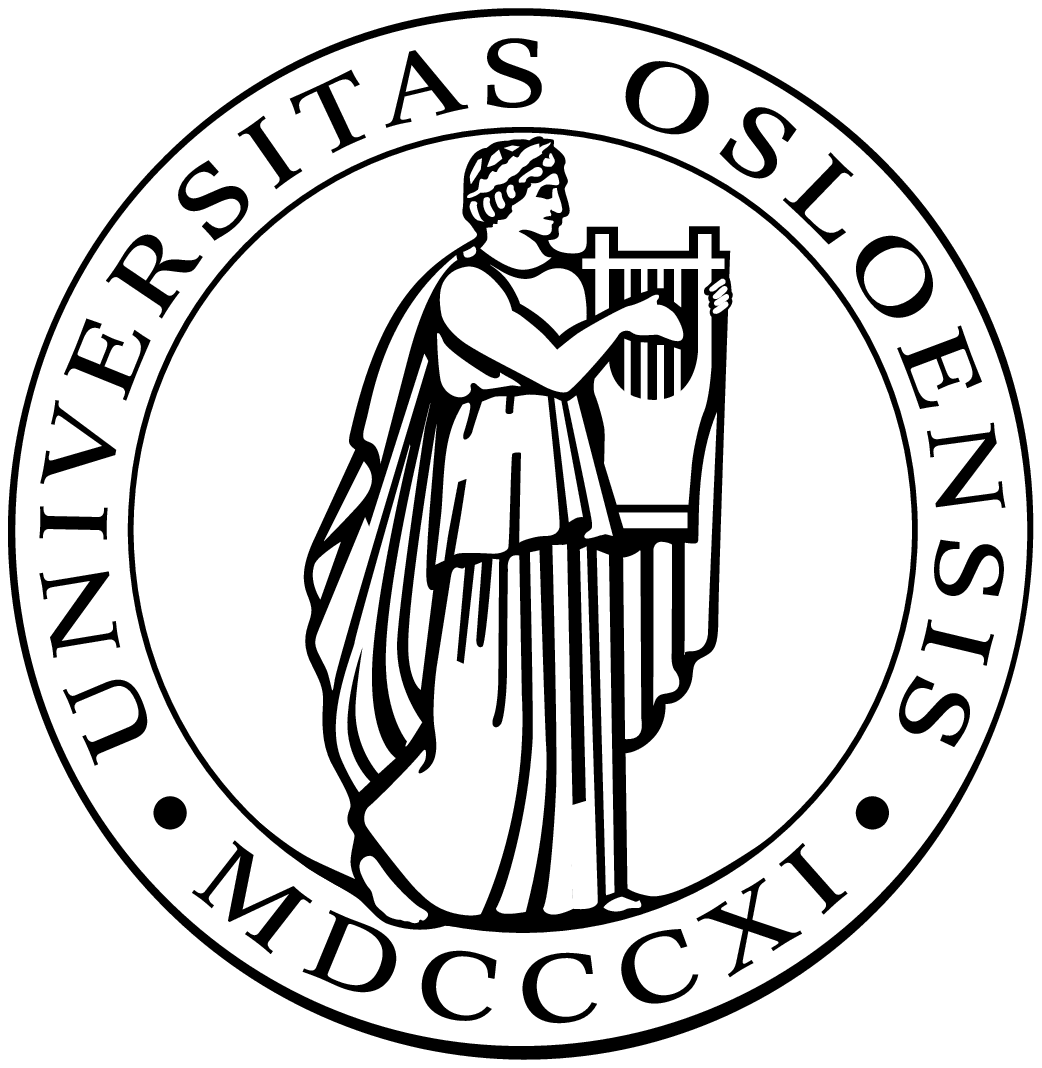,width=3.7cm}}
\vspace{2cm}
{\bf Thesis submitted for the Degree of } \\
{\bf Doctor Scientiarum} \\
\vspace{0.4cm}
{\bf Department of Physics} \\
{\bf University of Oslo} \\
{\bf June 1997} \\
\end{center}
\end{titlepage}

\cleardoublepage

\pagestyle{plain}
\pagebreak
\pagenumbering{roman}
\newpage
\cleardoublepage

\cleardoublepage
\clearpage 
\pagebreak
\chapter*{Acknowledgments}
\addcontentsline{toc}{chapter}{Acknowledgments}
First of all, I would like to thank my supervisor professor
Finn Ravndal for his
guidance during the work on this thesis. I am very grateful to him for
sharing his deep insight in physics with me. I have benefitted very much 
from our many discussions and his lectures. He 
has also patiently answered my numerous 
questions.

I would also like to thank my fellow students Tor Haugset and Jon {\AA}ge
Ruud, and Dr. H{\aa}rek 
Haugerud for many enlightening discussions, especially on effective field
theory. 
Special thanks to Tor for our cooperation and for allowing me to include our 
joint paper in my thesis. H{\aa}rek is also acknowledged
for carefully reading this thesis.

I should also thank Dr. Mark Burgess, 
for introducing me to ring corrections and resummation. I am also
thankful for his comments on this thesis.

It is also a pleasure to thank the Professors Andras Patk\'os,
Anton Rebhan and 
Eric Braaten for 
helpful communications.

I am very grateful to the Department of Physics at the University of Oslo
for financial support during four and half years, and thereby giving me the
opportunity to study quantum field theory. I also thank the theory group for
financial support to several conferences and workshops in Norway as well
as abroad. NORDITA is acknowledged for funding in connection with
a number of stays in Copenhagen.

I thank Anne-Cecilie Riiser, Nils Tveten and Heidi Kj{\o}nsberg for 
pleasant company during the time we have shared the office.

Finally, I am indebted to my mother for her moral support in tough 
periods.\\ \\
Oslo, June 1997\\ \\
Jens O. Andersen

\cleardoublepage
\chapter*{Abstract}
\addcontentsline{toc}{chapter}{Abstract}
This thesis is devoted to a study of quantum fields at finite temperature.
First, I consider Dirac fermions and bosons moving in a plane with a 
homogeneous static magnetic field orthogonal to the plane. 
The effective action for the gauge field is derived by integrating out the
matter field. The magnetization is calculated, and in the fermionic 
case it is demonstrated that the system exhibits de-Haas van Alphen
oscillations at low temperatures and weak magnetic fields.
I also briefly discuss the extension of the results to more
general field configurations.

Next, the breakdown of ordinary perturbation theory at high temperature
is studied. I discuss the need for an effective expansion 
and the resummation program of
Braaten and Pisarski in some detail.
The formalism is applied to Yukawa theory, and the screening mass 
squared and the free energy is derived to two and three loop 
order, respectively.

The main part of the present work is on effective field theories 
at finite temperature. I discuss the concepts of dimensional reduction,
modern renormalization theory, 
and renormalizable field theories (``fundamental theories'')
versus non-renormalizable theories (``effective theories''). 

Two methods for constructing effective three dimensional field theories
are discussed. The first is based on the effective potential, and is
applied to field theory with $N$ charged $U(N)$ symmetric scalar
coupled to an Abelian gauge field.
The effective theory obtained may be used to study phase transition
non-perturbatively as a function of $N$.
The second method is an effective field theory approach based on
diagrammatical methods, recently developed by Braaten and Nieto.
I apply the method to spinor and scalar QED, and the screening masses
as well the free energies are obtained.

\cleardoublepage
\tableofcontents
\cleardoublepage
\listoffigures
\cleardoublepage
\setcounter{page}{13}
\cleardoublepage

\chapter*{}
\chapter*{Preface}
\addcontentsline{toc}{chapter}{Preface}


The thesis is based upon the following papers
\begin{itemize}
\item Magnetization in (2+1)-dimensional QED at Finite Temperature and density.
Jens O. Andersen and Tor Haugset, Phys. Rev. {\bf D 51}, 3073, 1995.
\end{itemize}
\begin{itemize}
\item Effective Potentials and Symmetry Restoration in the 
Chiral Abelian Higgs Model, Jens O. Andersen, Mod. Phys. Lett {\bf A 10}
997, 1995.
\end{itemize}
\begin{itemize}
\item The Free Energy of High Temperature QED to Order $e^{5}$ From Effective
Field Theory, Jens O. Andersen, Phys. Rev. {\bf D 53}, 7286, 1996.
\end{itemize}
\begin{itemize}
\item The Electric Screening Mass in Scalar Electrodynamics
at High Temperature, Jens O. Andersen, To appear in Z. Phys. C. {\bf 75}, 1997.
\end{itemize}
\begin{itemize}
\item The Free Energy  in Scalar Electrodynamics at High Temperature.
Effective Field Theory versus Resummation,
Jens O. Andersen, in preparation.
\end{itemize}

\cleardoublepage
\pagestyle{myheadings}
\pagenumbering{arabic}

\chapter{Particles in External Fields}
\heading{Particles in External Fields}{Particles in External Fields}
\section{Introduction}
\heading{Particles in External Fields}{Introduction}
One of the oldest problems in nonrelativistic quantum mechanics is that
of a charged particle in  a constant magnetic field. This problem was solved
in 1930 by Landau~\cite{landau}, 
and the energy levels are called Landau levels.

More generally, particles in external fields have been studied extensively
since the early fifties, when
Schwinger~\cite{schwing} calculated the effective action for 
constant field strengths
in QED using the proper time method. 
The study of matter under extreme conditions such as very strong
electromagnetic fields is of interest in various systems, and the
applications range from condensed matter to astrophysics [3,4].

In the case of a constant magnetic field there exists another and perhaps 
simpler method for obtaining the effective action of the gauge 
field~\cite{per}. 
Integrating out the fermion fields in the path integral gives rise
to a functional determinant, that must be evaluated.
In order to do so we exploit
the fact that the propagator equals
the derivative of the effective Lagrangian with respect to the mass in the
fermionic case, and with respect to the squared of the mass in the bosonic 
case.
This requires the knowledge of the propagator, which can be constructed
explicitly, since we know the solutions to the 
Dirac equation or the Klein-Gordon equation
in the case of a constant magnetic field.

Moreover, this method immediately generalizes to finite temperature and
nonvanishing chemical potential. Thus, it becomes easy to study fermions and 
bosons at finite temperature and density.

So far the gauge field has been treated classically. However, one may of 
course consider quantum fluctuations around the classical background field.
With the propagators at hand, one would then compute the vacuum diagrams in the
loop expansion in the usual way.
At the one loop level this implies a contribution to the effective
action from the photons which equals $\frac{\pi^2T^4}{45}$. in 3+1$d$, and
$\frac{\zeta (3)T^3}{\pi}$ in $2+1d$. 
This is the usual contribution
to the free energy from a free photon gas at temperature $T$.
Beyond one loop the evaluation of the graphs becomes difficult.
Ritus has carried out one of the very few existing two-loop calculations 
in $3+1d$ QED at $T=0$, but finite density~\cite{ritus}.

Many of the phenomena that have been discovered in condensed matter 
physics
over the last few decades are to a very good approximation two 
dimensional.
The most important of these are the (Fractional) Quantum Hall effect and 
high $T_{c}$
superconductivity~\cite{qhe}.

Quantum field theories in lower dimensions have therefore become of 
increasing
interest in recent years. Both systems mentioned above have been 
modeled by anyons, which are particles or excitations that obey 
fractional
statistics. Anyons can be described in terms of Chern-Simons field 
theories [7-9].

Some ten years ago Redlich~\cite{reddik} considered fermions in a plane moving
in a constant electromagnetic field. Using Schwinger's proper time 
method~\cite{schwing}
to obtain the effective action for the gauge field, he demonstrated that 
a Chern-Simons
term is induced by radiative corrections. The Chern-Simons term is parity
breaking and is gauge-invariant modulo surface terms.

External electromagnetic fields may give rise to induced charges in the
Dirac vacuum if the energy spectrum is asymmetric with respect to some 
arbitrarily chosen zero point. The vacuum charge comes
about since the number of particles gets reduced (or increased) relative 
to
the free case. Furthermore, induced 
currents may appear and are attributed to the drift of the induced 
charges.
This only happens if the external
field does not respect the translational symmetry of the system. 
These interesting
phenomena have been examined  in detail by Flekk{\o}y and Leinaas~\cite{lein} 
in connection with
magnetic vortices and their relevance to the Hall effect has been studied
by Fumita and Shizuya~\cite{japs}.

In this chapter we re-examine the system considered by 
Redlich~\cite{reddik}. We shall restrict ourselves to the
case of a constant magnetic field, but we extend the analysis by including
thermal effects and we shall mainly focus on the magnetization of the 
system. For completeness, we also consider bosons in a constant magnetic
field. Our calculations resemble the treatment given by 
Elmfors {\it et al.}~\cite{per}
of the corresponding system in $3+1d$. However, interesting differences occur,
mainly connected with the asymmetry in the Dirac spectrum, and we shall comment
upon them as we proceed.
\section{Fermions in a Constant Magnetic Field}
\heading{Particles in External Fields}{Fermions in a Constant Magnetic Field}
We start our discussion of particles in external fields by considering
fermions in two dimensions in a constant magnetic field. 
\subsection{The Dirac Equation}
\heading{Particles in External Fields}{Dirac Equation}
In this subsection we shall discuss some properties of the
Dirac equation equation
in $2+1d$. We also solve it for the case of a constant magnetic field 
along the $z$-axis. The Dirac 
equation reads
\begin{equation}
\label{eq:dirac}
(i\gamma^{\mu}\partial_{\mu}-e\gamma^{\mu}A_{\mu}-m)\psi=0,
\end{equation}
where the gamma matrices satisfy the Clifford algebra
\begin{equation}
\label{eq:clif}
\{ \gamma^{\mu},\gamma^{\nu}\}=2g^{\mu\nu}. 
\end{equation}
In 2+1 dimensions the fundamental representation of the Clifford algebra 
is
given by $2\times 2$ matrices and these can be constructed from the 
Pauli matrices. They are
\begin{equation}\sigma^{1}=
\left(\begin{array}{cc}
0&1\\
1&0\\
\end{array}\right),\hspace{1cm}\sigma^{2}=
\left(\begin{array}{cc}
0&-i\\
i&0\\
\end{array}\right),\hspace{1cm}\sigma^{3}=
\left(\begin{array}{cc}
1&0\\
0&-1\\
\end{array}\right).
\end{equation}
Furthermore, in $2+1d$ there are two inequivivalent 
choices of 
the gamma matrices, which corresponds to 
$\gamma^{\mu}\rightarrow -\gamma^{\mu}$. From Eq.~(\ref{eq:dirac}) we see
that this extra degree of freedom may be absorbed in the sign of $m$.
These choices correspond to ``spin up'' and ``spin down'', 
respectively~\cite{lein}. 

The angular momentum operator 
{\bf $\sigma$} is a  pseudo vector
in $2+1d$, implying that the Dirac equation written in terms of these 
matrices
does not respect parity~\cite{markus2}. 
This is no longer the case if the Dirac equation
is expressed in terms of $4\times 4$ matrices. These matrices can be taken
as the standard representation in $3+1d$
\begin{equation}\gamma^{0}_{4}=
\left(\begin{array}{cc}
I&0\\
0&I\\
\end{array}\right),\hspace{1cm}\gamma^{i}_{4}=
\left(\begin{array}{cc}
0&-\sigma^i\\
\sigma^i&0\\
\end{array}\right),
\end{equation}
where we simply drop $\gamma^3_4$. This representation is reducible, and 
reduces to the two inequivalent fundamental representation mentioned 
above~\cite{markus2}.

\begin{figure}[htb]
\begin{center}
\mbox{\psfig{figure=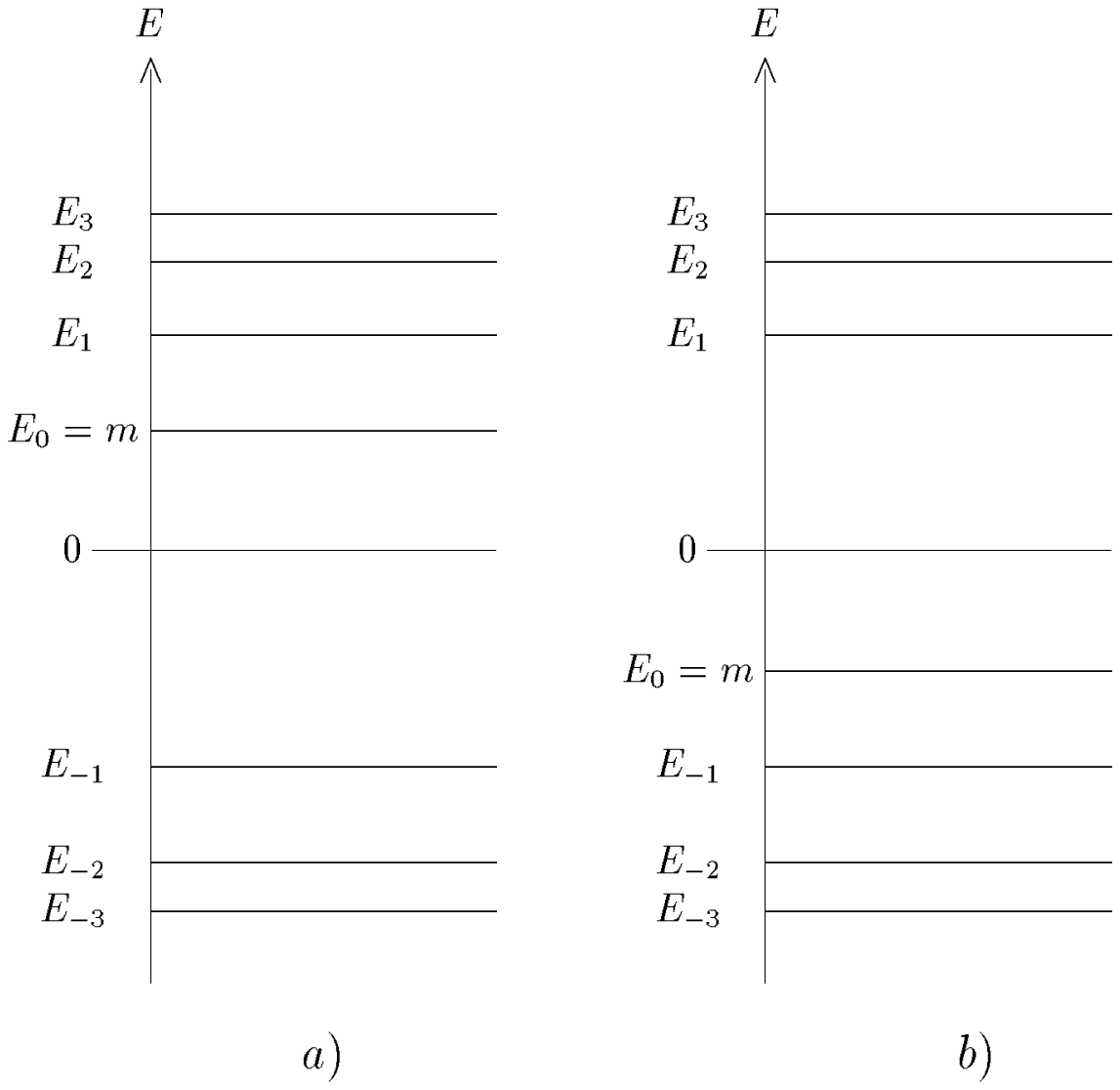}}
\end{center}
\caption[The energy spectra of Dirac fermions
in the presence of a constants magnetic field. a) $m>0$ and b) $m<0$.]{\protect The energy spectra of Dirac fermions
in the presence of a constants magnetic field. a) $m>0$ and b) $m<0$.}
\label{spec}
\end{figure}

In the following we make the choice $\gamma^{0}=\sigma^{3}$, 
$\gamma^{1}=-i\sigma^{2}$ and $\gamma^{2}=-i\sigma^{1}$. In this chapter
we use the real time formalism and the metric is $\mbox{diag}(1,-1,-1)$.
 For particles in a 
constant magnetic field, there are two convenient choices of the vector
potential. These are $A_{\mu}=(0,By/2,-Bx/2)$ and $A_{\mu}=(0,0,-Bx)$, and are
termed the symmetric and asymmetric gauge, respectively.
In the first case the Hamiltonian commutes with the angular momentum operator
and the solutions are given by Laguerre polynomials. 
In the second case we have $[H,p_{y}]=0$ and the solutions are 
Hermite polynomials (see below).
We have chosen the asymmetric gauge and
the Dirac equation then takes the form
\begin{equation}
\label{eq:d2}
\left(\begin{array}{cc}
i\frac{\partial}{\partial t}-m&i\frac{\partial}{\partial x}-ieBx+
\frac{\partial}{\partial y} \\
-i\frac{\partial}{\partial x}-ieBx+\frac{\partial}{\partial y} &
-i\frac{\partial}{\partial t}-m\\
\end{array}\right)\psi ({\bf x},t) =0.
\end{equation}
Here $\psi ({\bf x},t)$ is a two component spinor.
Since the Hamiltonian commutes with $p_{y}$, we can write the wave functions as
\begin{equation}\psi_{\kappa}({\bf x},t)=\exp (-iEt+iky)
\left(\begin{array}{c}
f_{\kappa}(x) \\
g_{\kappa}(x)  \\
\end{array}\right),\hspace{1cm}
\end{equation}
where ${\kappa}$ denotes all quantum numbers necessary in order to 
completely
characterize the solutions.
Inserting this into Eq.~(\ref{eq:d2}) one obtains
\begin{equation}
\label{eq:upd}
\left(\begin{array}{cc}
E-m&-\xi_{+} \\
\xi_{-}&-E-m\\
\end{array}\right)=
\left(\begin{array}{c}
f_{\kappa}(x) \\
g_{\kappa}(x) \\
\end{array}\right),
\end{equation}
where
\begin{equation}
\label{eq:xi}
\xi_{\pm}=-i\partial_{x}\mp i(k-eBx).
\end{equation}
The equation for $f_{\kappa}(x)$ is readily found from Eq.~(\ref{eq:upd}):
\begin{equation}
\label{eq:fx}
\left ( E^{2}-m^{2} -\xi_{+}\xi_{-}\right ) f_{\kappa}(x)=0.
\end{equation}
The eigenfunctions of $\xi_{+}\xi_{-}$, provided that $eB>0$, 
are~\cite{koba}
\begin{equation}
\label{eq:inn}
I_{n,k}=(\frac{eB}{\pi})^{\frac{1}{4}}\exp\left [\, -\frac{1}{2}
(x-\frac{k}{eB} )^{2}eB\,\right ]\frac{1}{\sqrt{n!}}H_{n}
\left [\,  \sqrt{2eB} (x-\frac{k}{eB})\,\right ].
\end{equation}
Here,
$H_{n}(x)$ is the $n$th Hermite polynomial.
Furthermore, $I_{n,k}(x)$ is normalized to unity and  satisfies
\begin{eqnarray} \nonumber
\xi_{-}I_{n,k}(x)&=&-i\sqrt{2eBn}I_{n-1,k}(x), \\  \nonumber
\xi_{+}I_{n,k}(x)&=&i\sqrt{2eB(n+1)}I_{n+1,k}(x).
\end{eqnarray}
Combining eqs. (\ref{eq:fx}) and (\ref{eq:inn}) yields
\begin{equation}
f_{\kappa}(x)=I_{n,k}(x),\hspace{1cm}E^{2}=m^{2}+2eBn.
\end{equation}
The function $g_{\kappa}(x)$ satisfies
\begin{equation}
g_{\kappa}(x)=\frac{\xi_{-}}{E+m}f_{\kappa}(x),
\end{equation}
implying that
\begin{equation}
g_{\kappa}(x)=-i\sqrt{2eBn}I_{n-1,k}(x).
\end{equation}
The normalized eigenfunctions become
\begin{equation}
\label{eq:eigenm}
\psi_{n,k}^{(\pm)} ({\bf x},t)=\exp (\mp iE_{n}t+iky)\sqrt{\frac{E_{n}
\pm m}{2E_{n}}}
\left(\begin{array}{c}
I_{n,k}(x) \\
\frac{\mp i\sqrt{2eBn}}{E_{n}\pm m}I_{n-1,k}(x) \\
\end{array}\right),
\end{equation}
where $n=0,1,2,...$, $E_{n}=\sqrt{m^{2}+2eBn}$ and $\psi_{n,k}^{(\pm)} 
({\bf x},t)$
are positive and negative energy solutions, respectively. 
Note that $\psi^{(-)}_{0,k}({\bf x},t)=0$ and that we have defined 
$I_{-1,k}(x)\equiv 0$.
The spectrum is therefore asymmetric and this asymmetry is intimately 
related
to the induced vacuum charge, as will be shown in subsection~\ref{ladningi}. 
In Fig.~\ref{spec} a) 
we have shown the spectrum for $m>0$ and  in Fig.~\ref{spec} b) for 
$m<0$. 
\\ \\
The field may now be expanded in the complete set of eigenmodes:
\begin{equation}
\label{expa}
\Psi ({\bf x},t)=\sum_{n=0}^{\infty}\int \frac{dk}{2\pi} \left [\,b_{n,k}
\psi_{n,k}^{(+)} ({\bf x},t)+ d_{n,k}^{\,\ast}\psi_{n,k}^{(-)} 
({\bf x},t)\,\right ]. 
\end{equation}
Quantization is carried out in the usual way by promoting the Fourier 
coefficients to operators satisfying
\begin{equation}
\{b_{n,k},b_{n^{\prime},k^{\prime}}^{\dagger}\}=\delta_{n,n^{\prime}}
\delta_{k,k^{\prime}},\hspace{1cm}\{d_{n,k},d_{n^{\prime},
k^{\prime}}^{\dagger}\}=\delta_{n,n^{\prime}}\delta_{k,k^{\prime}},
\end{equation}
and all other anti-commutators being zero.
\subsection{The Fermion Propagator}
\heading{Particles in External Fields}{The Fermion Propagator}
In the previous section we solved the Dirac equation and with the wave 
functions at hand, we can construct the propagator. In vacuum it is
defined by
\begin{equation}
iS_{F}(x^{\prime},x)=\langle 0\!\mid T\Big [\,\Psi ({\bf x}^{\prime},
t^{\prime})\overline{\Psi} ({\bf x},t) \,\Big ]\mid\!0\rangle,
\end{equation}
where $T$ denotes time ordering. By use of the expansion (\ref{expa}) 
one finds
\begin{equation}
iS_{F}(x^{\prime},x)=\sum_{n=0}^{\infty}\int \frac{dk}{2\pi}\left 
[\, \theta (t^{\prime}-t)\psi^{(+)}_{n,k}({\bf x}^{\prime},t^{\prime})
\overline{\psi}_{n,k}^{(+)}({\bf x},t)-\theta (t-t^{\prime})
\psi^{(-)}_{n,k}({\bf x}^{\prime},t^{\prime})
\overline{\psi}_{n,k}^{(-)}({\bf x},t)\,\right ].
\end{equation}
The step function has the following integral representation
\beq
\theta (t^{\prime}-t)=\frac{1}{2\pi i}\int\frac{e^{-\omega (t^{\prime}-t)}}{\omega-i\epsilon}d\omega.
\eeq
After some purely algebraic manipulations, we obtain
\bqa\nonumber
\label{fp}
S_{F}(x^{\prime},x)_{ab}&=&\frac{1}{4\pi^{2}}\sum_{n=0}^{\infty}
\int dkd\omega \frac{E_{n}+m}{2E_{n}}
\exp\left [\,-i\omega (t^{\prime}-t)+ik(y^{\prime}-y)\,\right ]\times\\
&&\frac{1}{\omega^{2}-E^{2}_{n}+i\varepsilon}S_{ab}(n,\omega,k).
\eqa
Here $S_{ab}(n,\omega,k)$ is the matrix
\begin{equation}
\label{eq:matrix}
\left(\begin{array}{cc}
(m+\omega )I_{n,k}(x^{\prime})I_{n,k}(x)&
-i\sqrt{2eBn}I_{n,k}(x^{\prime})I_{n-1,k}(x)\\
-i\sqrt{2eBn}I_{n-1,k}(x^{\prime})
I_{n,k}(x)&
(m-\omega )I_{n-1,k}(x^{\prime})I_{n-1,k}(x)\\
\end{array}\right).
\end{equation}
At finite temperature and chemical potential we write the thermal 
propagator as
(see Ref.~\cite{per} for details)
\begin{equation}
\langle S_{F}(x^{\prime},x)\rangle_{\beta ,\mu}=S_{F}(x^{\prime},x)+
S_{F}^{\beta ,\mu}(x^{\prime},x).
\end{equation}
The thermal part of the propagator is 
\begin{equation}
\label{eq:thermprop}
iS_{F}^{\beta,\mu}(x^{\prime},x)=-\sum_{n=0}^{\infty}\int 
\frac{dk}{2\pi}\left [\,f^{+}_{F}(E_{n})
\psi^{(+)}_{n,k}({\bf x}^{\prime},t^{\prime})
\overline{\psi}_{n,k}^{(+)}({\bf x},t)-f^{-}_{F}(E_{n})
\psi^{(-)}_{n,k}({\bf x}^{\prime},t^{\prime})
\overline{\psi}_{n,k}^{(-)}({\bf x},t)  \,\right ],
\end{equation}
where
\begin{equation}
f^{(+)}_{F}(\omega)=\frac{1}{\exp \beta(\omega-\mu)+1},
\hspace{1cm}f^{(-)}_{F}(\omega)=
1-f^{(+)}_{F}(-\omega)=\frac{1}{\exp \beta(\omega+\mu)+1}.
\end{equation}
This may be rewritten as
\begin{equation}
S_{F}^{\beta,\mu}(x^{\prime},x)=\frac{i}{2\pi}\sum_{n=0}^{\infty}\int 
dkd\omega\,\exp ik(y^{\prime}-y)
\exp i\omega(t^{\prime}-t)f_{F}(\omega)
\delta (\omega^{2}-E^{2}_{n}-i\varepsilon)S_{ab}
(n,\omega,k).
\end{equation}
Here
\beq
f_{F}(\omega)=\theta (\omega)F^{(+)}_{F}(\omega )
+\theta (-\omega)F_{F}^{(-)}(\omega ).
\eeq
As noted in Ref.~\cite{per}, one is not restricted to use equilibrium 
distributions in this approach. Single particle non-equilibrium  
distributions
may be more appropriate if e.g. an electric field has driven 
the system
out of equilibrium.
\subsection{The Effective Action}
\begin{figure}[htb]
\begin{center}
\mbox{\psfig{figure=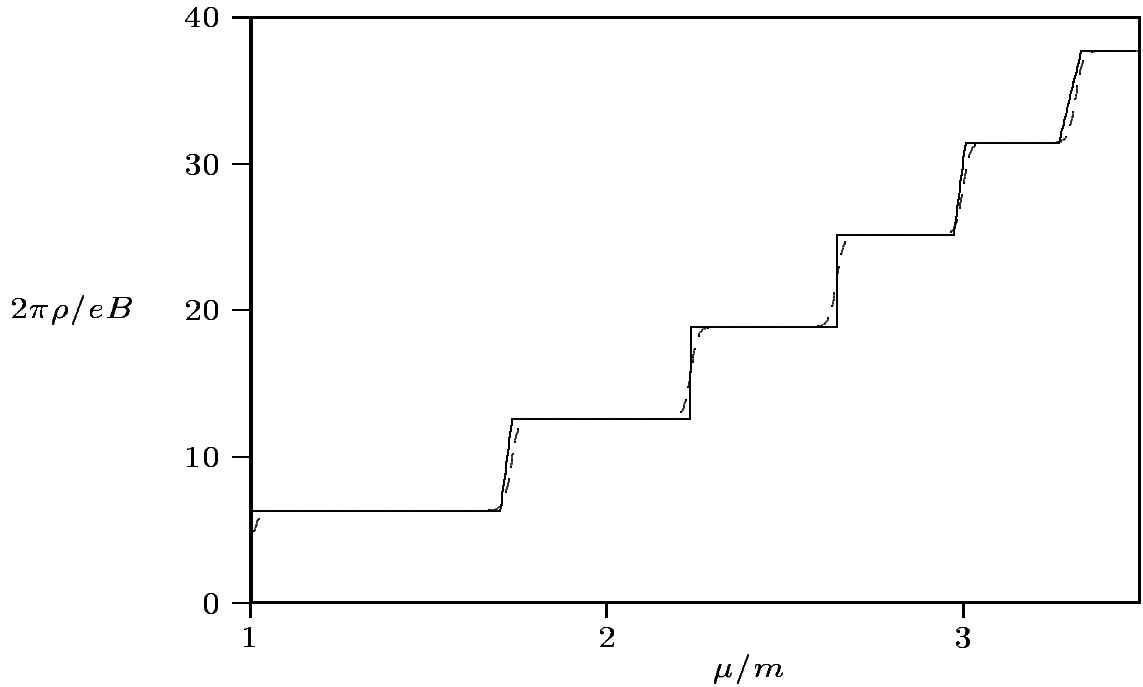,width=15cm,height=9cm}}
\end{center}
\caption[The density in units of $eB/2\pi$ as a 
function of $\mu/m$ for $T/m=1/1000$ (solid line) and for $T/m=1/100$ 
(dotted line). $eB/m^2=1$.]{\protect The density in units of $eB/2\pi$ as a 
function of $\mu/m$ for $T/m=1/1000$ (solid line) and for $T/m=1/100$ 
(dotted line). $eB/m^2=1$.}
\label{tettleik}
\end{figure}

\heading{Particles in External Fields}{The Effective Action}
The generating functional for fermionic Greens functions in an external 
magnetic
field may be written as a path integral:
\begin{equation}
Z(\eta,\overline{\eta},A_{\mu})=\int {\cal D}\psi\,{\cal D}
\overline{\psi}\exp 
\left [\,i\int d^{\,3}x \,\left (-\frac{1}{4}F_{\mu\nu}F^{\mu\nu}+
\overline{\psi}(iD\!\!\!\!/ -m)\psi-\overline{\eta}\psi+\overline{\psi}\eta 
\right ) 
\,\right ].
\end{equation}
The functional integral describes the interaction of fermions with a 
classical
electromagnetic field. It includes the effects of all virtual 
electron-positron
pairs, but virtual photons are not present. Taking this into account at
the one-loop simply amounts to including a temperature dependent, but field
independent term in ${\cal L}_{\mbox{\scriptsize eff}}$.\\ \\
The fermion field can be integrated over since the functional integral 
is Gaussian:
\begin{equation}
Z(\eta,\overline{\eta},A_{\mu})=\det [\,i(iD\!\!\!\!/-m)\,]\exp 
\Big [\,i\int d^{\,3}x\,\,\big [\,-\frac{1}{4}F_{\mu\nu}F^{\mu\nu}+
\int d^{\,3}y \overline{\eta}(x)S_{F}(x,y)\eta (y)\,\big ]\,\Big ].
\end{equation}
Taking the logarithm of $Z(\eta,\overline{\eta},A_{\mu})$ with vanishing 
sources gives the effective action 
\begin{equation}
\label{seff}
S_{\mbox{\scriptsize eff}}=
\int d^{\,3}x \left [\,-\frac{1}{4}F_{\mu\nu}F^{\mu\nu}\,\right ] 
-i\,\mbox{Tr}\log
\left[\,i(iD\!\!\!\!/-m)\, \right ].
\end{equation}
Note that we have written $\log\det =\mbox{Tr}\,\log$ by the use of a complete 
orthogonal
basis and that the trace is over space-time as well as spinor indices.  
Differentiating Eq.~(\ref{seff}) with respect to $m$ yields
\begin{equation}
\frac{\partial {\cal L}_{1}}{\partial m}=i\,\mbox{tr}S_{F}(x,x).
\end{equation}
The trace is now over spinor indices only.
By calculating the trace of the propagator and integrating this 
expression with
respect to $m$ thus yields the one-loop contribution to the effective 
action.
This method has been previously applied by Elmfors 
{\it et al.}~\cite{per} in 3+1 dimensions. \\ \\
The above equation may readily be 
generalized to finite temperature, where we separate the vacuum 
contribution in the effective action
\begin{equation}
{\cal L}={\cal L}_{0}+{\cal L}_{1}+{\cal L}^{\beta,\mu}
\equiv {\cal L}_{0}+{\cal L}_{\mbox{\scriptsize eff}}
\end{equation}
where ${\cal L}_{0}$ is the tree level contribution, and
\begin{equation}
\frac{\partial {\cal L}_{\mbox{\scriptsize eff}}}
{\partial m}=i\,\mbox{tr}\left [ \,S_{F}(x,x) 
+S_{F}^{\beta ,\mu}(x,x)\,\right ].
\end{equation} 
Using eqs. (\ref{fp}) and (\ref{eq:matrix}) a straightforward 
calculation
gives for the vacuum contribution
\begin{eqnarray}\nonumber
\mbox{tr}S_{F}(x,x)&=&\frac{1}{4\pi^2}\sum_{n=0}^{\infty}\int  
\frac{dk\,d\omega}{\omega^{2}-E^{2}_{n}+i\varepsilon}
\left [\, m\left (I_{n,k}^{2}(x)+I_{n-1,k}^{2}(x)\right )+
\omega\left (I_{n,k}^{2}(x)-I_{n-1,k}^{2}(x)\right ) 
\,\right ] \\ \nonumber
&=&-\frac{i}{2\pi}\sum_{n=1}^{\infty}\int dk \frac{m}{E_{n}}
I_{n,k}^{2}(x) - \frac{ieB}{4\pi}\\ 
&=&-\frac{ieB}{2\pi}\sum_{n=1}^{\infty}\frac{m}{E_{n}} - \frac{ieB}{4\pi}.
\end{eqnarray}
Integrating this expression with respect to $m$ yields
\begin{equation}
{\cal L}_{1}=\frac{eB}{2\pi}\sum_{n=1}^{\infty}\sqrt{m^{2}+2eBn}+
\frac{eBm}{4\pi}.
\end{equation}
The divergence may be sidestepped by using the integral representation 
of the
gamma function~\cite{tab} and
subtract a constant to make ${\cal L}_{1}$ vanish for $B=0$,
\begin{equation}
\label{eq:lvac}
{\cal L}_{1}=-\frac{1}{8\pi^{\frac{3}{2}}}\int_{0}^{\infty}
\frac{ds}{s^{\frac{5}{2}}}
\exp (-m^{2}s)\left [\, eBs\coth (eBs)-1\,\right ].
\end{equation}
This result 
calls for a few comments. We have chosen a gauge, where
$A_0=0$. However, we could equally well have chosen $A_{0}$ to be a nonzero
constant. This would
give rise to
an additional term in the effective action
\beq
\delta {\cal L}_1=-\frac{m}{|m|}\frac{e^{2}}{4\pi}A_0B. 
\eeq
This is simply the gauge dependent Chern-Simons term, whose existence
first was demonstrated by Redlich~\cite{reddik}.\\ \\
In the 
following, we shall only consider $m>0$,
except
for subsection~\ref{ladningi}. Similar results for $m<0$ 
can, of course, be obtained by the same methods. \\ \\
The finite temperature part of the effective action is calculated analogously
using the thermal part of the propagator~(\ref{eq:thermprop}). 
\begin{eqnarray}\nonumber
{\cal L}^{\beta, \mu}&=&\frac{TeB}{2\pi}\sum_{n=1}^{\infty}
\Big [\,\log [\,1+\exp -\beta(E_{n}-\mu)\,]+
\log [\,1+\exp -\beta (E_{n}+\mu )\,]\,
\Big ] \\ 
&&+\frac{TeB}{2\pi}\log \left [ \,1+\exp -\beta(m-\mu)\,\right].
\end{eqnarray}
Letting $B\rightarrow 0$ it can be shown that one obtains the pressure
of a gas of noninteracting electrons and positrons:
\bqa\nonumber
\label{eq:nullb}
{\cal L}_{0}^{\beta , \mu}&=&\frac{T}{2\pi}\int_{m}^{\infty}EdE
\Big [\,\log [\,1+\exp -\beta(E-\mu)\,]+
\log  [\,1+\exp -\beta(E+\mu)  \,]\,\Big ] \\ \nonumber
&=&-\frac{mT^2}{2\pi}\Big[\mbox{Li}_2(-\lambda e^{-\beta m})+
\mbox{Li}_2(-\lambda^{-1} e^{-\beta m})\Big]
-\frac{T^3}{2\pi}\Big[
\mbox{Li}_3(-\lambda e^{-\beta m})+
\mbox{Li}_3(-\lambda^{-1} e^{-\beta m})\Big].
\\&&
\eqa
Here, $\lambda=e^{\beta\mu}$ is the fugacity and
$\mbox{Li}_n(x)$ is the polylogarithmic function of order $n$:
\begin{equation}
\mbox{Li}_n(x) =\sum_{k=1}^{\infty}\frac{x^{k}}{k^{n}}.
\end{equation}
In the following we restrict ourselves to the case $\mu >0$.
Analogous
results can be obtained for $\mu <0$.  \\ \\
In the zero temperature limit of ${\cal L}^{\beta ,\mu}$ one gets
\begin{equation}
\label{eq:lo}
{\cal L}^{\beta,\mu}=\frac{eB}{2\pi}\sum_{n=0}^{\,\prime}(\mu-E_{n}),
\end{equation}
where the prime indicates that the sum is restricted to integers less than
$(\mu^{2}-m^{2})/2eB$. \\ \\
Similarly, one may derive the density 
\begin{eqnarray}
\label{eq:density}
\rho =\frac{\partial {\cal L}^{\beta, \mu}}{\partial\mu} 
&=&\frac{eB}{2\pi}\sum_{n=1}^{\infty}
\Big [\frac{1}{\exp \beta(E_{n}-\mu)\,+1}
-\frac{1}{\exp \beta(E_{n}+\mu)\,+1}\Big ] \\ \nonumber
&+&\frac{eB}{2\pi}\frac{1}{\exp \beta(m-\mu)\,+1}.
\end{eqnarray}
At $T=0$ this reduces to
\begin{equation}
\label{eq:rho}
\rho = 
\frac{eB}{2\pi}\Big [\mbox{Int}\,(\frac{\mu^2 - m^2}{2eB})+1\Big ],\hspace{1cm}\mu>m,
\end{equation}
in accordance with the result of Zeitlin~\cite{zeit}.
>From Eq.~(\ref{eq:rho}) one immediately finds that the density as a 
function of chemical potential
for fixed magnetic field is a step
function. This is intimately related to the integer Hall effect as noted
in Ref.~\cite{zeit2}.
In Fig.~\ref{tettleik} we have plotted the density as a function of chemical 
potential for low temperatures (dashes line: $T/m=1/100$, solid line:
$T/m=1/1000$). 
One observes that
the sharp edges get smeared out as the temperature increases.
\subsection{Magnetization and the de Haas-van Alphen Effect} 
\begin{figure}[htb]
\begin{center}
\mbox{\psfig{figure=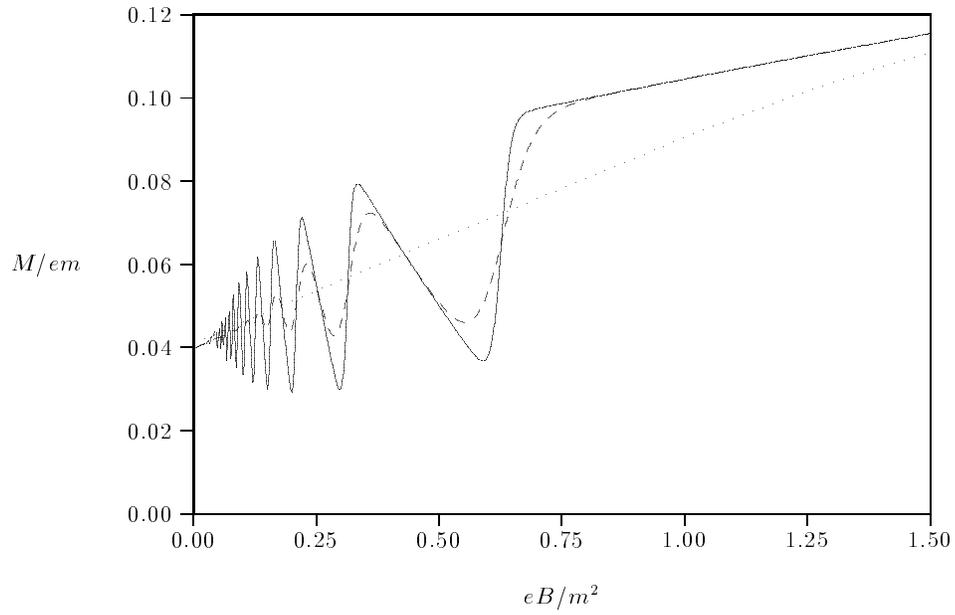,width=12.44cm,height=8.0cm}}
\end{center}
\caption[The magnetization in units of $em$ as a
function of $B$ in units of $m^2/e$ for different values of temperature. 
$\mu/m=3/2$.]{\protect The magnetization in units of $em$ as a
function of $B$ in units of $m^2/e$ for different values of temperature. 
$\mu/m=3/2$.}
\label{mag1}
\end{figure}

\begin{figure}[htb]
\begin{center}
\mbox{\psfig{figure=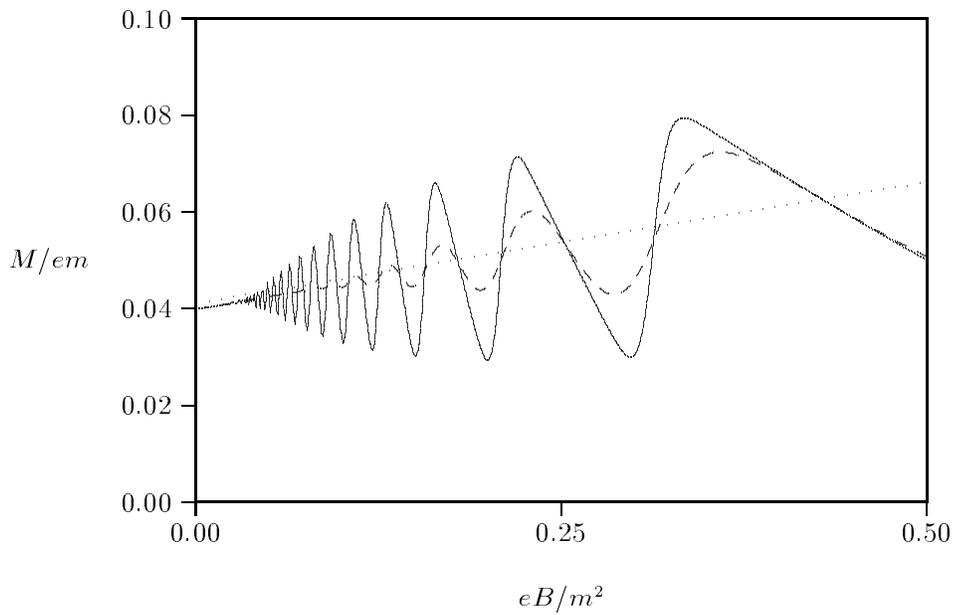,width=12.44cm,height=8.0cm}}
\end{center}
\caption[Magnification of the oscillatory region
in Fig. 1.4.]{\protect Magnification of the oscillatory region in Fig. 1.4.}
\label{mag2}
\end{figure}

\heading{Particles in External Fields}{Magnetization and the de Haas-van Alphen Effect}
In this section we study the physical content of the effective action
which was obtained in the previous section. In particular we investigate a
few limits to check the consistency of our calculations.\\ \\
The magnetization is defined by~\cite{per}
\begin{equation}
M=\frac{\partial {\cal L}_{\mbox{\scriptsize eff}}}{\partial B}.
\end{equation}
The vacuum contribution to the magnetization is obtained from 
Eq.~(\ref{eq:lvac})
\begin{equation}
\label{eq:magv}
M_{1}=-\frac{1}{8\pi^{\frac{3}{2}}}\int_{0}^{\infty}ds\frac{\exp \,
(-m^{2}s)}{s^{\frac{5}{2}}}\left [\,es\coth (eBs)- \frac{e^{2}Bs^{2}}
{\sinh^{2}(eBs)}\,\right ].
\end{equation}
For the thermal part of the magnetization we find 
\begin{eqnarray} \nonumber
\label{eq:magt}
M^{\beta,\mu}&=&\frac{Te}{2\pi}\sum_{n=1}^{\infty}\Big  [\, 
\log [\,1+\exp -\beta(E_{n}-\mu)\,]+\log [\,1+\exp -\beta(E_{n}+\mu)\,]
\,\Big] \\ \nonumber
&&+\frac{Te}{2\pi}\log \big [\, 1+\exp -\beta(m-\mu)\,\big] \\ 
&&-\frac{e^{2}B}{2\pi}\sum_{n=1}^{\infty}\frac{n}{E_{n}}
\Big [\, \frac{1}{\exp\beta (E_{n}-\mu)+1}+\frac{1}{\exp 
\beta (E_{n}+\mu)+1}\,\Big].
\end{eqnarray}
{\it Magnetization at zero temperature.}$\,\,$
In the zero temperature limit Eq.~(\ref{eq:magt}) reduces to
\begin{equation}
\label{eq:st}
M^{\beta,\mu}=\frac{e}{2\pi}\sum^{\prime}_{n=0}\left[\, 
\mu-E_{n}-\frac{eBn}{E_{n}} \, \right ], 
\end{equation}
where the sum again is restricted to integers less than
$(\mu^{2}-m^{2})/2eB$. The thermal part of the magnetization at zero 
temperature changes abruptly, when $(\mu^{2}-m^{2})/2eB$ increases by unity.
Thus, $M^{\beta,\mu}$ oscillates wildly, in particular is the limit
$B\rightarrow 0$ not well defined.
The strong field limit ($B\rightarrow\infty$) 
of $M^{\beta,\mu}$ is found to be
\beq
\frac{e}{2\pi}(\mu-m).
\eeq
In the weak $B$-field limit ($eB\ll \mu^{2}-m^{2}\ll m^{2}$)
the vacuum contribution becomes
\begin{equation}
M_{1}=-\frac{e^{2}B}{12\pi^{3/2}}\int_{0}^{\infty}ds\frac{
\exp (-m^{2}s)}{s^{\frac{1}{2}}}=-\frac{e^{2}B}{12\pi |m|}.
\end{equation}
This agrees with the results of Ref.~\cite{can}. 
In order to get the strong field limit $(\,eB\gg m^{2})$ of the 
vacuum contribution, we scale out
$eB$ and take $eB\rightarrow \infty$ in the remainder. This gives
\begin{equation}
{\cal L}_{1}\propto (eB)^{\frac{3}{2}}\Rightarrow M_{1}\propto 
e^{\frac{3}{2}}\sqrt{B}.
\end{equation}
Vacuum effects contribute to the magnetization proportional to the 
square root
of $B$. This should be compared 
with the corresponding result in $3+1d$, where the magnetization goes like 
$B\log (\frac{B}{m^{2}})$~\cite{per}. 
The thermal part of the magnetization was found to be 
$M^{\beta,\mu} = e(\mu - m)/2\pi$. Hence, the vacuum contribution dominates,
exactly as in $3+1d$. For $\mu\neq m$ we see that the thermal part of the 
magnetization is nonzero. From Eq.~(\ref{eq:rho}) one obtains 
$\rho =eB/2\pi$, so the nonzero magnetization is a consequence of the fact 
that the density increases
as the magnetic field increases (since all particles are in the ground state).
\\ \\
{\it Magnetization at finite temperature.}$\,\,$
In Fig.~\ref{mag1} we have displayed the total magnetization as a 
function 
of the external magnetic field for different values of the 
temperature ($\mu /m=3/2$,\, $T/m$ =1/150 solid line,\, 1/50
dashed line,\, 1/5 dotted line). 
Fig.~\ref{mag2} is a magnification
of Fig.~\ref{mag1} in the oscillatory region.

The fermion gas exhibits the de Haas-van Alphen oscillations for small 
values of 
the magnetic field. 
These oscillations have been observed in many
condensed matter systems~\cite{appli}, and they were first
observed experimentally in 1930~\cite{dehas}.
It is a direct consequence of the Pauli exclusion principle and the 
discreteness of the spectrum.

We also note that the magnetization approaches a nonzero value as 
$B\rightarrow 0$. More specifically, 
in Ref.~\cite{tortese} it is demonstrated that the limit equals
\beq
M^{\beta,\mu}=\frac{Te}{2\pi}\ln\Bigg[\left(1 + e^{-\beta(m-\mu)}\right) - \left(1 + e^{-\beta(m+\mu)}\right)\Bigg].
\eeq
Some comments are in order. It is perhaps somewhat surprising that the 
magnetization is nonzero in this limit. One should, however, bear in mind
that the sign of $m$ uniquely determines the spin of the particles 
(and antiparticles), implying that the system under investigation 
consists 
entirely of either spin up or spin down particles. This is not the case 
in $3+1d$, where the representations
characterized by the sign of $m$ are equivalent. 
By summing over $\pm m$, or equivalently, by using four component
spinors, one finds a vanishing magnetization as $B$ goes to zero, 
exactly as in $3+1$ dimensions.

\begin{figure}[htb]
\begin{center}
\vspace{-8.5cm}
\mbox{\psfig{figure=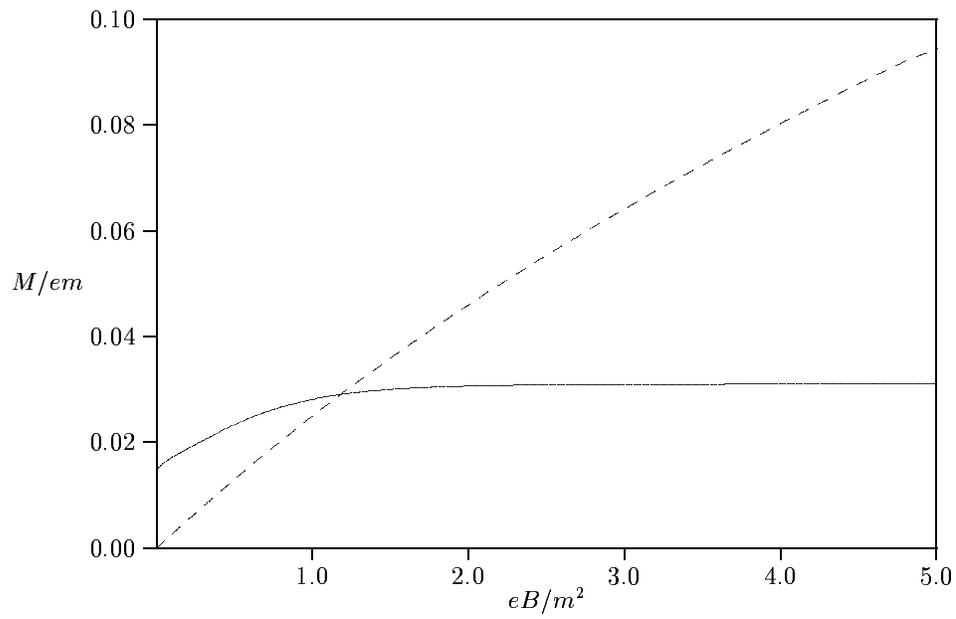,width=12.44cm,height=8cm}}
\end{center}
\caption[Thermal and vacuum contributions to the
magnetization for a Fermi gas.]{\protect Thermal and vacuum contributions to the
magnetization for a Fermi gas.}
\label{f4}
\end{figure}

Finally, we have displayed the modulus of the
vacuum part as well as the thermal part of the
magnetization over a rather broad interval of values of $eB/m^2$ in 
Fig.~\ref{f4}. ($T/m=1/5$ and
$\mu/m=1.1$). The thermal contribution saturates for values of the field
where the vacuum contribution starts to dominate. The reason is that all
particles are in the lowest Landau level for high values of $B$, and that
the energy of this level is independent of the magnetic field. 
 \\ \\
{\it High Temperature Limit $(\,T^{2}\gg m^{2}\gg eB,\mu=0)$.}$\,\,$ 
The high temperature limit is rather trivial. From a physical point of 
view, 
one expects that ${\cal L}^{\beta ,\mu}$ approaches the thermodynamic 
potential of a gas of noninteracting 
particles of mass $m$. Indeed, in this limit, one may recover 
Eq.~(\ref{eq:nullb}) by treating $n$ as a continuous variable.

\newpage
\clearpage
\vspace{-4cm}
We would also like to describe the system in terms of constant charge density.
At zero temperature it is not possible to invert Eq.~(\ref{eq:rho}) 
to write the chemical
potential as a function of density, since the step function is not one-to-one.
However, $\mu$ can be interpreted as the Fermi energy at $T=0$ (as long as the
highest occupied Landau level is not completely filled), so one can 
immediately write down the chemical potential as a function of density:
\begin{equation}
\label{eq:my}
\mu =\sqrt{m^{2}+2eB\,\mbox{Int}\,(\frac{2\pi\rho}{eB})}\,\,.
\end{equation}
We should also point out that at $T>0$ there is a one-to-one correspondence
between density and chemical potential (Fig.~\ref{tettleik}), so one can 
invert Eq.~(\ref{eq:density}) numerically.

We have used Eq.~(\ref{eq:my}) to make a plot of the thermal part of the 
magnetization as a function of magnetic field at constant density and at 
$T=1/100$.
The resulting curve is displayed in Fig.~\ref{Haas,T=0}.
 
\begin{figure}[htb]
\begin{center}
\mbox{\psfig{figure=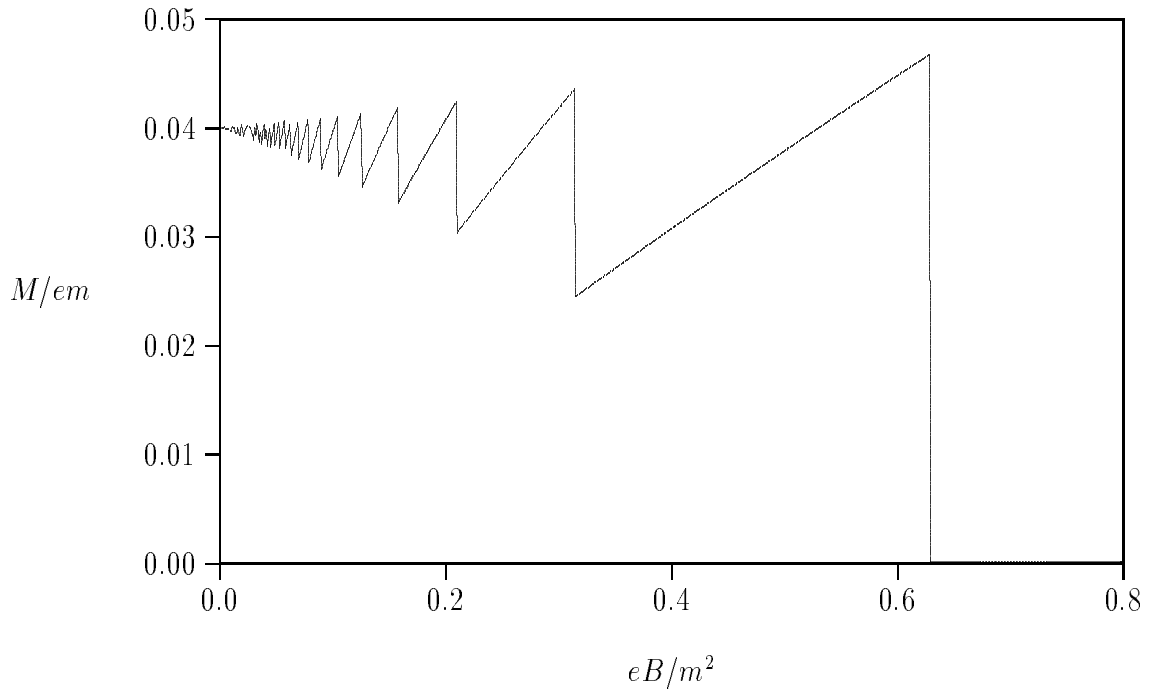,width=15cm,height=9cm}}
\end{center}
\caption[The magnetization in units of $em$ as a
function of $B$ in units of $m^2/e$ at $T=1/100$ and constant density $\rho/m^2=1$.]{\protect The magnetization in units of $em$ as a
function of $B$ in units of $m^2/e$ at $T=1/100$ and constant density $\rho/m^2=1$.}
\label{Haas,T=0}
\end{figure}
The de Haas-van Alphen oscillations are seen to be present for low
temperatures and weak magnetic fields. 
Furthermore, it is seen that the magnetization is zero
for large magnetic fields. This can be understood 
from the following physical argument: For large $B$-fields all particles 
are in the lowest Landau level and $\mu = m$ (recall that the degeneracy 
increases linearly with $B$). 
The energy of the single particle ground
state is independent of the external field 
($E_{0}=m$)
, so increasing 
$B$ cannot
lead to an increase in ${\cal L}^{\beta,\mu}$, when the charge density 
(and therefore the particle density) is held constant.
Hence, the contribution to the magnetization from real thermal particles
vanishes in the strong field limit.\\ \\
\subsection{Induced Vacuum Charges and Currents}
\heading{Particles in External Fields}{Induced Vacuum Charges and Currents}
In this subsection we calculate the vacuum expectation value of the induced
charge and 
current densities. Such calculations have been carried out in other 
contexts,
e.g. in connection with magnetic flux strings (see Ref.~\cite{lein}). 
We shall employ the most commonly used definition of the current operator
which can be shown to measure the spectral asymmetry  relative to the 
spectrum of free Dirac particles.
\begin{equation}
j^{\mu}\,(x)=\frac{e}{2}\left [\,\overline{\Psi}_{\alpha}(x),
\left ( \gamma^{\mu}\Psi (x)_{\alpha}\right) \,\right ].
\end{equation}
Using the complete set of eigenmodes as given by Eq.~(\ref{eq:eigenm}),
a straightforward calculation yields 
\begin{equation}
\label{eq:cs}
\langle\,\rho \,(x)\,\rangle =-\frac{m}{\mid\! m\!\mid}\frac{e^{2}B}{4\pi}.
\end{equation}
Eq.~(\ref{eq:cs}) is simply the Chern-Simons relation.
It has previously been obtained by e.g. Zeitlin~\cite{zeit} 
using the
proper time method. 
This  result has the following physical interpretation: As we turn the
magnetic field on, an unpaired energy level $E=m$ emerges 
(in the case $m>0$).
The number of positrons therefore gets reduced
relative to the free case. This can be interpreted as the appearance of
electrons and results in a negative charge density. For $m<0$ a 
similar argument applies.

A corresponding calculation of $\langle\, {\bf j}\,(x)\, \rangle$ reveals that 
the
induced current vanishes. This result should come as no surprise due to
translational symmetry of the system. A nonvanishing vacuum current 
would 
arise in the presence of an external electric field and is then 
attributed to the drift
of the induced vacuum charge.
\subsection{Conductivity and the Integer Quantum Hall Effect}
\heading{Particles in External Fields}{Conductivity and 
the Integer Quantum Hall Effect}
Let us next consider the conductivity. According to Ref.~\cite{zeit3}  
the expression for the components of the 
conductivity $\sigma_{ij}$ can be expressed in terms
of the polarization tensor $\Pi_{\mu\nu}(k_0,{\bf k})$:
\bqa
\sigma_{ij}&=&\left.i\frac{\partial \Pi_{0i}(0,{\bf k})}{\partial k_j}\right|_{{\bf k}\rightarrow 0},%
\eqa
and follows from linear response theory. Moreover, by considering
the functional derivative of the effective action with respect to $A_{\mu}$
one may deduce that~\cite{zeit3}
\beq
\Pi_{0i}(k_0,{\bf k}\rightarrow 0)=-ie\epsilon_{ij}k_j\frac{\partial \rho}{\partial B}.
\eeq
Combining the above equations, one may infer that
\beq
\sigma_{ij}=\sigma\epsilon_{ij}=\epsilon_{ij}e\frac{\partial \rho}{\partial B}.
\eeq
Thus, the conductivity is Hall like.
Using Eq.~(\ref{eq:density}) and including the contribution from the induced 
vacuum charge, which was calculated in the previous section, we obtain
\begin{eqnarray} \nonumber
\sigma &=&-\frac{e^{2}}{4\pi}+\frac{e^{2}}{2\pi}\sum_{n=1}^{\infty}
\Big [\frac{1}{1+\exp \beta(E_{n}-\mu)\,}
-\frac{1}{1+\exp \beta(E_{n}+\mu)\,}\Big ] +\frac{e^{2}}{2\pi}\frac{1}{1+\exp \beta(m-\mu)\,}\\
&&+\frac{e^{3}B}{2\pi T}\sum_{n=1}^{\infty}\frac{n}{E_{n}}\Big [\frac{\exp \beta(E_{n}-\mu)\,}{[1+\exp \beta(E_{n}-\mu)\,]^{2}}-\frac{\exp \beta(E_{n}+\mu)\,}{[1+\exp \beta(E_{n}+\mu)\,]^{2}} \Big ].
\end{eqnarray}
Letting $T\rightarrow 0$ one finds
\begin{equation}
        \sigma = -\frac{e^2}{4\pi}+ 
        \frac{e^2}{2\pi}
        \mbox{Int}\left[
        \frac{\mu^2 -m^2}{2eB}    \right].
\end{equation}
We thus see that the conductivity is a step function for $T=0$. The system
therefore contains the integer Quantum Hall effect. This was also noted by
Zeitlin~\cite{zeit}.
The generalization of Zeitlin's result to finite temperature is new.

%


\section{Bosons in a Constant Magnetic Field}
\heading{Particles in External Fields}{Bosons in a Constant Magnetic Field}
In this section we focus the attention on bosons in a constant magnetic
field. We calculate the effective action and derive the magnetization.
We point out the differences between the bosonic and fermionic results.
Finally, we generalize to  constant field strengths and study
pair production in a purely electric field.
\subsection{The Klein-Gordon Equation}
\heading{Particles in External Fields}{The Klein-Gordon Equation}
Let us for the convenience of the reader briefly discuss the solutions to 
the Klein-Gordon equation in an external constant magnetic field. It  reads
\begin{equation}
[D_{\mu}D^{\mu}+m^{2}]\phi ({\bf x},t)=0,
\end{equation}
where $D_{\mu}=\partial_{\mu}+ieA_{\mu}$ is the covariant derivative and the 
metric is diag\,($1$,\,$-1$,\,$-1$).
We have again chosen the asymmetric gauge $A_{\mu}=(0,0,-Bx)$ and 
assume that the wave functions are in the form
\begin{equation}
\phi ({\bf x},t)=e^{-iEt+iky}f(x).
\end{equation} 
The differential equation for $f(x)$ then becomes
\begin{equation}
\xi_{-}\xi_{+}f(x)=(E^{2}-m^{2}+eB)f(x),
\end{equation}
where $\xi_{\pm}$ were defined in Eq.~(\ref{eq:xi}) and the eigenfunctions
of $\xi_-\xi_+$ were defined in Eq.~(\ref{eq:inn}).
The normalized eigenfunctions of the Klein-Gordon equation are 
\begin{equation}
\label{eq:in}
\phi_{n,k} ({\bf x},t)=e^{-iEt+iky}(\frac{eB}{\pi})^{\frac{1}{4}}\exp\left [ -\frac{1}{2}(x-\frac{k}{eB} )^{2}eB\right ]\frac{1}{\sqrt{n!}}H_{n}\left [  \sqrt{2eB} (x-\frac{k}{eB})\right ],
\end{equation}
with corresponding eigenvalues $E_{n}=\sqrt{m^{2}+(2n+1)eB}$. 
The Klein-Gordon field can now expanded in the complete set of solutions:
\begin{equation}
\label{eq:expa}
\Phi ({\bf x},t)=\frac{1}{4\pi}\sum_{n=0}^{\infty}\int\frac{dk}{E_{n}}\Big[a_{n,k}\phi_{n,k}({\bf x},t)+b_{n,k}^{\ast}\phi_{n,k}^{\ast}({\bf x},t)\Big].
\end{equation}
Quantization is carried out as in the fermionic case by promoting the Fourier 
coefficients to operators. The only nonvanishing commutators are
\begin{equation}
[a_{n,k},a_{n^{\prime},k^{\prime}}^{\dagger}]=4\pi E_{n}\delta_{n,n^{\prime}}
\delta (k-k^{\prime}),\hspace{1cm}[b_{n,k},b_{n^{\prime},
k^{\prime}}^{\dagger}]=4\pi E_{n}\delta_{n,n^{\prime}}\delta (k-k^{\prime}).
\end{equation}
\subsection{Boson Propagators and the Effective Action}
\heading{Particles in External Fields}{Boson Propagators and the Effective Action}
The generating functional for bosonic Greens functions in an external magnetic
field may be written as a path integral in analogy with the fermionic case
\begin{equation}
Z(J,J^{\dagger},A_{\mu})=\int {\cal D}\phi\,{\cal D}\phi^{\dagger}\exp 
\left [i\int d^{\,3}x \left (-\frac{1}{4}F_{\mu\nu}F^{\mu\nu}+\phi^{\dagger}(D_{\mu}D^{\mu}+m^{2})\phi +J^{\dagger}\phi +\phi^{\dagger}J \right ) \right ].
\end{equation}
We integrate out the bosons in the functional integral and get a 
functional determinant:
\begin{equation}
Z(J,J^{\dagger},A_{\mu})=\det [i(D_{\mu}D^{\mu}+m^{2})]\exp 
\Big [i\int d^{\,3}x\,\,\big [-\frac{1}{4}F_{\mu\nu}F^{\mu\nu}+\int d^{\,3}y J^{\dagger}(x)\Delta_{F}(x,y)J (y)\big ]\Big ].
\end{equation}
Taking the logarithm of $Z(J,J^{\dagger},A_{\mu})$ with vanishing external
sources gives the effective action 
\begin{equation}
\label{eq:seff}
S_{\mbox{\scriptsize eff}}=\int d^{\,3}x \left [-\frac{1}{4}F_{\mu\nu}F^{\mu\nu}\right ] -i\,\mbox{tr}\log
\left[i(D_{\mu}D^{\mu}+m^{2}) \right ],
\end{equation}
where we have written $\log\det =\mbox{Tr}\,\log$ by the use of a complete orthogonal
basis.  The first term is denoted ${\cal L}_0$ and is the tree level 
contribution. For a constant magnetic field we have 
${\cal L}_0=-\frac{B^{2}}{2}$. 
Differentiating Eq.~(\ref{eq:seff}) with respect to $m^{2}$ yields
\begin{equation}
\label{eq:rel}
\frac{\partial {\cal L}_{1}}{\partial m^{2}}=-i\,\mbox{tr}\Delta_{F}(x,x).
\end{equation}
The next step is then to construct the boson propagator which in vacuum
is defined as
\begin{equation}
i\Delta_{F}(x^{\prime},x)=\langle 0\!\mid T\Big [\Phi ({\bf x}^{\prime},t^{\prime})
\Phi^{\dagger} ({\bf x},t) \Big ]\mid\!0\rangle.
\end{equation}
Here, $T$ denotes time ordering as usual. 
By use of the expansion~(\ref{eq:expa}) 
one finds
\begin{equation}
\Delta_{F}(x^{\prime},x)=-\frac{i}{4\pi^2}\sum_{n=0}^{\infty}\int \frac{dk}{E_{n}}\left [ \theta (t^{\prime}-t)\phi_{n,k}({\bf x}^{\prime},t^{\prime})\phi_{n,k}^{\ast}({\bf x},t)+\theta (t-t^{\prime})\phi_{n,k}({\bf x}^{\prime},t^{\prime})\phi_{n,k}^{\ast}({\bf x},t)\right ].
\end{equation}
After some algebraic manipulations and using the integral 
representation of the step function, we obtain
\begin{eqnarray}
\label{eq:fp}
\hspace{-0.8cm}\Delta_{F}(x^{\prime},x)&=&\frac{1}{4\pi^2}\sum_{n=0}^{\infty}\int \frac{dkd\omega}{\omega^{2}-E^{2}_{n}+i\varepsilon}\exp\left [-i\omega (t^{\prime}-t)+ik(y^{\prime}-y)\right ]I_{n}(x)I_{n}(x^{\prime}).
\end{eqnarray} 
The trace then becomes
\begin{eqnarray}
\mbox{tr}\Delta_{F}(x,x)&=&\frac{1}{4\pi^2}\sum_{n=0}\int\frac{dkd\omega}{\omega^{2}-E^{2}_{n}+i\varepsilon}I_{n}^{2}(x)\\ \nonumber
&=&-\frac{ieB}{4\pi}\sum_{n=0}^{\infty}\frac{1}{E_{n}}.
\end{eqnarray}
Integration with respect to $m^{2}$ gives the effective action:
\begin{equation}
{\cal L}_{1}=-\frac{Be}{2\pi}\sum_{n=0}^{\infty}\sqrt{m^{2}+(2n+1)eB}\,.
\end{equation}
Employing the integral representation of 
the $\Gamma$-function~\cite{tab}, we find
\begin{equation}
\label{eac}
{\cal L}_{1}=\frac{eB}{8\pi^{\frac{3}{2}}}\int_{0}^{\infty}\frac{ds}{s^{\frac{3}{2}}}e^{-m^{2}s}\Big [\frac{1}{\sinh (eBs)}-\frac{1}{eBs}\Big ].
\end{equation}
The above expression has been rendered finite by requiring
that ${\cal L}_{1}=0$ for $B=0$.
This result is in accordance with the leading term in the derivative 
expansion employed by Cangemi {\it et al.}~\cite{can}.\\ \\
At finite temperature and chemical potential we write the thermal propagator as
\begin{equation}
\langle \Delta_{F}(x^{\prime},x)\rangle_{\beta ,\mu}=\Delta_{F}(x^{\prime},x)+
\Delta_{F}^{\beta ,\mu}(x^{\prime},x).
\end{equation}
The thermal part of the propagator is 
\begin{equation}
\label{eq:thermpro}
i\Delta_{F}^{\beta,\mu}(x^{\prime},x)=-\frac{1}{4\pi}\sum_{n=0}^{\infty}\int \frac{dk}{E_{n}}\left [f^{+}_{B}(E_{n})\phi_{n,k}(x^{\prime})\phi^{\ast}_{n,k}(x)-f^{-}_{B}(E_{n})\phi_{n,k}(x^{\prime})\phi^{\ast}_{n,k}(x)  \right ].
\end{equation}
Here $f^{(+)}_{B}(\omega)$ and $f^{(-)}_{B}(\omega)$ are the bosonic
equilibrium distributions:
\begin{equation}
f^{(+)}_{B}(\omega)=\frac{1}{\exp \beta(\omega-\mu)-1},\hspace{1cm}f^{(-)}_{B}(\omega)=\frac{1}{\exp \beta(\omega+\mu)-1}.
\end{equation}
Eq. (\ref{eq:rel}) is easily generalized to finite temperature. 
Writing ${\cal L}={\cal L}_{0}+{\cal L}_{1}+{\cal L}^{\beta ,\mu}
\equiv {\cal L}_0+{\cal L}_{\mbox{\scriptsize eff}}$, we have
\begin{equation}
\frac{\partial {\cal L}_{\mbox{\scriptsize eff}}}{\partial m^{2}}=-i\,\mbox{tr}\Big [\Delta_{F}(x^{\prime},x)+\Delta_{F}^{\beta ,\mu}(x^{\prime},x)\Big ].
\end{equation}
Straightforward calculations give the thermal part of the effective action
\begin{equation}
{\cal L}^{\beta ,\mu}=-\frac{TeB}{2\pi}\sum_{n=1}^{\infty}
\Big [\,\log [\,1-\exp -\beta(E_{n}-\mu)\,]+
\log [\,1-\exp -\beta (E_{n}+\mu )\,]\,
\Big ]. 
\end{equation}
The limit $B\rightarrow 0$ is easily taken, and we find 
\bqa\nonumber
\!\!\!\!\!\!{\cal L}^{\beta ,\mu}&=&-\frac{T}{2\pi}\int_{m}^{\infty}DEE
\Big [\,\log [\,1-\exp -\beta(E-\mu)\,]+
\log [\,1-\exp -\beta (E+\mu )\,]\,
\Big ]\\
\!\!\!\!\!\!&=&\frac{mT^2}{2\pi}\Big[
\mbox{Li}_2(\lambda e^{-\beta m})+
\mbox{Li}_2(\lambda^{-1} e^{-\beta m})\Big]+
\frac{T^3}{2\pi}\Big[
\mbox{Li}_3(\lambda e^{-\beta m})+
\mbox{Li}_3(\lambda^{-1} e^{-\beta m})\Big]
\eqa
This is the minus the free energy for a gas of bosons, as expected. \\ \\
The limit $T\rightarrow 0$ is trivial in the bosonic case. There is no Fermi
energy, and all the particles are in the ground state. Hence
\beq
{\cal L}^{\beta ,\mu}=0.
\eeq
Recall that we work with the grand canonical ensemble, so the above result
implies that the pressure of the Bose gas vanishes. \\ \\
The high temperature limit equals the pressure of the Bose gas with 
$B=0$ as in the fermionic case. 
\subsection{Magnetization}
\begin{figure}[htb]
\begin{center}
\mbox{\psfig{figure=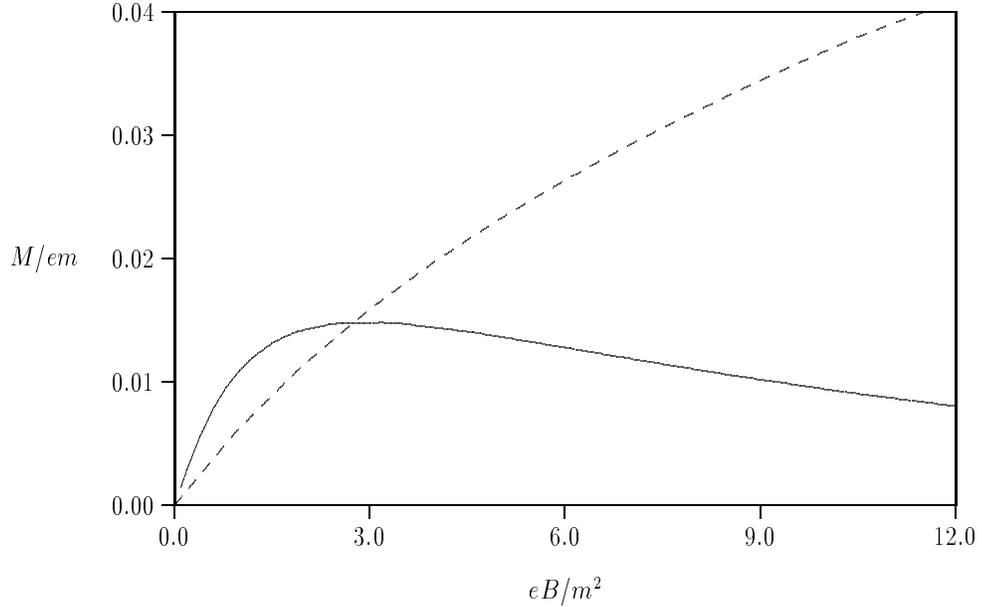,height=8cm,width=12.8cm}}
\end{center}
\caption[The vacuum and thermal contributions to the
magnetization for a Bose gas.]{\protect The vacuum and thermal contributions to the
magnetization for a Bose gas.}
\label{bose1}
\end{figure}

\heading{Particles in External Fields}{Magnetization}
The vacuum part of the magnetization becomes
\begin{equation}
\label{vacm}
M_{1}=\frac{1}{8\pi^{\frac{3}{2}}}\int_{0}^{\infty}\frac{ds}{s^{\frac{3}{2}}}e^{-m^{2}s}\Big [\frac{e}{\sinh (eBs)}-\frac{e^{2}Bs\cosh (eBs)}{\sinh^{2}(eBs)}\Big ].
\end{equation}
The thermal part is
\begin{eqnarray}\nonumber
\label{bmag}
M^{\beta\mu}&=&-\frac{Te}{2\pi}\sum_{n=1}^{\infty}
\Big [\,\log [\,1-\exp -\beta(E_{n}-\mu)\,]+
\log [\,1-\exp -\beta (E_{n}+\mu )\,]\,
\Big ] \\ 
&&-\frac{e^{2}B}{2\pi}\sum_{n=0}^{\infty}\frac{2n+1}{2E_{n}}\Big[\frac{1}{\exp [\beta(E_{n}-\mu)\,]-1}+\frac{1}{\exp [\beta(E_{n}+\mu)\,]-1} \Big].
\end{eqnarray}
Taking the weak field limit ($eB\ll m^{2}$) of Eq. (\ref{vacm}) yields
\begin{equation}
M_{1}=-\frac{e^{2}B}{24\pi^{3/2}}\int_{0}^{\infty}ds\frac{
\exp (-m^{2}s)}{s^{\frac{1}{2}}}=-\frac{e^{2}B}{24\pi |m|}.
\end{equation}
This is one half of the fermionic result.
In the strong field limit we find that the magnetization in the vacuum
sector goes like $e^{\frac{3}{2}}B^{\frac{1}{2}}$.
We have computed the vacuum and thermal parts of the magnetization numerically
for the  neutral Bose gas ($T/m=1$ and $\mu =0$).
The result is presented in Fig.~\ref{bose1}. Note that we have plotted the 
modulus
of the magnetization. 
We see that the thermal contribution to the
magnetization has a minimum, so the susceptibility changes sign. 
This was also observed in the corresponding system in $3+1d$ by Elmfors
{\it et al.} \cite{elm2}. The system thus changes from diamagnetic to 
paramagnetic behaviour. 
We also note that the thermal part
magnetization goes to zero as 
$B\rightarrow \infty$ as can be seen from Eq. (\ref{bmag}). This is in 
contrast with the fermionic case and stems from the
fact that the single-particle energies increases with the magnetic field.

\section{Bosons in Constant Electromagnetic Fields}
\heading{Particles in External Fields}{Bosons in Constant Electromagnetic Fields}
In the previous section we have considered the effective action for 
bosons in the presence of a constant magnetic field.
The formula can rather easily be generalized to the case of a 
constant electromagnetic field, without doing any actual calculations.
In 3+1 dimensions one can construct two independent Lorentz invariant
quantities of ${\bf E}$ and ${\bf B}$:
\beq
-\frac{1}{4}F_{\mu\nu}F^{\mu\nu}=\frac{1}{2}(E^{2}-B^{2}),
\hspace{1cm}^{\ast}\!F_{\mu\nu}F^{\mu\nu}={\bf E}\!\cdot\!{\bf B},
\eeq
where $^{\ast}\!F_{\mu\nu}=\frac{1}{2}
\epsilon_{\mu\nu\alpha\beta}F^{\alpha\beta}$. In 2+1 dimensions
there is only one such
quantity, namely $\frac{1}{2}(E^{2}-B^{2})$.
As a consequence of Lorentz invariance, one can simply let 
$B\rightarrow \sqrt{B^{2}-E^{2}}$ in the effective action.
Doing so, and expanding the formula in powers of 
$\sqrt{B^{2}-E^{2}}$, one finds
\beq
{\cal L}_{1}=\frac{e^{2}(E^{2}-B^{2})}{48\pi m}+
\frac{7e^{4}(E^{2}-B^{2})^{2}}{3840\pi m^{5}}+..
\eeq
The first term in the above expansion may be removed by redefining the 
gauge field.
This series expansion demonstrate the nonlinear
behaviour of the electromagnetic field, which is inherit in 
{\it quantum optics}, but is absent at the classical electromagnetism.
It is the two-dimensional (bosonic) counterpart of the famous
Euler-Heisenberg Lagrangian found as early as in 1936~\cite{euler}.

Effective Lagrangians and effective field theories is a major part of the
present thesis, and will be discussed at length in later chapters.
\subsection{Pair Production in an Electric Field}
\heading{Particles in External Fields}{Pair Production in an Electric Field}
In this section we calculate the effective action in the presence of a constant
external electric field $E$. 
In the case of $\mid\!E\!\mid \,>\,\mid\!B\!\mid$ and in particular
for $\mid \!B\!\mid\,=0$ the effective action has an imaginary part. 
The physical
interpretation of this imaginary part is an instability of the vacuum.
This may be seen by appealing to the definition of the effective action
as a vacuum to vacuum amplitude:
\begin{equation}
\langle\, \mbox{out}\mid \mbox{in}\,\rangle =e^{i\Gamma}\Rightarrow \,\,\mid \!\langle\, \mbox{out}\mid \mbox{in}\,\rangle \!\mid^{2}=e^{-2\mbox{\small
Im}\Gamma}.
\end{equation}
Thus, if the  effective action possesses an imaginary part the right hand side
of the above equation shows that the probability that the system remains
in the vacuum is less than unity and that pair production may take place.
Letting $B\rightarrow iE$ in Eq.~(\ref{eac}) we find
\begin{equation}
{\cal L}_{1}=\frac{eE}{8\pi^{\frac{3}{2}}}\int_{0}^{\infty}
\frac{ds}{s^{\frac{3}{2}}}\exp (-m^{2}s)\left [\frac{1}{\sin (eEs)}
-\frac{1}{eEs}\right ].
\end{equation}
${\cal L}_{1}$ now has poles along the real axis at 
$s_{n}=n\pi/eE$. The integration contour should now be considered to  
lie slightly above the real axis. 
The contribution to the imaginary part of the effective  action comes entirely
from the poles and using the usual prescription to handle 
them~\cite{greiner}, one finds:
\begin{equation}
\label{eq:dec}
\mbox{Im}\,{\cal L}_{1}=\frac{(eE)^{\frac{3}{2}}}{8\pi^3}\sum_{n=1}^{\infty}
(-1)^{n+1}\frac{\exp (-m^{2}n\pi/eE)}{n^{\frac{3}{2}}}=-\frac{(eE)^{\frac{3}{2}}}{8\pi^3}
\mbox{Li}_{\frac{3}{2}}\left[-\exp (-m^{2}\pi/eE)\right ].
\end{equation}
Eq. (\ref{eq:dec}) is then the probability per unit volume that the 
vacuum decays. 
Our result is very similar to that obtained by Schwinger \cite{schwing}.


\section{Concluding Remarks}
\heading{Particles in External Fields}{Concluding Remarks}
In the previous sections we have obtained the effective action for fermions
and bosons in the presence of a constant magnetic field. In the
fermionic case the most 
interesting findings were the de-Haas van-Alphen oscillations
and the nontrivial limit $B\rightarrow 0$ of the magnetization. 
We did not explicitly calculate the susceptibility, 
but it is straightforward to do so. However, for
bosons we noted that the susceptibility changed sign. 
In the fermionic case, the susceptibility has been calculated and analyzed
by Haugset in Ref.~\cite{tortese}, and it exhibits interesting
structures as the magnetization itself.

Our treatment could be extended in various ways. First, one should consider
doing a two-loop calculation to incorporate the effects of virtual photons
as well. At finite temperature, this is not trivial. 
The problem is infrared
divergences due to the fact that the zeroth component of the polarization
tensor, $\Pi_{\mu\nu}(k_0,{\bf k})$, is nonvanishing when $k_0=0$ in the 
limit ${\bf k}\rightarrow 0$. 
A careful study of the polarization tensor together with some resummation
approach would be valuable.

Secondly, one could consider more general field configurations than
a constant magnetic field, as has been done by Elmfors and Skagerstam
in Ref.~\cite{skagge}.

\cleardoublepage
\chapter{Resummation and Effective Expansions}
\heading{Resummation and Effective Expansions}{Resummation and Effective Expansions}
\section{Introduction}
\heading{Resummation and Effective Expansions}{Introduction}
In the previous chapter we have seen how quantum field theory can be used to
compute the expectation values of physical quantities in relatively simple
systems.
We also saw that the introduction of finite temperature complicated 
matters and made it necessary to compare scales $m$, $\sqrt{eB}$ and $T$.
In this chapter we turn our attention to finite $T$ calculations
in more ambitious field theories, where calculations are far less 
trivial than those of chapter one. 

Quantum chromodynamics (QCD) is today widely believed to be the theory which
correctly describes strong interactions~\cite{guidry}. 
As long as the number of quarks 
is sufficiently low, this theory is asymptotically free. This means that
the coupling constant decreases with the energy scale, and that perturbative
calculations can be carried out at high momentum transfer.
Moreover, at long distances QCD becomes a strongly interacting theory, and 
lattice QCD indicates that there is a linear potential between two quarks
in the strong coupling limit of QCD~\cite{guidry}.
This potential is responsible for confinement, which is the fact that one
never observes free quarks or gluons. The physical states are all colour
singlets, which are bound states of quarks. These states are the familiar
hadrons such as pions, nucleons and kaons.
Lattice QCD also suggests that hadronic matter undergoes a deconfinement
phase transition at sufficiently high temperature
or high density~\cite{poly}.
This state of
matter is simply a plasma which consists of free quarks and gluons.
The early universe may very well have provided such extreme  conditions
(high temperatures), and so the study of the quark-gluon plasma is 
important in understanding the early universe [28,29].
Furthermore, a quark-gluon plasma may also be produced in future colliders
in heavy-ion collisions, and these experiments then makes it
possible to study the deconfined phase directly~\cite{heavyion}. 

There are several physical quantities of interest in a QCD 
plasma. One of them is the plasmon which is a longitudinal collective 
excitation. The real part of the longitudinal part of the gluon propagator
gives the plasmon mass. The plasmon mass is equal to the plasma frequency
which is the lowest frequency in the medium. The imaginary part of the
longitudinal part of the gluon propagator yields information about 
the decay or lifetime of these excitations~\cite{bellac}.
Historically, this quantity was extremely important for the development
of a consistent perturbative expansion for quantum field theories
at finite temperature.

The first calculations of the damping rate $\gamma$ were based on 
conventional perturbation theory. 
However, it was soon realized that the result for $\gamma$ in naive 
perturbation
theory is dependent on the gauge fixing condition, while the
plasmon mass is
gauge fixing independent (see Ref.~\cite{kobes} and Refs. therein). 
Moreover, in some gauges it also turned out to
be negative, which has been interpreted as a plasma instability.

The situation was, of course, unsatisfactory, since the damping rate
is a physical quantity and if computed correctly it must be gauge invariant.
The puzzle was around during the eighties, and
led people to consider new propagators, which by construction
are gauge fixing independent, and derive the damping rate using linear
response theory.
However, the results were mutually disagreeing and it was
certainly not easy to discriminate between the various values of $\gamma$.
The resolution of this problem was given by Pisarski, who discovered
that the one-loop result is incomplete~\cite{piss2}.
The leading order result 
receives contributions from all order in the loop 
expansion (The first example of this was provided by Gell-Mann and 
Br\"uckner in the late fifties in nonrelativistic QED~\cite{gell}. 
In order to get
the leading order contribution to the free energy, one had to sum an infinite
series of graphs, which were called plasmon diagrams).
In other words, the naive perturbative theory
breaks down and must be 
replaced by an effective expansion in which loop corrections are 
suppressed by powers of $g$.
Pisarski was able to isolate this infinite subset of diagrams, which gave the
leading order result, and resum them into an effective expansion that
includes all effects to leading order in $g$. 

These results is a part of the so-called resummation program
mainly due to Braaten and Pisarski~\cite{pis} and 
is the topic of this chapter. 
The effective expansion involves effective propagators and in nonabelian
gauge theories also effective vertices, and is mandatory to use
at high temperature, in order to obtain complete results.
Resummed perturbation theory restores the connection between the number
of loops in the loop expansion and powers of the coupling constant.  
Moreover, it cures the problem of gauge fixing dependence of
quantities that should be independent. In particular, 
Braaten and Pisarski, demonstrated the gauge invariance and also the
positivity of the gluon
damping rate.
This settled the controversy of the gauge dependence of the damping rate, and
initiated intense studies of thermal field theory.
There are many important contributions, and also improvements of the original
approach, and we shall comment upon them as we move along.

A major application of resummed perturbation theory in recent years
has been the calculation of free energies and effective potentials.
The effective potential is an important tool in the investigation of phase 
transitions at finite $T$, for theories where some symmetry has been 
spontaneously broken at $T=0$. Since the calculation of the effective 
potential, which is a static quantity, normally is carried out in the 
imaginary time formalism, it turns out that one only needs effective 
propagators for the bosons. Fermions need not resummation, and it is also
sufficient to use bare vertices.
The most important phase transition is the electro-weak phase transition
in the standard model or extensions thereof, which took place in the early
universe~\cite{extension}. Its possible role for the baryon asymmetry that we observe
today is basically a question of the nature of the phase transition.
In order to generate any baryon asymmetry, the universe must have been out of
equilibrium, and so the phase transition must be of first order.
The electro-weak phase transition has been studied independently by several
groups using resummed perturbation theory. 
Fodor and Hebecker~\cite{fodor1} have calculated the two-loop effective 
potential in the standard model using Landau gauge. Arnold and 
Espinosa~\cite{arnold} have also investigated this phase transition as well as
the Abelian Higgs model, applying a simplified resummation approach in which
one uses an effective propagator for the $n=0$ bosonic mode only.
The Abelian Higgs model is of interest in its own right, since this is a
model of a relativistic superconductor (See Ref.~\cite{super}).\

The literature on calculation of free energies of quantum field theories
at high temperature ($T$ well above $T_c$) is now vast, and we shall
comment upon only a few selected papers. Frenkel, Saa and Taylor
were the first to push calculations beyond two loop in resummed 
perturbation theory. They computed the free energy to order $\lambda^2$ in 
$\phi^4$-theory~\cite{frenkel}, 
and Parwani and Singh have extended this to order 
$\lambda^{5/2}$~\cite{parw3}. 
Arnold and Zhai have computed the free energy in QED and QCD to fourth
order in the coupling constant~\cite{arnold1}. 
The free energy of high temperature QED
has also been computed independently by Corian\`o and Parwani~\cite{par2}.
In their papers Arnold and Zhai develop
the machinery to deal with complicated multi-loop sum-integrals 
analytically, and this represents significant progress in
perturbative calculations. 
Zhai and Kastening \cite{kast}
have since extended the computations
through fifth order. 

The outline of the chapter is as follows. In the next section
we discuss the breakdown of perturbation theory and
the resummation program of Braaten and Pisarski.
In the following sections we apply resummed perturbation theory
to Yukawa theory, and calculate the screening mass squared and the
pressure to two and three loop order, respectively. 

In the Feynman diagrams, a dashed line denotes a scalar field and 
a solid line denotes a fermion. Our notation and conventions are summarized
in the beginning of Appendix A and B.

\section{The Breakdown of Perturbation Theory}
\begin{figure}[htb]
\begin{center}
\mbox{\psfig{figure=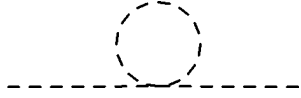}}
\end{center}
\caption[One-loop correction to the two-point function
in scalar theory.]{\protect One-loop correction to the two-point function
in scalar theory.}
\label{1ltwo}
\end{figure}

\heading{The Breakdown of Perturbation Theory}{Resummation and Effective Expansions}
We shall first list some definitions, introduced by Braaten and Pisarski
in Ref.~\cite{pis}:
\begin{itemize}
\item By high temperature (or hot field theories), 
we mean $T\gg m$, where $m$ is any zero temperature
mass.
\end{itemize}
\begin{itemize}
\item A momentum $(k_0,{\bf k})$ is called
soft when both $k_0$ and $k=|{\bf k}|$ are of the order $gT$.
\end{itemize}
\begin{itemize}
\item A momentum $(k_0,{\bf k})$ is 
termed hard when at least one of the components is of the
order $T$.
\end{itemize}
We shall now discuss the breakdown of perturbation theory at finite 
temperature. Consider massless $\lambda\phi^{4}$-theory, for which the
Euclidean Lagrangian reads
\begin{equation}
{\cal L}_{E}=\frac{1}{2}(\partial_{\mu}\phi )^{2}+\frac{\lambda}{24}\phi^{4}.
\end{equation}
The one-loop contribution to the two-point function is depicted in 
Fig.~\ref{1ltwo} 
and
is independent of the external momentum.
It gives a contribution
\beq
\Sigma_{1}(k_0,{\bf k}) =\frac{\lambda}{2}
\hbox{$\sum$}\!\!\!\!\!\!\int_P\frac{1}{P^{2}}.
\eeq
This sum-integral is defined in Appendix A, and in dimensional
regularization it is finite. 
The renormalized inverse propagator at one-loop is then
\beq
\label{inv}
\Gamma^{(2)}_1(k_0,{\bf k})=k_0^2+k^{2}+\frac{\lambda T^{2}}{24}.
\eeq

From this expression we see that the one-loop correction to the full
propagator
is of the same order
as the bare propagator
for soft external momenta. Thus, the above calculation reveals
that naive perturbation theory breaks down for $k\sim \sqrt{\lambda}T$,
and that one must use some kind of effective expansion in which loop 
corrections are down by powers of the coupling constant.
As a part of the resummation program, we define an effective propagator
which is the inverse of Eq.~(\ref{inv}). 

We may also write the one-loop correction as $\frac{\lambda T^2}{24}
\frac{1}{P^2}\times${\it the tree amplitude}, 
where $P$ is the external momentum.
More generally, loop diagrams that can be written in this way 
(where $P$ characterizes the external momenta) are termed {\it 
hard thermal loops}.
By definition then, hard thermal loops (HTL) are equally important as tree 
diagrams for soft external momenta.
Moreover, the hard thermal loops receive their main contribution from
a small region in  momentum space, where the momentum is hard.

\begin{figure}[htb]
\begin{center}
\mbox{\psfig{figure=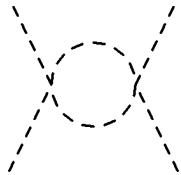}}
\end{center}
\caption[One-loop correction to the four-point function in
scalar theory.]{\protect One-loop correction to the four-point function in
scalar theory.}
\label{fourp}
\end{figure}

Are there other hard thermal loops in $\lambda\phi^4$-theory than the tadpole?
Or in other words, do we need resummed vertices as well as a resummed
propagator?
The answer is no, and below we shall demonstrate that the four-point function
receives a one-loop correction which only depends logarithmically on the
temperature at soft momenta.
Hence, it suffices to use the bare vertex in perturbative calculations.
The one-loop correction to the four-point functions
is depicted in Fig~\ref{fourp} and the expression is.
\beq
\Gamma^{(4)}_1(k_0,{\bf k})=\frac{3}{2}\lambda^2
\hbox{$\sum$}\!\!\!\!\!\!\int_P\frac{1}{P^2(P-K)^2}.
\eeq

The first term in the integrand is now written as
\beq
\label{rep}
\frac{1}{E_p^2+\omega_n^2}=\frac{1}{2E_p}\Big[\frac{1}{i\omega_n-E_p}
-\frac{1}{i\omega_n+E_p}\Big],
\eeq
and correspondingly for the second one. Here, $E_p=p$ and so on. 
One will then encounter terms such as
\beq
\hbox{$\sum$}\!\!\!\!\!\!\int_P
\frac{1}{[i\omega_n-E_p]}\frac{1}{[i(\omega_n -\omega)-E_{p-k}]},
\eeq
where $\omega=k_0$.
The frequency sum is carried out by
rewriting it as a contour integral in the complex plane~\cite{kapusta}. 
The results are then expressed in terms of Bose-Einstein distribution 
functions, which is
\beq
n(E_p)=\frac{1}{e^{\beta E_p}-1}.
\eeq
The term in Eq.~(\ref{rep}) is then replaced by
\beq
\frac{1+n(E_p)+n(E_{p+k})}{i\omega-E_p-E_{p+k}},
\eeq
and similarly for the others. This gives
\bqa\nonumber
\label{hight}
&&\frac{3}{2}\lambda^2\int\frac{d^3p}{(2\pi )^3}\frac{1}{4E_{p}E_{p+k}}
\Big[[1+n(E_{p})+n(E_{p+k})][\frac{1}{ik_0-E_{p}-E_{p+k}}
-\frac{1}{ik_0+E_{p}+E_{p+k}}]\\
&&-[n(E_{p})-n(E_{p+k})][\frac{1}{ik_0-E_{p}+E_{p+k}}
-\frac{1}{ik_0+E_{p}-E_{p+k}}]\Big].
\eqa
Let us now discuss the various terms in the above equation.
The term that is independent of the Bose-Einstein distribution function 
represents the diagram at $T=0$. It is logarithmically ultraviolet
divergent, and after renormalization it goes like $\ln\frac{k^2}{\mu^2}$,
where $\mu$ is the renormalization scale. 
For the $T$-dependent terms it is convenient to distinguish between 
soft and hard loop momenta.
When both the external and the loop momenta is soft, the contribution to the
integral is down by factors of $\lambda$, since the soft momentum is the
only scale in the integral~\cite{pis}.

The contribution from hard loop momenta can be estimated as follows.
We use the approximations 
\bqa
ik_0\pm (E_{p}+E_{p-k})\approx\pm 2E_p&&ik_0\pm (E_{p}-E_{p-k})\approx
ik_0\pm k\cos\theta,\\
n(E_p)+n(E_{p-k})\approx 2n(E_p)&&n(E_p)-n(E_{p-k})\approx
-n(E_p)(1+n(E_p))\frac{k\cos\theta}{T},\\
\hspace{-1cm}\!\!\!\!\!E_{p-k}\approx E_p-k\cos\theta.
\eqa
These approximations are straightforward to derive 
by Taylor expansions in $k/p$.
Here, $\theta$ is the angle between $p$ and $k$.
Plugging this into the first term in Eq.~(\ref{hight}) reads
\beq
\frac{3}{8}
\lambda^2\int\frac{d^3p}{(2\pi )^3}\frac{n(E_p)}{p^2(p-k\cos\theta)}.
\eeq
The angular integral decouples from the radial integral and one 
finds
\beq
\frac{3}{8}
\lambda^2\int\frac{dp}{(2\pi )^2}\frac{1}{k}\ln\Big(\frac{p-k}{p+k}\Big)
n(E_p).
\eeq
Noting that the distribution function cuts off the integral at $p\sim T$,
we see that this term contributes only logarithmically in $T$ to the
in the one-loop diagram. This is also the case for the remaining terms, and
we can conclude that loop corrections are down by powers of the coupling.

\begin{figure}[htb]
\begin{center}
\mbox{\psfig{figure=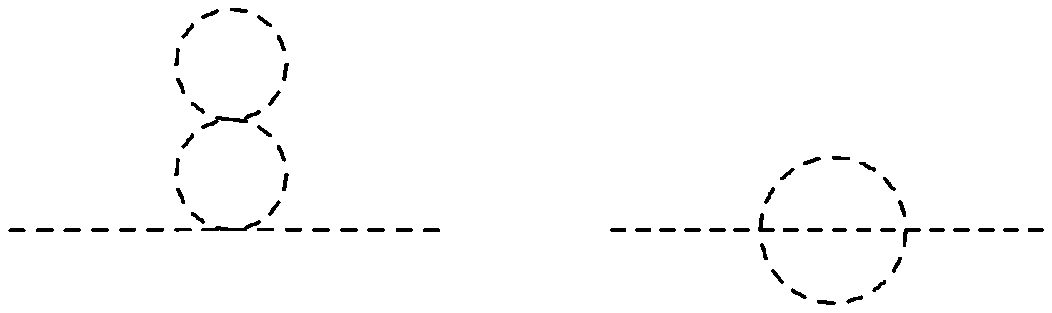}}
\end{center}
\caption[The two-loop graphs for the two-point function.]{\protect The two-loop graphs for the two-point function.}   
\label{skalar2}
\end{figure}

There is yet another way to see the breakdown of perturbation theory
due to infrared divergences. 
Assume that we would like to compute the screening mass of the scalar field.
Generally, it is given by the pole position of the propagator and to leading 
order this is simply $m^2=\lambda T^2/24$, 
which follows from the above calculations.
Beyond leading
order it becomes more complicated. 
Naively, one would expect that the contribution at
next-to-leading order goes like $\lambda^2T^2$ and is given by the
two-loop graphs shown in Fig.~\ref{skalar2}

\begin{figure}[htb]
\begin{center}
\mbox{\psfig{figure=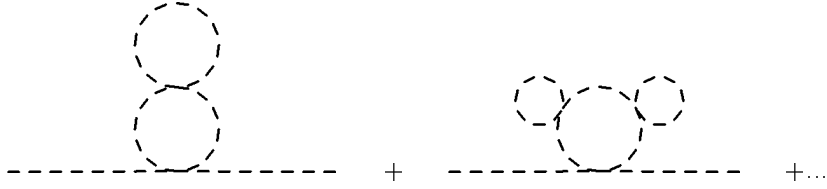}}
\end{center}
\caption[Ring diagrams in scalar theory.]{\protect Ring diagrams in scalar theory.}   
\label{scrin}
\end{figure}

However, this is incorrect. The first two-loop
diagram reads
\beq
-\frac{\lambda^2}{4}\hbox{$\sum$}\!\!\!\!\!\!\int_{PQ}\frac{1}{P^2Q^4}.
\eeq
The term in which $q_0=0$ is linearly divergent\footnote{The setting sun
diagram has a logarithmic divergence in the infrared. Hence, after curing
the infrared divergence it contributes first at order $\lambda^2\ln\lambda$.} 
in the infrared and assuming
an infrared cutoff order $m$, the two-loop goes like $\lambda^{3/2}T^2$.
This diagram is the first in an infinite series of diagrams which 
are increasingly infrared divergent. They are called
ring diagrams or daisy diagrams, and are displayed in Fig.~\ref{scrin}. 
One can easily demonstrate that they all contribute at $\lambda^{3/2}$
to the screening mass, and hence the screening mass gets contributions from
all orders in perturbation theory, analogous to the damping rate discussed
in the introduction.

Due to the similarity in this series of diagrams, one is able to sum it and 
although every term is IR-divergent, the sum turns out to be convergent
both in the infrared and in the ultraviolet. Taking the symmetry factors
into account one finds that a diagram with $m$ loops yields a contribution
\bqa
\frac{(-1)^{m-1}}{2^m}\lambda^m\Big[
\hbox{$\sum$}\!\!\!\!\!\!\int_{P}\frac{1}{P^2}\Big]^{m-1}
\hbox{$\sum$}\!\!\!\!\!\!\int_{Q}\frac{1}{(Q^2)^{m}}.
\eqa
We now restrict ourselves to the $n=0$ mode in the sum over $q_0$, since
the other terms are down by powers of the coupling. Summing over
$m$ produces
\bqa
\frac{\lambda T}{2}\int_q
\sum_{m=1}^{\infty}\frac{1}{q^2}\Big(\frac{-1}{q^2}\Big)^m
\Big(\frac{\lambda T^2}{24}\Big)^m=
\frac{\lambda T}{2}\int_p\frac{1}{q^2+m^2}.
\eqa
Here, $m^2$ is the thermal mass at leading order. This integral is 
listed in Appendix B, and one finds
\beq
\label{leadring}
-\frac{\lambda Tm}{8\pi}.
\eeq
The nonanalyticity in $\lambda$ shows that it is 
nonperturbative with respect to ordinary perturbation theory or that 
we receive contributions from all orders in perturbation theory.
Alternatively, the infrared divergences that we have encountered reflects 
the fact we need to use an improved propagator in the perturbative expansion.

\begin{figure}[htb]
\begin{center}
\mbox{\psfig{figure=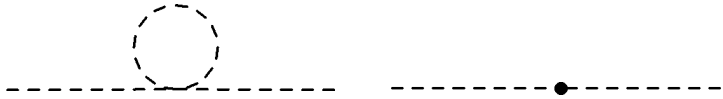}}
\end{center}
\caption[Diagrams contributing in the recalculation of the
screening mass in scalar theory.]{\protect Diagrams contributing in the recalculation of the
screening mass in scalar theory.}   
\label{recal}
\end{figure}

It is straightforward to demonstrate that the improved propagator defined
by the inverse of Eq.~(\ref{inv}), actually resums this infinite set
of diagrams.
The improved propagator is then given 
\beq
\Delta (k_0,{\bf k})=\frac{1}{K^{2}+m^{2}}.
\eeq
However, in order to avoid double counting of diagrams, we must
also subtract a mass term in the Lagrangian and treat this term as an
interaction. We then split the Lagrangian
into a free piece and an interacting piece
according to
\bqa
{\cal L}_{0}&=&\frac{1}{2}(\partial_{\mu}\phi )^{2}+
\frac{1}{2}m^{2}\phi^{2},\\
{\cal L}_{\mbox{\scriptsize int}}&=&-\frac{1}{2}m^2\phi^2+\frac{\lambda}{24}\phi^4.
\eqa
We can now 
recalculate the 
self-energy  and demonstrate that it is really a 
perturbative correction
to $m^{2}=\frac{\lambda T^{2}}{24}$. The self-energy 
is now given by the
usual one-loop contribution as well as the new vertex 
which are shown in figure~\ref{recal}. We get
\bqa\nonumber
\Sigma_{1}(k_0,{\bf k})&=&-m^{2}+\frac{\lambda}{2}
\hbox{$\sum$}\!\!\!\!\!\!\int_{P}\frac{1}{P^2+m^2}\\
&&-m^2+\frac{\lambda}{2}\hbox{$\sum$}\!\!\!\!\!\!\int_{P}\frac{1}{P^2}
+\frac{\lambda T}{2}\int_p\frac{1}{p^2+m^2}-\frac{\lambda m^2}{2}
\hbox{$\sum$}\!\!\!\!\!\!\int_P^{\prime}
\frac{1}{P^4}+....
\eqa
The prime indicates that the $n=0$ mode has been left out from the sum.
Since this contribution is set to zero in dimensional regularization, we can
still use the expression for the sum-integral listed in Appendix A.
The self-energy is rendered finite by adding the mass counterterm
$\frac{\lambda m^2}{32\pi^2\epsilon}$, and using the appendices, one finds
\bqa
M^2&=&m^2+\Sigma_{1}(0,0) \\ \nonumber
&=&\frac{\lambda T^2}{24}-\frac{mT}{8\pi},
\eqa
which is down by $\sqrt{\lambda}$, as promised, and also
reproduces the leading part of the sum of the ring diagrams, given by 
Eq.~(\ref{leadring}). 

\begin{figure}[htb]
\begin{center}
\mbox{\psfig{figure=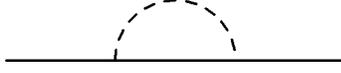}}
\end{center}
\caption[One-loop fermion self-energy correction in Yukawa theory.]{\protect One-loop fermion self-energy correction in Yukawa theory.}
\label{fer1}
\end{figure}

Let us continue our discussion of the resummation program of 
Braaten and Pisarski by considering more complicated theories. The simplest
theory containing fermions is Yukawa theory, which has been studied by Thoma
in Ref.~\cite{thoma}. The Euclidean Lagrangian is
\beq
{\cal L}_{E}=\frac{1}{2}(\partial_{\mu}\phi )^{2}
+\frac{\lambda}{24}\phi^{4}+
\bar{\psi}\partial\!\!\!/\psi+g\bar{\psi}\psi\phi.
\eeq
It is straightforward to show that one also needs an effective fermion 
propagator in Yukawa theory, and that is a common feature of all theories
involving fermions. As in the pure scalar case one may show that the 
one-loop correction to the $\bar{\psi}\psi\phi$ also has a logarithmic
dependence on $T$ and it is not necessary to resum the vertex.
One can demonstrate that this is the case for all other $n-$point functions
in Yukawa theory, and hence we may conclude 
that only propagators need to be resummed. Let us take a closer look at this.
The one-loop fermionic self-energy is shown in Fig.~\ref{fer1} and we have
\bqa
\label{fers}
\Sigma_{f}
(k_0,{\bf k})=g^2\hbox{$\sum$}\!\!\!\!\!\!\int_{P}
\frac{P\!\!\!\!/-K\!\!\!\!/}{P^2(P-K)^2}
\eqa

The sum over Matsubara frequencies are carried out as in the bosonic case.
The main difference is that the result involves both Bose-Einstein and
Fermi-Dirac distribution functions, which is
\beq
\tilde{n}(E_p)=\frac{1}{e^{\beta E_p}+1}.
\eeq 
After the appropriate substitutions
and noting that we may neglect $K$ in comparison with $P$ in the
numerator in Eq.~(\ref{fers}), we arrive at
\bqa\nonumber
\Sigma_{f}(k_0,{\bf k})&=&
\frac{i\gamma_{0}}{8\pi^2}g^2\int\frac{d^3p}{(2\pi )^3}\frac{1}{4E_{p-k}}
\Big[[1+n(E_{p})-\tilde{n}(E_{p-k})][\frac{1}{ik_0-E_{p}-E_{p-k}}
\\ \nonumber
&&
-\frac{1}{ik_0+E_{p}+E_{p-k}}]+[n(E_{p})+\tilde{n}(E_{p-k})][\frac{1}{ik_0+E_{p}-E_{p-k}}\\ \nonumber
&&
-\frac{1}{ik_0-E_{p}+E_{p-k}}]\Big]-\frac{\gamma_{i}}{8\pi^2}g^2
\int\frac{d^3p}{(2\pi )^3}\frac{p_i}{4E_pE_{p-k}}
\Big[[1+n(E_{p})-\tilde{n}(E_{p-k})]\\ \nonumber 
&&
[\frac{1}{ik_0-E_{p}-E_{p-k}}
-\frac{1}{ik_0+E_{p}+E_{p-k}}] \\ 
&&+[n(E_{p})+\tilde{n}(E_{p-k})][\frac{1}{ik_0+E_{p}-E_{p-k}}
-\frac{1}{ik_0-E_{p}+E_{p-k}}]\Big].
\eqa
Now, 
one may infer that it is only the terms which involve the sum of 
two distribution functions that
contribute of order $T^2$; The first term which is independent
of the distribution function, represent the $T=0$ contribution to the
fermion self-energy and is 
therefore linearly divergent. The others are non-leading in $T$.
Using the approximations above for hard loop momenta (the same approximations
are valid for Fermi-Dirac distributions), 
we find that the
angular integral decouples from the radial integral. The calculations
are then straightforward and the fermion self-energy 
takes the form
\beq
\Sigma_{f}(k_0,{\bf k})=\frac{im_{f}^2}{k}
\gamma_{0}Q_{0}(\frac{k_0}{k})+\frac{m_{f}^2}{k}
\gamma_{i}\hat{k}_{i}
\Big[1-Q_{0}(\frac{k_0}{k})\Big].
\eeq
Here, we have introduced 
the fermion mass $m_{f}^2$ and the Legendre function of the
second kind, $Q_0(x)$:
\beq
m_{f}^2=
\frac{g^2T^2}{16}, 
\hspace{1cm}
Q_0(x)=\frac{1}{2}\ln\Big[\frac{x+1}{x-1}\Big].
\eeq
The effective inverse fermion propagator then takes the form
\beq
\label{efferm}
\tilde{\Delta}^{-1}(k_0,{\bf k})=-iK\!\!\!\!/+i\Sigma_{f}(k_0,{\bf k}).
\eeq
At this point we would like to comment upon the fermion propagator,
which is given by the inverse of Eq.~(\ref{efferm}).
As first pointed out by Klimov and Weldon, the effective fermion propagator
has two poles [47,48]. 
The first corresponds to eigenstates where helicity equals
chirality and is the usual mode, known from $T=0$.
The second pole corresponds to eigenstates where helicity is minus chirality.
This mode is a collective excitation and is occasionally referred to as the
{\it plasmino.}

Using the methods we have discussed in this chapter, the reader may
convince herself that there are no other hard thermal loops in Yukawa
theory. In particular, the $\bar{\psi}\psi\phi$ receives a one-loop
correction which depends on the temperature, only through logarithms, exactly
as the quartic vertex in $\phi^4$-theory~\cite{thoma}.

In QED, the only hard thermal loops are in the amplitude between a pair
of fermions (fermion self-energy) and between a pair of photons 
(polarization tensor or the photon self-energy).
The effective photon propagator is rather involved due to its non-trivial
momentum dependence, in contrast with the local mass term in pure scalar
theory.
In QCD, there are hard thermal loops in all multi-gluon amplitudes, and also
in the amplitude between a pair of quarks and any number of 
gluons~\cite{pis}. Hence, effective vertices are required, too.
The hard thermal loops have a remarkable property, namely that of gauge fixing
independence. 
Klimov and Weldon were the first to show this property for the self-energy
in QCD [47,48], and these results were extended by Braaten and Pisarski to
all hard thermal loops by explicit calculations~\cite{pis}.
Moreover, Kobes {\it et al.}
have given a general field theoretic argument
of this property~\cite{kobes}.
After the discovery of the gauge fixing independence, Braaten and Pisarski
were able to construct an effective Lagrangian that generates the hard thermal
loops for all amplitudes, and this effective Lagrangian is 
gauge invariant~\cite{effact}.
Taylor and Wong constructed independently another equivalent effective
Lagrangian with the same properties~\cite{wongleik}.
An effective Lagrangian that generates the hard thermal loops in Yukawa
theory naturally also exists, and it reads
\beq
\label{ea}
{\cal L}_{\mbox{\scriptsize eff}}={\cal L}+m^2_{f}\bar{\psi}
\int\frac{d\Omega}{4\pi}
\frac{\partial\!\!\!/}{(\partial\cdot \hat{K})}\psi
+m^2_{s}\phi\int\frac{d\Omega}{4\pi}
\frac{\partial^2}{(\partial\cdot \hat{K})^2}\phi.
\eeq
Here, we have introduced the four-vector $\hat{K}=(-i,\hat{k})$. The integral
represents the average over the sphere. The fact that this effective
Lagrangian generates the effective propagators follows easily from 
doing the angular integrals. In particular, the last term in Eq.~(\ref{ea}) 
is reduced
to a local mass term.

Above, we have seen how the improved expansion screens the infrared 
singularities that appeared in bare perturbation theory.
However, in some applications it turns out that the HTL action does
not screen all IR-singularities. A well-known example of this is
the calculation of the electric screening mass in QCD 
beyond leading order~\cite{anton1}.
At next-to-leading order mass-shell singularities arise due to 
unscreened magnetic modes. A similar problem arises in scalar electrodynamics
in the calculation of the scalar screening mass beyond leading order [52,53].
Another example of the breakdown of the original approach, is in the
calculation of the production rate of real soft photons in a quark-gluon
plasma. For this problem,  Flechsig and Rebhan have invented an improved
effective action which removes these singularities~\cite{imhtl}. 
It can also be written in a gauge invariant way and reduces to the
conventional HTL action, where the latter is valid.
\section{A Simplified Resummation Scheme}
\heading{Resummation and Effective Expansions}{A Simplified Resummation Scheme}
In the previous sections we have discussed the resummation program of
Braaten and Pisarski and demonstrated that we need to use effective 
propagators and, in some cases, effective vertices too. 
However, in the calculation of static quantities such
as free energies (or effective potentials) and screening masses there exists
a simplified resummation scheme due to Arnold and Espinoza~\cite{arnold}.
The point is that for calculating Greens functions with zero external
frequency, this is most conveniently carried out in the imaginary
time formalism, without analytic continuation to real energies.
We also know that in the imaginary time formalism, 
the Matsubara frequencies act as masses.
For $n\neq 0$ bosonic modes and fermionic modes they 
provide an IR cutoff of order $T$. Hence, for
these modes, thermal corrections are truly perturbative 
(down by a factor of $g$),
and it should be sufficient to dress the zero modes. 
Although it seems at first sight that the distinction
between light and heavy modes may complicate things, it actually 
simplifies calculations a lot.
We shall apply this approach in the next section to compute the screening mass
and the free energy to order to order $\lambda^2$, $\lambda g^2$ and $g^4$
in Yukawa theory.

Finally, we would emphasize that this simplified approach can not be applied
in the calculations of dynamical Greens functions with soft external
frequencies, as demonstrated by Krammer {\it et al.} in Ref.~\cite{kram}.
For instance, they show that it predicts an incorrect value of the
plasma frequency at next-to-leading order in scalar electrodynamics,
and they identify the problem to be that of an ambiguity in the
analytic continuation to real energies.
\section{The Screening Mass to Two-loop Order}
\begin{figure}[htb]
\begin{center}
\mbox{\psfig{figure=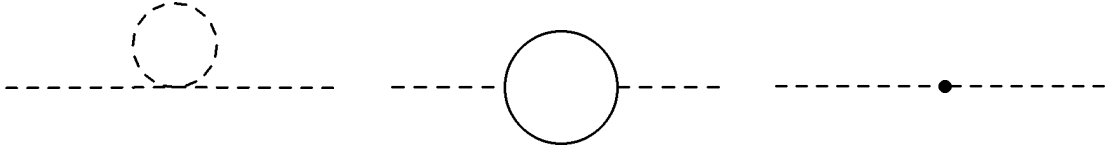}}
\end{center}
\caption[Leading order contributions to the screening mass in Yukawa theory.]{\protect Leading order contributions to the screening mass in Yukawa theory.}   
\label{yse1l}
\end{figure}

\heading{Resummation and Effective Expansions}{The Screening Mass to Two-loop Order}
As we have seen in the pure scalar case we must rearrange our Lagrangian 
according to
\beq
{\cal L}=\frac{1}{2}(\partial_{\mu}\phi )^{2}+\frac{1}{2}m^2\delta_{k,0}
+\frac{\lambda}{24}\phi^{4}+\bar{\psi}\partial\!\!\!/\psi+
g\bar{\psi}\psi\phi-\frac{1}{2}m^2\delta_{k,0}.
\eeq
Here, the mass parameter $m^2$ is simply the bosonic self-energy at
one-loop at zero external momentum. 
The relevant graphs are the first two Feynman diagrams in 
Fig.~\ref{yse1l}, and they give
\bqa
\Sigma_1(0,0)&=&\frac{\lambda}{2}\hbox{$\sum$}\!\!\!\!\!\!\int_{P}\frac{1}{P^2}
-g^2\hbox{$\sum$}\!\!\!\!\!\!\int_{\{P\}}{\mbox{Tr}}
\Big[\frac{P\!\!\!\!/P\!\!\!\!/}{P^4}\Big].
\eqa
Using Appendix A we find
\beq
m^2=\frac{\lambda}{24}T^2+\frac{g^2}{6}T^2.
\eeq
The resummed 
bosonic propagator is then
\beq
\Delta (k_0,{\bf k})=\frac{1-\delta_{k,0}}{K^2}+\frac{\delta_{k,0}}{k^2+m^2}.
\eeq
The screening mass is given by the location 
of the pole of the propagator at spacelike momentum~\cite{anton1}. 
At the one-loop level, $M^2$ is then given in terms of the
infrared limit of the self-energy function of the scalar field:
\beq
M^2=m^2+\Sigma_1(0,0).
\eeq
Here, $\Sigma_n(0,{\bf k})$ denotes the $n$th order contribution to 
$\Sigma (0,{\bf k})$ in the resummed loop expansion. 
At the two-loop level we must take into account the momentum dependence
of the self-energy function $\Sigma (0,{\bf k})$. 
The screening mass is then given by
\beq
\left.M^2=m^2+\Big[\Sigma_1(0,{\bf k})+\Sigma_2(0,{\bf k})\Big]\right|_{k=im}.
\eeq
Consider the  leading order contribution to $\Sigma (0,k)$, which is
depicted in Fig.~\ref{yse1l}. The second diagram is momentum dependent. 
Since the fermionic loop momentum is 
always hard, we can expand $\Sigma_{1}(0,k)$ in powers of the 
(soft) external momentum:
\bqa\nonumber
\label{resum}
\Sigma_{1}(0,{\bf k})&=&\frac{\lambda }{2}\hbox{$\sum$}\!\!\!\!\!\!\int_{P}
\Big[\frac{1-\delta_{p,0}}{P^2}+\frac{\delta_{p,0}}{p^2+m^2}\Big]
-g^2\hbox{$\sum$}\!\!\!\!\!\!\int_{\{P\}}{\mbox{Tr}}
\Big[\frac{P\!\!\!\!/(P\!\!\!\!/+K\!\!\!\!/)}{P^2(P+K)^2}\Big]-m^2\\
&=&
-\frac{\lambda mT}{8\pi}+
2k^2g^2\hbox{$\sum$}\!\!\!\!\!\!\int_{\{P\}}\frac{1}{P^4}+{\cal O}(k^4/T^2).
\eqa
The sum-integral above is divergent and the divergence is removed by
the field strength renormalization counterterm. To leading order we 
have~\cite{schub}:
\beq
\label{vaag1}
Z_{\phi}=1-\frac{g^{2}}{8\pi^{2}\epsilon}.
\eeq
Thus, one finds 
\beq
\Sigma_1 (0,{\bf k})=-\frac{\lambda mT}{8\pi}+\frac{2g^2k^2}{(4\pi )^2}
(2\ln\frac{\Lambda}{4\pi T}+2\gamma_{E}+4\ln 2).
\eeq
Here, $\Lambda$ is the renormalization scale introduced by dimensional
regularization (see Appendix A)

Let us next consider the two-loop diagrams from the scalar sector.
The first is independent of the external momentum $k$,
and reads 
\bqa\nonumber
-\frac{\lambda^2}{4}\hbox{$\sum$}\!\!\!\!\!\!\int_{PQ}
\Big[\frac{1-\delta_{p,0}}{P^2}+\frac{\delta_{p,0}}{p^2+m^2}\Big]
\Big[\frac{1-\delta_{q,0}}{Q^2}+\frac{\delta_{q,0}}{q^2+m^2}\Big]^2
&=&-\frac{\lambda^2}{4}\hbox{$\sum$}\!\!\!\!\!\!\int_{PQ}^{\prime}
\frac{1}{P^2Q^4}
-\frac{\lambda^2T^3}{384\pi m}+\\
&&\frac{\lambda^2T^2}{128\pi^2}+
\frac{\lambda^2mT}{16\pi}\hbox{$\sum$}\!\!\!\!\!\!\int_{P}^{\prime}
\frac{1}{P^4}.
\eqa
Note that here and in the following the prime indicates that the $n=0$
mode is left out in the sum (This does not affect the value of the 
sum-integral since the integral for the zero-frequency mode is set
to zero).
For the present calculation, the last term in the above equation is not needed,
since it is of higher order in the couplings, and it is consequently dropped
in the following. 

The second two-loop graph, the sunset diagram, is  
also the most complicated one. 
The terms in the sum for which at least one Matsubara frequency is 
nonvanishing are IR-safe, so that $m$ is not a relevant infrared
cutoff to order $\lambda^2$. Thus, we may put $m=0$ here. 
The remaining part where $p_0=q_0=0$ is infrared divergent and so we must
keep the mass $m$. Hence, to order $\lambda^2$ we can write
\beq
-\frac{\lambda^2}{6}\hbox{$\sum$}\!\!\!\!\!\!\int_{PQ}
\frac{1-\delta_{p_0,0}\delta_{q_0,0}}{P^2Q^2}
-\frac{\lambda^2}{6}
\int_{pq}\frac{1}{(p^2+m^2)(q^2+m^2)[({\bf p}+{\bf q}+{\bf k})^2+m^2]}.
\eeq
Using the methods of Appendix C, one can demonstrate that the first
term above is zero in dimensional regularization, and we are left with
the second term.
This integral is dependent on the external momentum $k$, and in order to 
calculate the screening mass consistently we must
compute it at $k=im$. In order to see that this in fact is necessary, one
can perform an expansion in the external momentum $k$, and verify that all
terms are equally important for soft $k\sim\sqrt{\lambda}T\sim gT$.

\begin{figure}[htb]
\begin{center}
\mbox{\psfig{figure=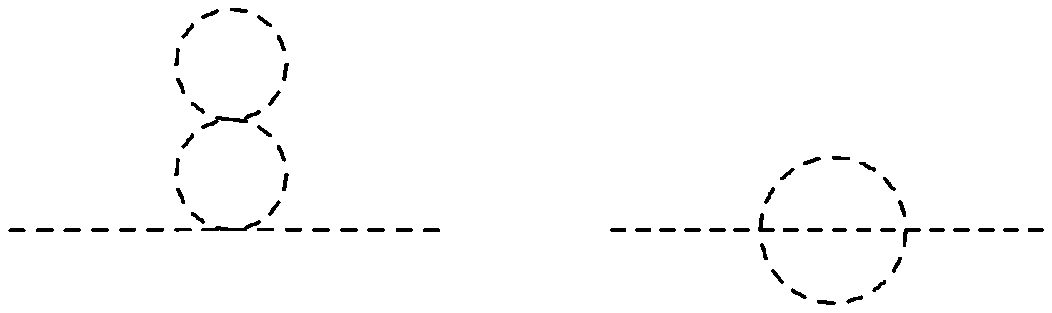}}
\end{center}
\caption[Two-loop scalar diagrams.]{\protect Two-loop scalar diagrams.}
\label{skalar22}
\end{figure}

To this order we must also consider the tadpole with
a thermal counterterm insertion. This is calculated 
the same way and one finds a contribution
\bqa\nonumber
\frac{\lambda m^2}{2}
\hbox{$\sum$}\!\!\!\!\!\!\int_{P}
\Big[\frac{1-\delta_{p,0}}{P^2}+\frac{\delta_{p,0}}{P^2+m^2}\Big]^2
\delta_{p_0,0}=\frac{\lambda mT}{16\pi}.
\eqa

\begin{figure}[htb]
\begin{center}
\mbox{\psfig{figure=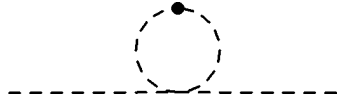}}
\end{center}
\caption[One-loop graph with a thermal counterterm insertion.]{\protect One-loop graph with a thermal counterterm insertion.}
\label{skins}
\end{figure}

\begin{figure}[htb]
\begin{center}
\mbox{\psfig{figure=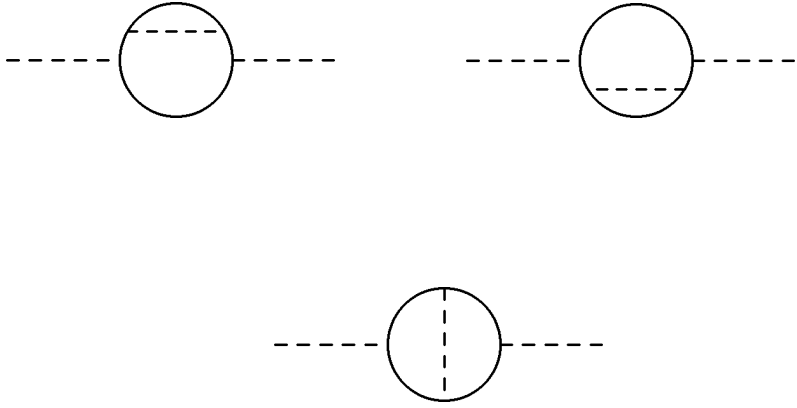}}
\end{center}
\caption[Two-loop self-energy diagrams from the Yukawa sector.]{\protect Two-loop self-energy diagrams from the Yukawa sector.}   
\label{ysc2}
\end{figure}

Let us now turn to the the two-loop diagrams which come entirely from the
Yukawa interaction. These are depicted in Fig.~\ref{ysc2} 
and are all infrared safe
when the mass $m$ is set to zero. This implies that the leading order
contribution, ${\cal O}(g^4)$ can be found using the unresummed propagator.
Moreover, due to the IR-convergence, $m$ is not a relevant
infrared cutoff and one
may expand in the external momentum $k$. Thus, to ${\cal O}(g^4)$ 
it sufficient to consider the diagrams at vanishing external momenta.
The first two diagrams obviously contribute equally and yield
\bqa
2g^4\hbox{$\sum$}\!\!\!\!\!\!\int_{\{P\}Q}\mbox{Tr}\Big[\frac{P\!\!\!\!/P\!\!\!\!/P\!\!\!\!/
(P\!\!\!\!/+Q\!\!\!\!/)}{P^6Q^2(P+Q)^2}\Big]
=
4g^4\hbox{$\sum$}\!\!\!\!\!\!\int_{\{P\}Q}\frac{1}{P^4Q^2}
-4g^4\hbox{$\sum$}\!\!\!\!\!\!\int_{\{PQ\}}\frac{1}{P^4Q^2}.
\eqa
The next diagram is treated in a similar fashion:
\bqa
\label{tildi}
g^4\hbox{$\sum$}\!\!\!\!\!\!\int_{\{PQ\}}\mbox{Tr}\Big[
\frac{P\!\!\!\!/P\!\!\!\!/Q\!\!\!\!/Q\!\!\!\!/}{P^4Q^4(P+Q)^2}\Big]
=4g^4\hbox{$\sum$}\!\!\!\!\!\!\int_{\{PQ\}}
\frac{1}{P^2Q^2(P+Q)^2}.
\eqa
This diagram actually vanishes in dimensional regularization, and this is
demonstrated in Appendix C.

The only mixed two-loop diagrams is infrared divergent when the mass
is set to zero, so we keep the resummed propagator and find
\beq
\frac{\lambda g^2}{2}\hbox{$\sum$}\!\!\!\!\!\!\int_{\{P\}Q}
{\mbox{Tr}}\Big\{\frac{P\!\!\!\!/(P\!\!\!\!/+Q\!\!\!\!/)}{P^{2}(P+Q)^2}
\Big[\frac{1-\delta_{q,0}}{Q^2}+\frac{\delta_{q,0}}{q^2+m^2}\Big]^2\Big\}.
\eeq
After the usual tricks we have
\beq
\frac{\lambda g^2}{2}\Big[4\hbox{$\sum$}\!\!\!\!\!\!\int_{\{P\}Q}^{\prime}
\frac{1}{P^2Q^4}
+\frac{T}{2\pi m}\hbox{$\sum$}\!\!\!\!\!\!\int_{\{P\}}\frac{1}{P^2}+...
\Big].
\eeq
Here, the ellipsis indicate higher order terms, which can be dropped in the 
present calculations.

\begin{figure}[htb]
\begin{center}
\mbox{\psfig{figure=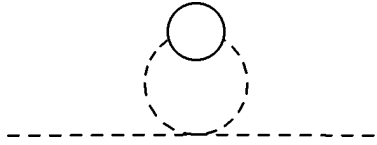}}
\end{center}
\caption[Mixed two-loop diagram contributing to the
screening mass.]{\protect Mixed two-loop diagram contributing to the
screening mass.}
\label{mixed}
\end{figure}

\begin{figure}[htb]
\begin{center}
\mbox{\psfig{figure=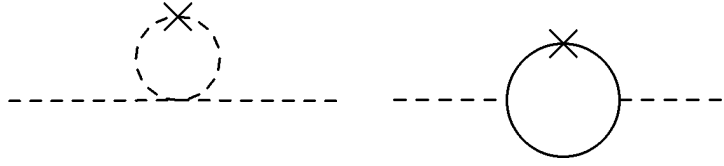}}
\end{center}
\caption[One-loop graphs with wave function counterterm insertion
in Yukawa theory.]{\protect One-loop graphs with wave function counterterm insertion
in Yukawa theory.}   
\label{ysc1c}
\end{figure}

Finally, we include the one-loop diagrams with counterterm insertions.
These are depicted in Fig.~\ref{ysc1c}.
At leading order we may again use the bare propagator and they contribute,
respectively
\bqa
8(Z_{\psi}-1)g^{2}\hbox{$\sum$}\!\!\!\!\!\!\int_{\{P\}}\frac{1}{P^{2}}
,\hspace{1cm}
(1-Z_{\phi})\frac{\lambda}{2}\hbox{$\sum$}\!\!\!\!\!\!\int_{P}
\frac{1}{P^2}.
\eqa
$Z_{\phi}$ was given in Eq.~(\ref{vaag1}), while the other renormalization 
counterterm reads~\cite{schub}
\beq
Z_{\psi}=1-\frac{g^{2}}{32\pi^{2}\epsilon}.
\eeq

The renormalization of the vertices are carried out by the replacements
\beq
\lambda\rightarrow Z_{1}\lambda=\lambda+\frac{3\lambda^2-48g^4}{32\pi^2\epsilon}
,\hspace{1cm}
g^2\rightarrow Z_{2}g^2=g^2+\frac{2g^4}{16\pi^2\epsilon}.
\eeq
Evaluating $\Sigma_1(0,{\bf k})$ at $k=im$ and collecting our two-loop
contributions, one finally obtains for the screening mass
\bqa\nonumber
\label{scsm}
M^2&=&\frac{\lambda T^2}{24}+\frac{g^2T^2}{6}-\frac{\lambda mT}{8\pi}-
\frac{\lambda^2}{16\pi^2}\frac{T^2}{12}\Big[\ln\frac{\Lambda}{4\pi T}
+2\ln\frac{\Lambda}{2m}+\frac{1}{2}\gamma_{E}-4\ln 2
+\frac{1}{2}-\frac{\zeta^{\prime}(-1)}{\zeta (-1)}
\Big]
\\ 
&&-\frac{\lambda g^2}{16\pi^2}\frac{T^2}{12}\Big[4\ln\frac{\Lambda}{4\pi T}
+4\gamma_{E}+2\ln2
\Big]
+\frac{g^4}{16\pi^2}\frac{T^2}{12}\Big[4\ln\frac{\Lambda}{4\pi T}
+4\gamma_{E}-8\ln2\Big].
\eqa
This result here is new and it
is easy to check that our result is renormalization group invariant
by using the RG-equations for the couplings $\lambda$ and $g^2$~\cite{ford}:
\bqa
\mu\frac{d\lambda}{d\mu}&=&\frac{3\lambda^2+8\lambda g^2-48g^4}{16\pi^2},\\
\mu\frac{dg^2}{d\mu}&=&\frac{5g^4}{8\pi^2}.
\eqa
Moreover, setting $g=0$ our result is in accordance with the one obtained by
Braaten and Nieto using the effective field theory approach that we shall
discuss thoroughly in the next chapters~\cite{braaten}. 
We have also checked that this approach
yields the same result also for $g\neq 0$ (as well as the original approach
by dressing all modes).
We should also mention that the contribution from the one-loop with a
thermal counterterm has canceled the 
two-loop contributions which individually
were of the form $\lambda^2T^3/m$ and $\lambda g^2T^3/m$. 

Before closing this section we would like to make a few remarks.
Firstly, consider the double-bubble, the mixed two-loop graph
and the tadpole with a mass insertion. Individually, thse diagrams contribute
to  lower order than $\lambda^2$, $\lambda g^2$ and $g^4$, but these cancel 
in the sum. This is of course necessary in order for resummed perturbation 
theory to work, and the reason for this is that the particular combination
of these diagrams is infrared finite. This implies that to order
$\lambda^2$, $\lambda g^2$ and $g^4$, one may use the bare propagator.
Above, we use the resummed propagator and calculated the diagrams, one by one,
in order to demonstrate this cancelation explicitly. In the next section
we shall use this observation to simplify the calculation of some three-loop
graphs contributing to the free energy.

\begin{figure}[htb]
\begin{center}
\mbox{\psfig{figure=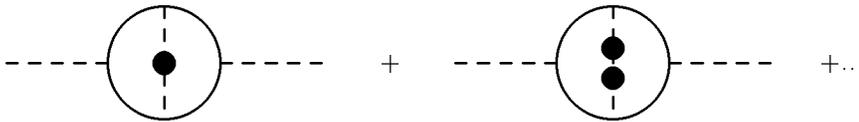}}
\end{center}
\caption[Infinite string of diagrams 
with scalar self-energy insertions.]{\protect Infinite string of diagrams 
with scalar self-energy insertions.}   
\label{ysring}
\end{figure}

Secondly, imagine that we would compute subleading contributions to the
screening mass coming from the diagrams in the Yukawa sector.
It is then mandatory to use the resummed propagator, and we now 
show how to extract such contributions. Let us for simplicity confine
ourselves to the second diagram which now reads
\bqa\nonumber
I&=&g^4\hbox{$\sum$}\!\!\!\!\!\!\int_{\{PQ\}}\mbox{Tr}
\frac{P\!\!\!\!/P\!\!\!\!/Q\!\!\!\!/Q\!\!\!\!/}{P^4Q^4}
\Big[\frac{1-\delta_{(p+q)_0,0}}{(P+Q)^2}
+\frac{\delta_{(p+q)_0,0}}{(p+q)^2+m^2}\Big]\\
&=&4g^4\hbox{$\sum$}\!\!\!\!\!\!\int_{\{PQ\}}
\frac{1}{P^2Q^2}\Big[\frac{1-\delta_{(p+q)_0,0}}{(P+Q)^2}
+\frac{\delta_{(p+q)_0,0}}{(p+q)^2+m^2}\Big].
\eqa
As previously explained, this diagram is IR-safe in the limit 
$m\rightarrow 0$. The leading term is given by Eq.~(\ref{tildi}), and the 
subleading term is then found by subtracting the leading part
(In this case the leading part accidentally vanishes, but that is besides
the point).
After changing variables, one finds
\beq
I_{\mbox{\scriptsize sub}}=-
4m^2g^4\hbox{$\sum$}\!\!\!\!\!\!\int_{\{P\}Q}\frac{1}{P^2(P+Q)^2}
\frac{\delta_{q_0,0}}{q^2(q^2+m^2)}.
\eeq
This expression is infrared divergent when the mass is set to zero.
Hence, the second integral 
picks up its main contribution when $q$ is of order
$m$. Since $P$ is always hard, one may to leading order in the couplings
neglect $q$ in comparison with
$P$. Hence the integrals decouple, and the leading part of 
$I_{\mbox{\scriptsize sub}}$ is
\bqa\nonumber
I_{\mbox{\scriptsize sub}}^{\mbox{\scriptsize lead}}&=&-
4m^2g^4T\int_q\frac{1}{q^2(q^2+m^2)}\hbox{$\sum$}\!\!\!\!\!\!\int_{\{P\}}
\frac{1}{P^4}\\
&=&
-\frac{mT}{\pi}g^4\hbox{$\sum$}\!\!\!\!\!\!\int_{\{P\}}\frac{1}{P^4}.
\eqa
The other diagram yields a contribution at this order which is twice as
large. The ultraviolet divergences are
canceled by renormalization
of the quartic vertex in the third term in 
Eq.~(\ref{scsm})).
This contribution to the screening mass has a simple interpretation
in terms of bare perturbation theory. It corresponds to multiple scalar
self-energy insertions on the bosonic line in these graphs. These
diagrams can be summed ad infinitum, and the situation is indicated in 
Fig.~\ref{ysring}.
In the last chapter we consider QED, and similar diagrams are present there
(namely insertions of the photon self-energy). It is rather amazing
that the corresponding contribution can be obtained by an almost trivial
one-loop calculation in three dimensions!

\section{Free Energy in Yukawa Theory to order $\lambda^2$, $\lambda g^2$ and $g^4$}
\heading{Resummation and Effective Expansions}{Free Energy in Yukawa Theory to order $\lambda^2$, $\lambda g^2$ and $g^4$}
In this section we shall compute the free energy to order 
$\lambda^2$, $\lambda g^2$ and $g^4$ using the methods from last
section.
The one-loop contribution is given by the following expression
\beq
\frac{1}{2}\hbox{$\sum$}\!\!\!\!\!\!\int_{P}^{\prime}\ln P^2
+\frac{1}{2}T\int_p\frac{1}{p^2+m^2}
-2\hbox{$\sum$}\!\!\!\!\!\!\int_{\{P\}}\ln P^2
=-\frac{9\pi^2T^4}{180}-\frac{Tm^3}{12\pi}.
\eeq

\begin{figure}[htb]
\begin{center}
\mbox{\psfig{figure=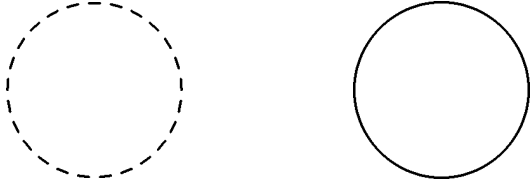}}
\end{center}
\caption[One-loop contributions to the free energy
in Yukawa theory.]{\protect One-loop contributions to the free energy
in Yukawa theory.}    
\label{yuk1loop}
\end{figure}

The scalar two-loop yields
\bqa
\label{sc2}
\frac{Z_{1}\lambda}{8}\Big[\hbox{$\sum$}\!\!\!\!\!\!\int_{P}
\frac{1-\delta_{p_0,0}}{P^2}+\frac{\delta_{p_0,0}}{p^2+m^2}\Big]^2=
\frac{Z_1\lambda}{8}\hbox{$\sum$}\!\!\!\!\!\!\int_{PQ}^{\prime}
\frac{1}{P^2Q^2}
-\frac{\lambda mT^3}{192\pi}+\frac{\lambda m^2T^2}{128\pi^2}.
\eqa
Note that to order we calculate we need not renormalize the second and third
term in the above equation.
The theta-diagram reads
\bqa\nonumber
\label{theta}
\hspace{-4cm}-\frac{Z_{2}g^2}{2}\hbox{$\sum$}\!\!\!\!\!\!\int_{P\{Q\}}
\mbox{Tr}\Big\{
\frac{(P\!\!\!\!/+Q\!\!\!\!/)Q\!\!\!\!/}{(P+Q)^2Q^2}
\Big[\frac{1-\delta_{p_0,0}}{P^2}+\frac{\delta_{p_0,0}}{p^2+m^2}\Big]\Big\}
&&=\\ \nonumber
Z_{2}g^2\hbox{$\sum$}\!\!\!\!\!\!\int_{\{PQ\}}\frac{1}{P^2Q^2}
-2Z_{2}g^2\hbox{$\sum$}\!\!\!\!\!\!\int_{P\{Q\}}^{\prime}\frac{1}{P^2Q^2}
+\frac{mT}{2\pi}g^2\hbox{$\sum$}\!\!\!\!\!\!\int_{\{P\}}\frac{1}{P^2}
-m^2g^2\hbox{$\sum$}\!\!\!\!\!\!\int_{\{P\}Q}
\frac{\delta_{q_0,0}}{P^2Q^2(P+Q)^2}.&&
\eqa
Note again, that renormalization of the vertex is only necessary for the
leading terms above. We therefore set $Z_2=1$ for the remaining terms.

\begin{figure}[htb]
\begin{center}
\mbox{\psfig{figure=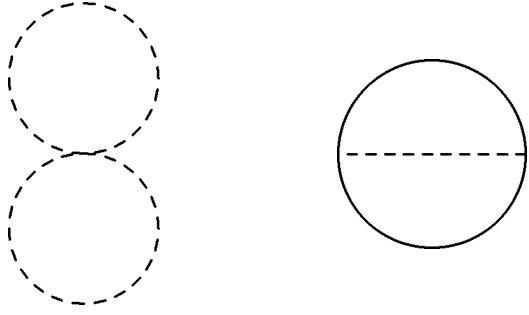}}
\end{center}
\caption[Two-loop vacuum graphs.]{\protect Two-loop vacuum graphs.}
\label{yuk2loop}
\end{figure}

The one-loop diagram with a thermal counterterm insertion is
\beq
-\frac{1}{2}m^2\hbox{$\sum$}\!\!\!\!\!\!\int_{P}
\Big[\frac{1-\delta_{p_0,0}}{P^2}+\frac{\delta_{p_0,0}}{p^2+m^2}\Big]
\delta_{p_0,0}=\frac{m^3T}{8\pi}.
\eeq
Note that we may combine the terms from Eqs.~(\ref{sc2}) and (\ref{theta})
which go like $\lambda mT^3$ and $g^2mT^3$ to obtain 
$\frac{m^3T}{8\pi}$. Hence it cancels the contribution from the 
one-loop diagram with a mass insertion.

\begin{figure}[htb]
\begin{center}
\mbox{\psfig{figure=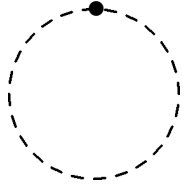}}
\put(-31,65){\makebox(0,0){\large$\bullet$} }    
\end{center}
\caption[One-loop diagram with a thermal counterterm.]{\protect One-loop diagram with a thermal counterterm.}
\label{yct}
\end{figure}

\begin{figure}[htb]
\begin{center}
\mbox{\psfig{figure=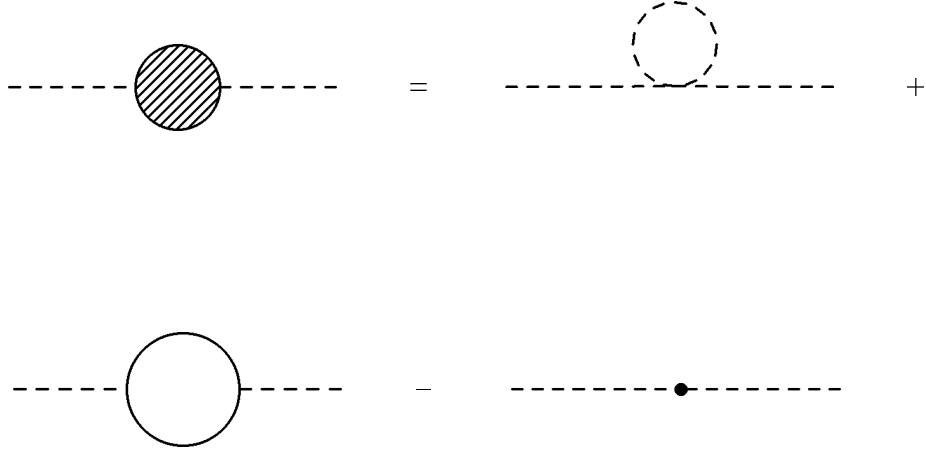}}
\end{center}
\caption[Definition of the shaded blob.]{\protect Definition of the shaded blob.}
\label{ysum}
\end{figure}

Let us move on to the higher order contributions. At the
three-loop level there are three diagrams which are infrared divergent.
However, we also have a one-loop diagram with two mass insertions and a 
two-loop graph with one thermal counterterm insertion. These may 
formally be combined
to a single graph, where the shaded blob denotes the one-loop self-energy
for the scalar field minus the thermal counterterm. This is schematically
displayed in Fig.~\ref{ysum}. The point here is that
this particular combination is infrared finite.
So we may use the bare propagator here, if we are to extract the 
leading contribution, which goes like $\lambda^2$, $\lambda g^2$ 
and $g^4$. For these diagrams, we can then write
\beq
-\frac{1}{4}\hbox{$\sum$}\!\!\!\!\!\!\int_{P}\frac{1}{P^4}[\Delta\Sigma (P)]^2,
\eeq
where
\bqa\nonumber
\label{deltas}
\Delta\Sigma (P)&=&\frac{\lambda}{2}\hbox{$\sum$}\!\!\!\!\!\!\int_{Q}
\frac{1}{Q^2}
-g^2\hbox{$\sum$}\!\!\!\!\!\!\int_{\{Q\}}{\mbox{Tr}}
\Big[\frac{(P\!\!\!\!/+Q\!\!\!\!/)Q\!\!\!\!/}{(P+Q)^2Q^2}\Big]
-m^2\delta_{p_0,0}\\
&=&\frac{\lambda}{2}\hbox{$\sum$}\!\!\!\!\!\!\int_{Q}\frac{1}{Q^2}-
4g^4\hbox{$\sum$}\!\!\!\!\!\!\int_{\{Q\}}\frac{1}{Q^2}+
2g^2\hbox{$\sum$}\!\!\!\!\!\!\int_{\{Q\}}\frac{P^2}{Q^2(P+Q)^2}
-m^2\delta_{p_0,0}.
\eqa

In order to proceed, we must distinguish between the light and the heavy 
modes. We first consider the heavy modes, and  
using the definition of the mass $m^2$ we easily find
\bqa\nonumber
&&-\frac{\lambda^2}{16}\hbox{$\sum$}\!\!\!\!\!\!\int_{PQK}^{\prime}
\frac{1}{P^4Q^2K^2}+\lambda g^2\hbox{$\sum$}\!\!\!\!\!\!\int_{PQ\{K\}}^{\prime}
\frac{1}{P^4Q^2K^2}
-4g^4\hbox{$\sum$}\!\!\!\!\!\!\int_{P\{QK\}}^{\prime}
\frac{1}{P^4Q^2K^2}
\\
\label{pn3}
&&
-g^4\hbox{$\sum$}\!\!\!\!\!\!\int_{P\{QK\}}^{\prime}
\frac{1}{Q^2K^2(P+Q)^2(P+K)^2}-
m^2\hbox{$\sum$}\!\!\!\!\!\!\int_{P\{Q\}}^{\prime}
\frac{1}{P^2Q^2(P+Q)^2}.
\eqa
For the $p_0=0$ mode one obtains
\beq
\label{p03}
-g^4\hbox{$\sum$}\!\!\!\!\!\!\int_{P\{QK\}}
\frac{\delta_{p_0,0}}{Q^2K^2(P+Q)^(P+K)^2}.
\eeq
We see that the fourth term in Eq.~(\ref{pn3}) combines with the
term in Eq.~(\ref{p03}) to the fermionic basketball.
Note also that the last term in Eq.~(\ref{pn3})
cancels the last term in Eq~(\ref{theta}), which follows from the 
fact that the fermionic setting sun diagram vanishes.

\begin{figure}[htb]
\begin{center}
\mbox{\psfig{figure=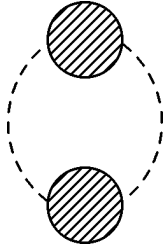}}
\end{center}
\caption[Infrared finite combination of diagrams.]{\protect Infrared finite combination of diagrams.}   
\label{yself3}
\end{figure}

\begin{figure}[htb]
\begin{center}
\mbox{\psfig{figure=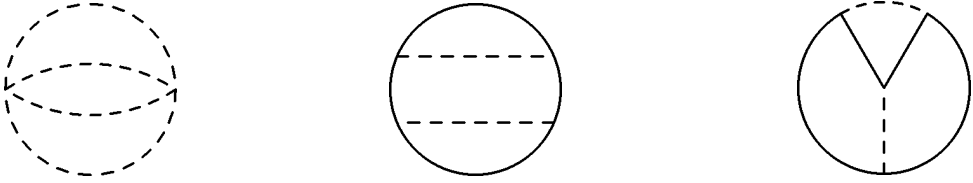}}
\end{center}
\caption[Infrared safe three-loop diagrams in Yukawa theory.]{\protect Infrared safe three-loop diagrams in Yukawa theory.}   
\label{yk3}
\end{figure}

The next task is to consider the infrared-safe three-loop diagrams.
These are depicted in Fig.~\ref{yk3}. 
To leading order in the couplings, we can again
put $m=0$ in the propagators.
The first one is simply the bosonic basketball:
\beq
-\frac{\lambda^2}{48}
\hbox{$\sum$}\!\!\!\!\!\!\int_{PQK}\frac{1}{P^2Q^2K^2(P+Q+K)}.
\eeq
The others stem from the Yukawa sector, and the first one reads
\bqa\nonumber
\frac{1}{2}\hbox{$\sum$}\!\!\!\!\!\!\int_{\{P\}}\mbox{Tr}
\Big[\frac{P\!\!\!\!/}{P^2}\Sigma_{f}(P)\Big]^2
&=&
g^4\hbox{$\sum$}\!\!\!\!\!\!\int_{\{P\}}\frac{1}{P^4}\Big[
\hbox{$\sum$}\!\!\!\!\!\!\int_{\{Q\}}\frac{1}{Q^2}
-\hbox{$\sum$}\!\!\!\!\!\!\int_{Q}\frac{1}{Q^2}\Big]^2\\ \nonumber
&&
-2g^4\hbox{$\sum$}\!\!\!\!\!\!\int_{\{P\}QK}
\frac{QK}{P^2Q^2K^2(P+K)^2(P+Q)^2}\\ 
&&
+
g^4\hbox{$\sum$}\!\!\!\!\!\!\int_{PQ\{K\}}\frac{1}{P^2Q^2K^2(P+Q+K)^2}.
\eqa
Here, $\Sigma_{f}(P)$ is the fermionic self-energy function defined in
Eq.~(\ref{fers}).\\ \\
The second graph yields
\bqa\nonumber
\frac{1}{4}g^4\hbox{$\sum$}\!\!\!\!\!\!\int_{P\{QK\}}\mbox{Tr}\Big[
\frac{Q\!\!\!\!/(P\!\!\!\!/-Q\!\!\!\!/)(P\!\!\!\!/-K\!\!\!\!/)K\!\!\!\!/}{P^2Q^2K^2(P-Q)^2(Q-K)^2(P-K)^2}\Big]&=&\\ 
g^4\hbox{$\sum$}\!\!\!\!\!\!\int_{PQ\{K\}}\frac{1}{P^2Q^2K^2(P+Q+K)^2}
-\frac{1}{2}g^4\hbox{$\sum$}\!\!\!\!\!\!\int_{\{PQK\}}
\frac{1}{P^2Q^2K^2(P+Q+K)^2}.
\eqa

\begin{figure}[htb]
\begin{center}
\mbox{\psfig{figure=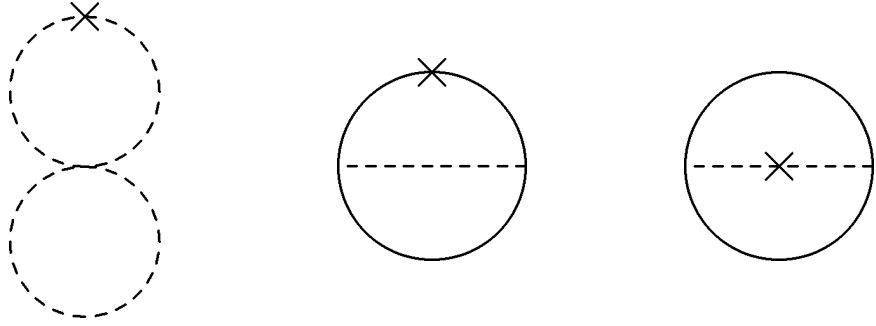}}
\end{center}
\caption[Two-loop vacuum graphs with wave function
counterterm insertions in Yukawa theory.]{\protect Two-loop vacuum graphs with wave function
counterterm insertions in Yukawa theory.}   
\label{yuc22}
\end{figure}

Finally, we have to include the two-loop diagrams with wave function
renormalization counterterm insertions. The mass in the propagator is dropped
for reason that should now be well-known.
The first diagram is the double bubble, which reads
\bqa
\frac{(1-Z_{\phi})\lambda}{4}\hbox{$\sum$}\!\!\!\!\!\!\int_{PQ}\frac{1}{P^2Q^2}.
\eqa
The theta diagram with wave function renormalization counterterms
are displayed in Fig.~\ref{yuc22}. They contribute
\bqa
\Big[(1-Z_{\phi})g^2+2(1-Z_{\psi})g^2\Big]
\Big[2\hbox{$\sum$}\!\!\!\!\!\!\int_{P\{Q\}}
\frac{1}{P^2Q^2}-
\hbox{$\sum$}\!\!\!\!\!\!\int_{\{PQ\}}\frac{1}{P^2Q^2}\Big].
\eqa
The renormalization of the vertices are carried out by using the
expressions for $Z_1$ and $Z_2$ given earlier. 
Alternatively, all ultraviolet divergences at three loops can be canceled
by renormalizing the coupling constants through the 
substitution $\lambda\rightarrow Z_{\lambda}\lambda $
and $g^2\rightarrow Z_{g^2}g^2$ in the two-loop diagrams. Here
\beq
Z_{\lambda}\lambda=\lambda+\frac{3\lambda^2 +8\lambda g^2-48g^4}
{32\pi^2\epsilon},\hspace{1cm}
Z_{g^2}g^2=g^2+\frac{5g^4}{16\pi^2\epsilon}.
\eeq

Putting our results together, using Appendix A,
we finally obtain the free energy through order
$\lambda^2$, $\lambda g^2$ and $g^4$:
\bqa\nonumber
\label{elg} 
\!\!\!\!\!\!\!\!\!\!\!\!\!\hspace{-2.0cm}
{\cal F}&=&-\frac{9\pi^2T^4}{180} -\frac{m^3T}{12\pi}
+\frac{\lambda}{8}\Big(\frac{T^2}{12}\Big)^2
+\frac{5g^2}{4}\Big(\frac{T^2}{12}\Big)^2\\ \nonumber
&&-\frac{\lambda^2}{16\pi^2}\Big(\frac{T^2}{12}\Big)^2
\Big[\frac{3}{8}\ln\frac{\Lambda}{4\pi T}+\frac{1}{8}\gamma_{E}
-\frac{59}{120}-\frac{2}{3}\frac{\zeta^{\prime}(-3)}{\zeta (-3)}
+\frac{1}{2}\frac{\zeta^{\prime}(-1)}{\zeta (-1)}
\Big]\\ \nonumber&
&-\frac{\lambda g^2}{16\pi^2}\Big(\frac{T^2}{12}\Big)^2
\Big[\ln\frac{\Lambda}{4\pi T}+\gamma_{E}
-\ln2\Big]\\ 
\!\!\!&&-\frac{g^4}{16\pi^2}\Big(\frac{T^2}{12}\Big)^2
\Big[\frac{13}{2}\ln\frac{\Lambda}{4\pi T}+2\gamma_{E}
-\frac{675}{80}-\frac{127}{10}\ln2
-\frac{9}{2}\frac{\zeta^{\prime}(-3)}{\zeta (-3)}
+9\frac{\zeta^{\prime}(-1)}{\zeta (-1)}
\Big].
\eqa
This is the main result of the chapter on resummation and has not appeared
in the literature before.
It is easily checked that our result is renormalization group invariant
as usual. Moreover, it coincides to order $\lambda^2$ with previous
results when $g=0$~\cite{frenkel}.


\cleardoublepage
\chapter{Effective Field Theory Approach I}
\heading{Effective Field Theory Approach I}{Effective Field Theory Approach I}
\section{Introduction}
\heading{Effective Field Theory Approach I}{Introduction}
The ideas of effective field theory or low energy Lagrangians have a rather
long history and dates back to the early work of Euler and Heisenberg
in the thirties, where they constructed an effective Lagrangian for QED,
which could be used to compute low energy photon-photon 
scattering~\cite{euler}.
In recent years, the applications of effective field theory ideas
in various branches of physics have exploded.
Most applications have been to systems at zero temperature 
(Refs. [59-65] and Refs. therein), but there is an 
increasing number of papers devoted to the study of effective field theories
at finite temperature [66-76].
It is the purpose of this section to introduce the basic ideas of effective
field theory in a rather general setting, and use the Euler-Heisenberg as
a concrete example. We will also briefly discuss some major applications
of effective Lagrangians at $T=0$, that have appeared in the literature
in the last couple of years.
The introduction to effective field theories at finite temperature is
deferred to the next section.

The improved understanding
of effective Lagrangian and the modern developments in renormalization theory
have also led to some nice introductory papers to the 
subject, and we recommend the articles by Kaplan~\cite{kaplan1},
by Manohar~\cite{manohar} and by Lepage~\cite{lepage}.

Now, what are the ideas of effective field theory and when can they be 
applied?
Assume that we have a field theory which contains light particles of mass
$\sim m$, and heavy particles of mass $\sim M$, where $m\ll M$ so that
one can speak of a {\it mass hierarchy}.
Consider a process, e.g. scattering, which is characterized by energies
far below $M$, so that no real heavy particles can be produced.
It is then reasonable to believe that there exists an effective field theory
for the light fields, which yields identical predictions for physical
quantities as the full theory in the low energy domain.
This is an example of a general idea that pervades all physics;
The detailed dynamics at high energies is irrelevant for the understanding of low energy
phenomena. Of course, the high energy fields do affect the low energy
world, but their effects may be fully absorbed into the parameters of the
low energy Lagrangian.

It is then the purpose of effective field theory methods to construct this
effective
Lagrangian which reproduces the full theory in the low energy
domain, without its full complexity. 
Effective field theory  
ideas can, loosely speaking, 
be applied to any physical system with two or more distinct
energy scales. One takes advantage of the separation of scales in the problem
and treats each scale separately. This streamlines calculations, since we
do not mix them. 

The effective field theory program can conveniently be summarized in the
following points~\cite{br2}:

\begin{itemize}
\item Identify the low energy fields (the particle content) from which
the effective Lagrangian is built.
\end{itemize}

\begin{itemize}
\item Identify the symmetries which are present at low energies. 
\end{itemize}

\begin{itemize}
\item Write down the most general local effective Lagrangian, which consists
of all terms that can be built from the low energy fields, consistent
with the symmetries.
\end{itemize}

\begin{itemize}
\item The effective field theory can reproduce the full theory to any desired
accuracy in the low energy domain by including sufficiently many operators
in the effective Lagrangian. Specify this accuracy.
\end{itemize}

\begin{itemize}
\item Determine the coefficients in the effective Lagrangian by calculating
physical quantities at low energies in the two theories and demand that they
be the same. This procedure is called {\it matching}.
\end{itemize}

The fourth point above is in some sense a generalization of the
Appelquist-Carrazone theorem~\cite{appel}:
Consider correlators or Greens functions in the full theory
with only light fields on the
external legs, which are characterized by momenta $k\ll M$.
The Appelquist-Carrazone decoupling theorem then says that the Greens
functions of the full theory can be reproduced up to corrections of
order $k/M$ and $m/M$ by a renormalizable field theory involving only the
light fields~\cite{appel}. 
By adding more and more operators to the effective Lagrangian and tuning
the parameters, Greens functions can be reproduced to any specified
accuracy. 

Schematically, the effective field theory can be written as 
\beq
{\cal L}_{\mbox{\scriptsize eff}}={\cal L}_{0}+\sum_n\frac{O_n}{M^n},
\eeq
where we explicitly have isolated the renormalizable part, ${\cal L}_0$, of 
${\cal L}_{\mbox{\scriptsize eff}}$, and $O_n$ are operators of dimension $n$.
The expansion in powers of $k/M$ and $m/M$ is referred to as the
{\it low energy expansion}.

The form of the effective Lagrangian can be very well understood in terms
of the Wilsonian approach to the renormalization group~\cite{wilson}.
The starting point is the Euclidean path integral representation of the
generating functional of the Greens functions in the full theory.
We exclude the high energy modes in the path integral
by using a cutoff $\Lambda$. This should be chosen to be much
larger than $M$.
The effects of the scale $M$ have two sources; high energy modes of the
light fields and all modes of the heavy fields.
One way to isolate the effects of the scale $M$ is to introduce a 
new cutoff $\Lambda^{\prime}$, 
so that $m\ll\Lambda^{\prime}\ll M$, and to integrate over
all modes larger than $\Lambda^{\prime}$ for the light fields
and over {\it all} modes for the heavy fields.
The latter integration means that we actually eliminate
the heavy fields from the path integral (integrating out the heavy
fields).
One is then left with a theory for the small momentum or long distance
modes of the light field, 
and this has infinitely many terms. In this process the 
coupling constants get modified by the high energy modes, which is nothing
but a renormalization of the parameters. The physics on the scale
$M$ is now encoded in the coupling constants.  
The fact that it 
is a local field theory follows from the Heisenberg uncertainty principle.
The modes with momentum of order $M$ are highly virtual
and can  only propagate over a distance
of the order $1/M$. So, at the scale $m$, this looks local and
the effects of these
virtual states can be mimicked by local interactions~\cite{lepage}. 

Let us take QED as an example. In this case the low energy field is the
photon field. The symmetries are Lorentz invariance, gauge invariance,
charge conjugation symmetry, parity, and time reversal.
The next task is to specify the precision of the low energy theory.
Since we are interested in describing photon-photon scattering we must include
interaction terms in the effective Lagrangian. Let us for simplicity
confine ourselves to the first nontrivial order in the low energy expansion.
The most general Lagrangian 
to order $k^4/m^4$ (note that the electron mass is denoted by $m$, which
is the heavy scale),
satisfying the above requirements is the famous
Euler-Heisenberg Lagrangian~\cite{euler}:
\beq
{\cal L}_{\mbox{\scriptsize eff}}=-\frac{1}{4}F_{\mu\nu}F^{\mu\nu}
+\frac{a}{m^2}F_{\mu\nu}\Box F^{\mu\nu}+\frac{b}{m^4}
(F_{\mu\nu}F^{\mu\nu})^2
+\frac{c}{m^4}(^{\ast}F_{\mu\nu}F^{\mu\nu})^2.
\eeq
Here, $a$, $b$ and $c$ are dimensionless constants. 
The parameter $a$ is determined by considering the one-loop correction
to the photon propagator in QED to leading order in $k^2/m^2$.
The parameters $b$ and $c$ are 
determined by calculating the scattering amplitude
for photon-photon scattering at energies well below the electron mass $m$
in full QED and in the effective theory, and require that they be the same.
These parameters can be written as power series in $\alpha$, and the  
first term in this expansion was found by Euler and Heisenberg in 
Ref.~\cite{euler}. 
Recently, $b$ and $c$ have been determined to order $\alpha^3$ 
by Reuter {\it et al.} using string inspired methods~\cite{string} (See
also ref.~\cite{string2}).

\begin{figure}[htb]
\begin{center}
\mbox{\psfig{figure=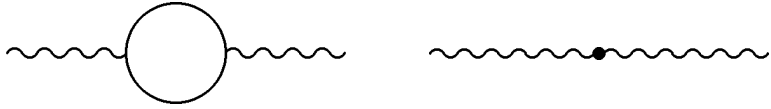}}
\end{center}
\caption[Matching two-point functions at low energy.]{\protect Matching two-point functions at low energy.}
\label{correct}
\end{figure}

\begin{figure}[htb]
\begin{center}
\mbox{\psfig{figure=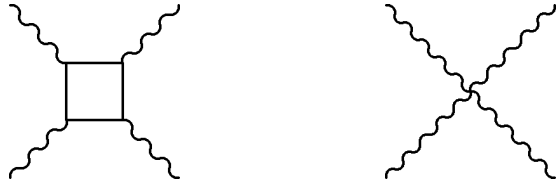}}
\end{center}
\caption[Light-by-light scattering 
in full QED, which can be mimicked
by a local interaction on the scale $\Lambda\ll m$.]{\protect Light-by-light scattering 
in full QED, which can be mimicked
by a local interaction on the scale $\Lambda\ll m$.}
\label{low}
\end{figure}

\begin{figure}[htb]
\begin{center}
\mbox{\psfig{figure=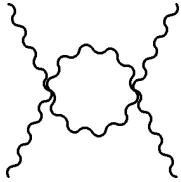}}
\end{center}
\caption[One-loop correction to $\gamma$-$\gamma$ scattering
in the Euler-Heisenberg Lagrangian.]{\protect One-loop correction to $\gamma$-$\gamma$ scattering
in the Euler-Heisenberg Lagrangian.}
\label{onecorr}
\end{figure}

After the constants $a$, $b$ and $c$ have been determined, we should be able
to use it in the study of photon-photon scattering at low energy.
At the tree level this presents no problem, but according to the traditional
view on nonrenormalizable field theories it cannot be used at the loop level.
Lets us consider this in some detail, and explain why the old view is 
incorrect. The operators $\frac{b}{m^4}
(F_{\mu\nu}F^{\mu\nu})^2$ and 
$\frac{c}{m^4}(^{\ast}\!F_{\mu\nu}F^{\mu\nu})^2$ give rise to 
$\gamma$-$\gamma$ scattering, and they are nonrenormalizable, since their
coupling constants have negative mass dimension.
At the tree level, these operators reproduce the scattering amplitude
in full QED up to corrections of 
order $k^6/m^6$ by construction, and all is well.
The one-loop correction to photon-photon scattering is depicted in 
Fig.~\ref{onecorr} and was first computed by Halter in 
Ref.~\cite{halter}\footnote{Halter's computation is correct, but it is does
not provide the 
complete result to order $k^8/m^8$. The amplitude in the tree approximation
is proportional to $b$ and $c$, and consistency requires that these
constants be determined to order $\alpha^4$.}.

The amplitude is dimensionless, and
each vertex gives a factor $1/m^4$. Hence, the integral must have dimension
eight. If we use dimensional regularization, power divergences
are set to zero, and logarithmic divergences show up as poles in 
$\epsilon$~\cite{priv}.

The only mass scale in the integral is the external momenta $k_i$, and
one will encounter divergent terms with more than four powers of $k_i$.
In order to render the amplitude finite, we must renormalize and absorb
the divergences in the counterterms. However, there is no operator
in the Euler-Heisenberg Lagrangian, which has dimension
five or more.
So, to get rid of the infinities, we must add
one or more operators to the low energy Lagrangian
which have the right dimension.
Thus, we must introduce one or more coupling constants that must be 
determined from QED. Or, if we did not know the underlying theory, we had
to carry out experiments to determine them. 
This argument can be repeated for any operator of a given dimension, and
to any order in the loop expansion. This implies 
that ${\cal L}_{\mbox{\scriptsize eff}}$ must in principle
contain infinitely many operators, and that we have to determine infinitely
many coupling constants. This is, of course, an impossible task, and has led
people to the conclusion that the effective Lagrangian is useless and 
without predictive power beyond the tree approximation.
This conclusion is not quite correct for the following reason:
In order to reproduce the values of physical quantities to some desired
accuracy, there is only a limited number of interaction terms 
in ${\cal L}_{\mbox{\scriptsize eff}}$, which has to be retained in the
low energy expansion. The others are simply too suppressed by the heavy scale.
So, if we are satisfied with finite precision, we only have to know a limited
number of coupling constants in the effective Lagrangian, and this is 
naturally possible to calculate.
This is relevant, because experiments are always performed with a finite
precision.
Hence, the effective Lagrangian is as good as any other quantum field
theory, and can be used in practical calculations.

The reader might nevertheless
wonder: what is the point of constructing a low energy
theory of photons? After all we do know the underlying theory, and full
QED is probably not more difficult to use than 
${\cal L}_{\mbox{\scriptsize eff}}$. This may be so in the scattering
example, but it does serve as an illustration of the general philosophy.
Moreover, the Euler-Heisenberg Lagrangian has other applications.
A recent example can be found in Ref.~\cite{kong}, where Kong and Ravndal
calculate the lowest radiative correction to the energy density
for an interaction photon gas at temperature $T\ll m$. 
The relevant vacuum diagram being the photon double-bubble.
The idea is again that of two widely 
separated mass scales $m$ and $T$, and the correction goes like
$T^8/m^4$.

There is another important point we wish to make at this stage. We have
argued that renormalizability is no longer a requirement  of a useful and
consistent quantum field theory. Why, then, is the very succesfull
standard model renormalizable? 
This reflects the fact that the new physics first enters at a scale
well above the one tested in this generation of accelerators.
Moreover, there is today a large activity in the search for physics beyond
the standard model, and future precision tests may very well
reveal the presence of nonrenormalizable interactions which are
signs of new physics. 
The bottom line is that before we have the final theory of everything
(string theory or whatever) which applies to any energy scale,
every field theory should be viewed as an effective field theory
valid and with predictive power at a certain scale.
Furthermore, even if we did know the theory of everything, it is 
unlikely that it will be of any use in low energy physics e.g. 
condensed matter.
Reductionism in physics is more a principal question than a practical one.
Thus, effective field theory methods are likely to be with us in the
future as one of the most important tools 
for practical calculations.

The full theory may not always be used at low energies, the most prominent
example of this is quantum chromodynamics. QCD is a strongly interacting
theory at low energies, and perturbation theory is useless in this domain.
Hence, effective field theory methods are mandatory to apply. 
The essential ingredients in the construction of the effective Lagrangian
is again symmetries and particle content. At the lowest energies only
pions are present, and the symmetry is (an approximate) chiral symmetry
(in addition to the space-time symmetries etc).
The pions are approximate Goldstone bosons and in the chiral limit,
they are massless. Chiral symmetry puts severe restrictions on the
possible terms in the Lagrangian.
Although the nonlinear sigma model has been around for many
years, it is only in the last decade or so, it has been applied
beyond tree level~\cite{chiral}.

We shall give yet a few examples of effective field theory,
which have received much attention in recent years.
These are nonrelativistic QED and nonrelativistic QCD, which are effective
field theories that are applied to bound states and were formulated
by Caswell and Lepage~\cite{caswell}.
This is an elegant alternative to the traditional approaches such as the
Bethe-Salpeter equation, which do not take advantage of the nonrelativistic
nature of the problem. Instead, it mixes contributions from different scales,
making explicit calculations unnecessarily difficult.
Let us for simplicity consider NRQED, which has been applied by several authors
the last few years [63,64].
The form of the Lagrangian is again uniquely determined by particle content
and symmetries. The fields present in NRQED are the two-component
electron field, the electromagnetic field, and other fermion fields necessary
for the actual problem. The symmetries are Galilean invariance, gauge 
invariance, time reversal symmetry and parity.
The coupling constants in NRQED are determined by the requirement that
it reproduces full QED at low energy. This can be obtained by calculating
scattering amplitudes in the two theories e.g. at threshold and demand
that they be the same. 
This implies that the effects from nonrelativistic momenta are taken care
of by NRQED, and that the effects of relativistic momenta are encoded in the
parameters of the theory~\cite{caswell}.
After the determination of the parameters in NRQED, one uses time ordered
perturbation theory with the usual Schr\"odinger wave functions as the 
unperturbed states.

Finally, there is general relativity.
The modern ideas of effective field theory and
renormalization theory have recently been applied to general relativity
by Donoghue in a series of papers~\cite{gr}. 
Conventional wisdom says that it is impossible to construct a
meaningful quantum theory of general relativity, since the Lagrangian
is nonrenormalizable. This apparent incompatibility of gravitation
and quantum mechanics
was considered as one of the greatest
problems in theoretical physics. This is no longer so.
As long as we are well below the
heavy scale, which in this case probably is the Planck scale, it 
is perfectly possible
to quantize gravity and it is a completely consistent theory at 
present energies.
The fact that classical general relativity is in accordance with 
measurements simply reflects that the non-leading terms in the Lagrangian
are strongly suppressed by the heavy scale $M_p$.

Of course there exist systems where the effective field theory program
does not apply. This is the case if the energy scales in a 
system under consideration 
are not widely separated. Consider e.g. a hydrogenic ion with a large
nuclear charge $Z$. This system is relativistic, which means
the momentum of the electron is not much smaller than its mass.
NRQED is therefore
not a particularly useful approach and the low energy expansion converges 
very poorly.  

\section{Finite Temperature}\label{fin}
\heading{Effective Field Theory Approach I}{Finite Temperature}
Let us now begin our discussion of effective field theories at finite 
temperature.\\ \\
In the imaginary time formalism (ITF) of quantum field theory, there
exists a path integral representation of the partition function~\cite{gins}:
\begin{equation}
{\cal Z}=\int {\cal D}\phi\exp\Big[-\int_{0}^{\beta}d\tau\int d^{3}x\,
{\cal L}_E\Big ].
\end{equation}
Here, $\phi$ is a generic field, and ${\cal L}_E$ is the Euclidean Lagrangian
which is obtained from the usual Lagrangian after Wick rotation,
$t\rightarrow -i\tau$.
Moreover, the boundary conditions in imaginary time are that
bosonic fields are
periodic in the time direction with 
period $\beta$ and that fermionic fields are
antiperiodic with the same period. \\ \\
The (anti)periodicity implies that we can
decompose the fields into Fourier components characterized by their 
Matsubara frequencies:
\bqa
\label{exp}
\Phi ({\bf x},\tau)&=&\beta^{-\frac{1}{2}}\Big [\,\phi_{0}({\bf x})
+\sum_{n\neq 0}\phi_{n}({\bf x})e^{2\pi in\tau/\beta}\,\Big ],\\
\Psi ({\bf x},\tau)&=&\beta^{-\frac{1}{2}}\sum_{n}\psi_{n}({\bf x})
e^{\pi i(2n+1)/\beta}\,\,.
\end{eqnarray}
The $n=0$ bosonic mode is called a {\it light} or {\it static} mode, 
while the $n\neq 0$ modes
as well as the fermionic modes are termed {\it heavy} or {\it nonstatic}. 
In the ITF, we can therefore
associate a free propagator
\beq
\Delta_n(k_0,k)=\frac{1}{k^2+\omega_n^2}
\eeq
with 
the $n$th Fourier mode. Hence, a quantum field theory at finite temperature
may be viewed as an infinite tower of fields in three dimensions, where
the Matsubara frequencies act as tree-level masses of order $T$ for the
heavy modes, while the light mode is actually {\it massless}.

In the preceding chapter we have seen that the scalar field acquires a thermal
mass of order $gT$ 
at the one-loop level. For the heavy modes this represents
a perturbative correction which is down by a power of the coupling, and these
modes are still characterized by a mass of order $T$. However, the light
mode is no longer massless, but its mass is of order $gT$.
Hence, we conclude that we have two widely separated mass scales at high
temperature, which are $T$ and $gT$. 
The Appelquist-Carrazone decoupling
theorem then suggests that the heavy modes decouple
on the scale $gT$, and that we are left with an effective Lagrangian of
the static mode. The process of going from a full four dimensional theory
to an effective three dimensional Lagrangian is called 
{\it dimensional reduction}.
This is the key observation and the starting point for
the construction of effective field theories at finite temperature.
These ideas were first applied in the eighties, and the 
main contributions from this period can be found in the papers of
Ginsparg~\cite{gins}, Jourjine~\cite{jour}, and Landsman~\cite{lands}.
The parameters in the effective Lagrangian were determined by considering
the one-loop corrections from the nonstatic modes to the static $n$-point
functions. In other words, only nonstatic modes circulate around in the loops
and one speaks about {\it integrating out} the heavy modes.
Thus, the effects of the scale $T$ is now encoded in the parameters of
the ${\cal L}_{\mbox{\scriptsize eff}}$, while the low energy
effects should be fully accounted for by the effective field theory.

Now, it was realized by Landsman that the effective Lagrangian does
not completely reproduce the underlying theory~\cite{lands}. This was taken
as an indication that dimensional reduction only takes place 
approximately.
However, this apparent failure of dimensional reduction
reflects the fact that only renormalizable field theories were considered.
If one exploits the effective field theory program fully, and allows
for nonrenormalizable interactions, dimensional reduction does take place
in accordance with expectations~\cite{agus}.

Let us discuss the effective Lagrangian in the case where the
underlying theory is $\lambda\phi^4$-theory. In the Feynman graphs
below, light modes are indicated by dotted lines, while heavy modes
are denoted by solid lines.

First, we must identify the symmetries. We have a Z$_2$ symmetry 
$\phi\rightarrow -\phi$, which follows from the corresponding symmetry in
the full theory. There is also a three dimensional rotational symmetry.
Hence, we can write
\beq
{\cal L}_{\mbox{\scriptsize eff}}=
\frac{1}{2}(\partial_i\phi_0)^2+\frac{1}{2}m^2(\Lambda)\phi^2_0
+\frac{\lambda_3(\Lambda)}{24}\phi^4_0
+\frac{g(\Lambda)}{6!}\phi^6_0+h_1(\Lambda)
\phi^2_0\nabla^2\phi^2_0+\frac{h_2(\Lambda)}{8!}\phi^8_0+...
\eeq
Of the operators we have listed above, only the last one is nonrenormalizable.
The coupling constants generally depend on the ultraviolet cutoff
or the renormalization scale $\Lambda$. This is also the case for the
field, but normally we shall suppress this dependence for notational ease.
In our calculations, we use dimensional 
regularization, which, by definition sets the power divergences to zero, and
where the logarithmic divergences show up as poles 
in $\epsilon$~\cite{priv}\footnote{The power divergences are unphysical in the sense that
the depend upon the regulator, and so dimensional regularization is a
particular convenient choice.}.

Now, the coupling constants in the ${\cal L}_{\mbox{\scriptsize eff}}$ are
not arbitrary but determined by our {\it matching condition}, namely the
requirement that the {\it static correlators} $\Gamma^{(n)}(0,{\bf k})$ 
in the full theory
are reproduced to some desired
accuracy by the correlators in the effective theory at distances $R\gg 1/T$.
This matching requirement was first introduced by Braaten and 
Nieto~\cite{braaten}, and independently by Kajantie {\it et al.}~\cite{laine}.
Below, we shall comment on the connection between the matching of Greens
function and the old way of integrating out the nonstatic modes.

The parameters in the effective Lagrangian are determined by ordinary 
perturbation theory in $\lambda$, or $g^2$, if we consider gauge theories.
In the underlying theory this naturally corresponds to the following
partition of the Lagrangian into a free part and an interacting part
\bqa
{\cal L}_{0}&=&\frac{1}{2}(\partial_{\mu}\Phi)^2\\
{\cal L}_{\mbox{\scriptsize int}}&=&\frac{\lambda}{24}\Phi^4.
\eqa
In the effective theory $m^2(\Lambda)$, $\lambda_3(\Lambda)$ 
and $h_1(\Lambda)$ are all of order 
$\lambda$, while other constants are of order $\lambda^2$ or even higher
(although the operator $\phi^2\nabla^2\phi^2$ first contributes to the
screening mass at order $\lambda^{5/2}$ and to the free energy at order 
$\lambda^3$). This implies that we split the Lagrangian according to
\bqa
({\cal L}_{\mbox{\scriptsize eff}})_{0}&=&\frac{1}{2}(\partial_i\phi_0)^2\\
({\cal L}_{\mbox{\scriptsize eff}})_{\mbox{\scriptsize int}}&=&
\frac{1}{2}m^2(\Lambda)\phi^2_0
+\frac{\lambda_3(\Lambda)}{24}\phi^4_0+...
\eqa
This way of carrying out perturbative calculations to determine the 
coupling constants was first
introduced by Braaten and Nieto in Ref.~\cite{braaten}, 
and they refer to it as 
``strict perturbation theory'' (see also chapter four).
Now, we know from our previous discussion that strict perturbation theory
is afflicted with infrared divergences, which are due to the masslessness of 
the fields. These divergences become more and more severe in the loop 
expansion, but we can nevertheless use it as a device to determine the
coupling constants in ${\cal L}_{\mbox{\scriptsize eff}}$.
Below, we shall demonstrate that identical infrared divergences appear
in the perturbation expansion in the effective theory. Thus perturbation
theory breaks down in exactly the same way in the two theories, and
the infrared divergences in the matching equations cancel.
The coupling constants in ${\cal L}_{\mbox{\scriptsize eff}}$
are only sensitive to the scale $T$ at which perturbation theory works fine.
We are namely integrating out the scale $T$, since the heavy modes
have masses of this order, and the parameters encode the physics at this
scale.
So it does not matter that we make some incorrect assumptions about 
the physics on the
scale $gT$ (we dot use resummation).
However, when we use the effective three-dimensional theory in real
calculations, we must 
take the screening effects properly into account. This amounts to
including the mass parameter in the free part of the Lagrangian, and 
treat the other operators as perturbations. 
 
Let us now carefully demonstrate how this approach works in practical
calculations. We shall outline how one determines the mass parameter
at the two-loop level. We must start by determining the tree level Lagrangian.
This is carried out by substituting the expansion of the scalar field in 
Eq.~(\ref{exp}) into the path integral, and integrating over $\tau$, using the
orthogonality of the modes. The expression we obtain contains terms
which are made exclusively up of the static modes, terms that contain
only nonstatic modes, and products of light and heavy modes. 
We can then read off the coefficients of the operators in
${\cal L}_{\mbox{\scriptsize eff}}$ by comparing it to the part of the
Lagrangian in the full theory that contains only the zero-frequency 
modes~\cite{agus}.
The reader may convince herself that the only nonzero coefficient in the
tree approximation, is the parameter
in front of the quartic coupling; $\lambda_3(\Lambda)=\lambda T$. 

\begin{figure}[htb]
\begin{center}
\mbox{\psfig{figure=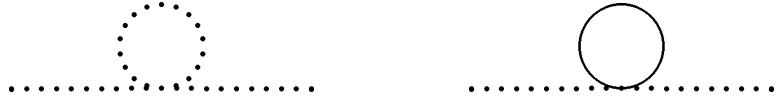}}
\end{center}
\caption[The tadpole graph, whose contributions from light and
heavy modes have been separated.]{\protect The tadpole graph, whose contributions from light and
heavy modes have been separated.}
\label{ys1e}
\end{figure}

At the one-loop level in the full theory, there is one contributing
diagram, namely the tadpole. 
This is depicted in Fig.~\ref{ys1e}, where we have 
explicitly separated the contributions to the static two-point function
from the static mode and the heavy modes. One finds
\beq
\Gamma^{(2)}(0,{\bf k})=k^2+\frac{\lambda T}{2}\int_p\frac{1}{p^2}
+\frac{\lambda}{2}\hbox{$\sum$}\!\!\!\!\!\!\int_P^{\prime}\frac{1}{P^{2}}.
\eeq
Here, the prime indicates that the $n=0$ mode has been left out from the sum.

\begin{figure}[htb]
\begin{center}
\mbox{\psfig{figure=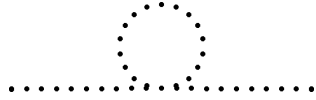}}
\end{center}
\caption[The one-loop diagrams in the effective theory appearing
in the matching procedure.]{\protect The one-loop diagrams in the effective theory appearing
in the matching procedure.}
\label{yssc1}
\end{figure}

In the effective theory, the only relevant operators are $m^2(\Lambda)\phi_0^2$
and $\lambda_3(\Lambda)\phi_0^4$. The corresponding contributions to the 
two-point function are shown in Fig.~\ref{yssc1}. 
The corresponding expression is
\beq
\Gamma^{(2)}_{\mbox{\scriptsize eff}}(k)=k^2+m^2(\Lambda)+
\frac{\lambda_3(\Lambda)}{2}
\int_{p}\frac{1}{p^2}.
\eeq
Demanding that these expressions
be the same, determines the mass parameter to order $\lambda$:
\bqa\nonumber
m^2(\Lambda)+\frac{\lambda_3(\Lambda)}{2}\int_p\frac{1}{p^2}&=&
\frac{\lambda T}{2}\int_p\frac{1}{p^2}+
\frac{\lambda}{2}
\hbox{$\sum$}\!\!\!\!\!\!\int_P^{\prime}\frac{1}{P^{2}}.
\eqa
Exploiting the fact that $\lambda_3(\Lambda)=\lambda T$ at leading order,
we see that the second term on the left hand side cancels the
first term on the right hand side. 
(Incidentally, this term is set to zero in dimensional regularization,
since there is no scale in the integral, but that is besides the point).
Hence
\beq
\label{massone}
m^2(\Lambda)=
\frac{\lambda}{2}
\hbox{$\sum$}\!\!\!\!\!\!\int_P^{\prime}\frac{1}{P^{2}}.
\eeq
From Eq.~(\ref{massone}), we conclude that the mass parameter is
determined by the effects of the nonstatic modes circulating in the loop.
This is actually a general feature of our matching procedure at the one-loop 
level; the first quantum correction to a coupling constant (which is the 
leading term in the expansion if the corresponding operator is not present
at the classical level), is determined by the effects of the heavy modes
in the loop.
The matching procedure at the one-loop order coincides,
not unexpectedly, with the original approach to dimensional reduction and 
effective field theories at finite temperature~\cite{lands}. 

At two-loop order, it becomes more complicated. Integrating out the heavy modes
in the above meaning of the word, implies that one considers the effects
of heavy modes in the loops. This is problematic, since it has been 
demonstrated by Jakov\'ac that it generally
produces non-local operators that cannot be expanded in powers of 
$k/T$~\cite{jakoleik}.
Thus, it is difficult to construct a local effective field theory.
A similar problem appeared in the study of QED many years ago. Ovrut
and Schnitzer
investigated 
QED with both light and heavy fermions~\cite{ovrut}. 
They wanted to construct a low-energy
theory containing the photon and the light fermion. Beyond one loop, they 
realized that one had to include both light and heavy fermions in two-loop
graphs in order to obtain an effective theory, which reproduced
the full theory at low energy.
The solution to the problem at finite temperature is similar; One must be 
careful and consider diagrams with both static and nonstatic modes on
internal lines.

\begin{figure}[htb]
\begin{center}
\mbox{\psfig{figure=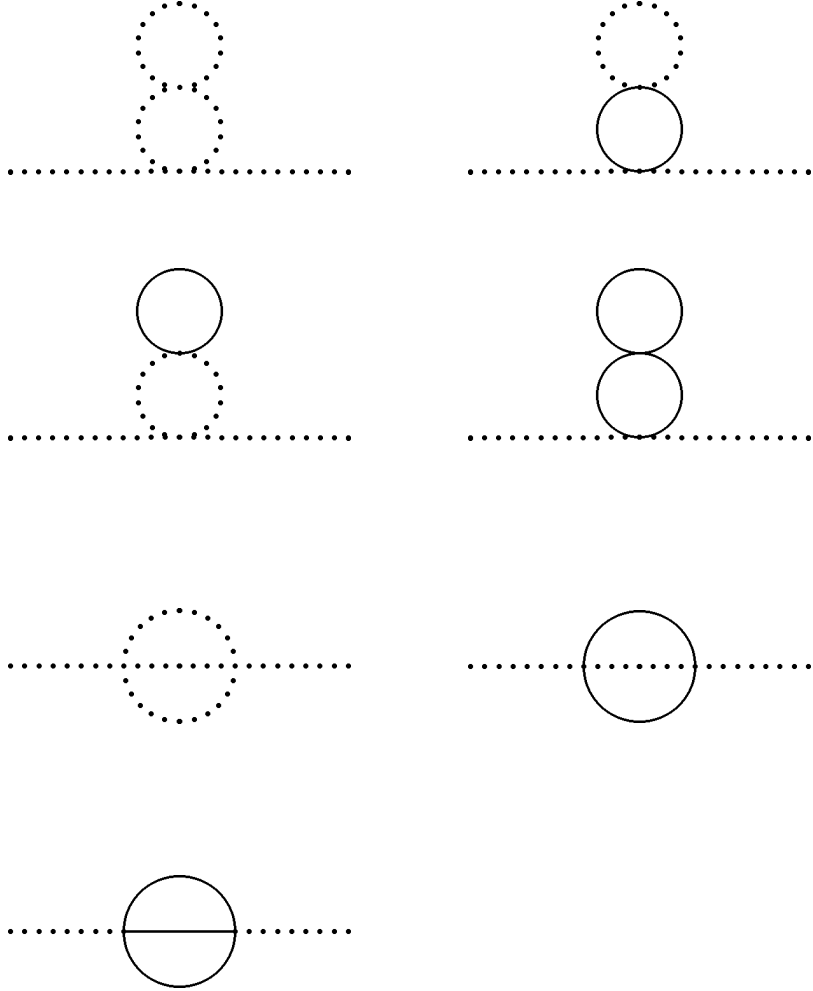}}
\end{center}
\caption[The two-loop graphs for the two-point function, 
where the contributions from the light and heavy particles have been
separated  explicitly.]{\protect The two-loop graphs for the two-point function, 
where the contributions from the light and heavy particles have been
separated  explicitly.}   
\label{yuc2}
\end{figure}

This is illustrated in Fig~\ref{yuc2}, where
we have displayed the graphs that contribute to the
scalar self-energy function at two loops.
The two-point function receives contributions from diagrams with only light
lines, only heavy lines, with both light and heavy particles.

Firstly, we would like to point out that the momentum dependence of the 
setting sun diagram is irrelevant for the present calculation. The momentum
dependence of loop diagrams contributing to $\Gamma^{(2)}(0,{\bf k})$
gives rise to the renormalization of the fields in the effective theory.
However, since this occurs at the two-loop level, this redefinition or
renormalization first comes into play at the three-loop level 
(order $\lambda^3$).

\begin{figure}[htb]
\begin{center}
\mbox{\psfig{figure=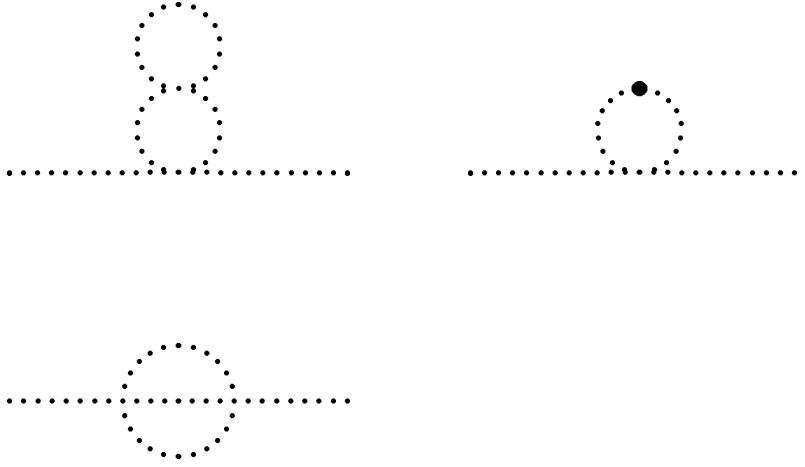}}
\end{center}
\caption[The two-loop graphs for the two-point function, 
in the three-dimensional theory.]{\protect The two-loop graphs for the two-point function, 
in the three-dimensional theory.}
\label{ys2e}
\end{figure}

We note that the two diagrams which contain only light 
modes (diagrams one and five) cancel against the corresponding graphs in the
effective theory (see Fig~\ref{ys2e}), exactly as at one loop.
Moreover, the first of these diagrams in linearly infrared divergent, while
the second has a logarithmic divergences in both 
the infrared and the ultraviolet. 
\newpage

The second diagram is zero, since the upper loop vanishes, while the
third diagram is infrared divergent and reads
\beq
-\frac{\lambda^2T}{4}
\int_p\frac{1}{p^4}\hbox{$\sum$}\!\!\!\!\!\!\int_Q^{\prime}\frac{1}{Q^{2}}.
\eeq
However, this diagram is canceled by the diagram in the effective theory
with a mass insertion. This graphs reads
\beq
-\frac{m^2(\Lambda)\lambda_3(\Lambda)}{2}
\int_p\frac{1}{p^4}.
\eeq
Consistency in the matching procedure, requires that we use the 
parameters $m^2(\Lambda)$ and 
$\lambda_3(\Lambda)$ at leading order in $\lambda$, and the
cancelation then follows.
The fourth diagrams gives a contribution
\beq
-\frac{\lambda^2}{4}
\hbox{$\sum$}\!\!\!\!\!\!\int_{PQ}^{\prime}\frac{1}{P^2Q^{4}},
\eeq
while the sum of the sum of the last two diagrams is
\beq
-\frac{\lambda^2}{6}\hbox{$\sum$}\!\!\!\!\!\!\int_{PQ}\frac{1-\delta_{p_0,0}\delta_{q_0,0}}{P^2Q^{2}(P+Q)^2}.
\eeq
Using the methods of appendix C, one can
demonstrate that the above sum-integral vanishes
in dimensional regularization. We can then summarize our discussion
in the following matching equation:
\beq
m^2(\Lambda)
+\delta m^2=\frac{\lambda Z_{\lambda}}{2}
\hbox{$\sum$}\!\!\!\!\!\!\int_{P}^{\prime}
\frac{1}{P^2}-\frac{\lambda^2}{4}
\hbox{$\sum$}\!\!\!\!\!\!\int_{PQ}^{\prime}\frac{1}{P^2Q^{4}}.
\eeq
Note that we have included a mass counterterm on the left hand, which
is necessary to cancel the divergence which is still left after the
renormalization
of the coupling constant\footnote{This divergence is related to the 
logarithmic UV-divergence of the setting sun graph in three dimensions.}. 
The mass counterterm and the 
renormalization constant for $\lambda$ are, respectively~\cite{braaten},
\beq
\delta m^2=\frac{T^2}{24}\frac{\lambda^2}{16\pi^2\epsilon},\hspace{1cm}
Z_{\lambda}=1+\frac{3\lambda}{32\pi^2\epsilon}.
\eeq
Using appendix A, one obtains the mass parameter to second order in $\lambda$:
\beq
m^2(\Lambda)=\frac{\lambda T^2}{24}\Big[1+\frac{\lambda}{16\pi^2}
[\ln\frac{\Lambda}{4\pi T}-\gamma_E
+2+2\frac{\zeta^{\prime}(-1)}{\zeta (-1)}
]\Big].
\eeq
The mass parameter is not renormalization group invariant, and this is 
a consequence of the $\Lambda$-dependence of the two-point function
at two loop in the effective theory.

We close this section by two comments. We have seen that the infrared
divergences that arise at the two-loop in the underlying theory match
the infrared divergence in the effective theory, as promised. This is a general
feature of the present approach, although we shall not explicitly demonstrate 
this cancelation in the following calculations.

Secondly, we saw that the sum of the last two diagrams
in Fig~\ref{yuc2} vanishes, but individually they do not. 
This fact shows the difference between the matching procedure and the old
way of integrating out the heavy modes. As previously noted, the latter
method generates non-local operators. More precisely, the effects of the
diagram with both light and heavy modes are incorporated by introducing
a momentum dependent four-point vertex~\cite{jakoleik}.
\section{Spontaneously Broken Gauge Theories}
\heading{Effective Field Theory Approach I}{Spontaneously Broken Gauge Theories}
In the previous section, we studied the determination of the mass parameter
in pure scalar theory at two-loop order. At the one-loop level this was 
straightforward, but the complexity increased at two-loop order.
This will be even more dramatic in more complicated theories, which involves
gauge fields. There exists a nice alternative to the direct evaluation
of the Feynman graphs, namely the use of the effective potential.
The main advantage of this method is that there are normally 
fewer diagrams involved in the calculations, and that symmetry factors
are easier to figure out.
This approach is due to Kajantie {\it et al.}~\cite{laine}, 
who in series of papers
study the construction of effective three dimensional field 
theories [68,73,74].

These effective theories are then used in perturbative studies
and lattice simulations of the phase transition in spontaneously broken
gauge theories. This includes investigation of the electro-weak
phase transition in the standard model~\cite{laine} 
or supersymmetric extensions 
thereof by B\"odeker {\it et al.}~\cite{extension}, 
as well
as $SU(5)$ by Rajantie~\cite{raja}, and the Abelian Higgs model by 
Karjalainen and Peisa~\cite{peisa}. 

The idea is that the effective potential is the generator of one-particle
irreducible Greens functions at zero external momentum. Now, assume that
we split the Higgs field into a background field and quantum field in the
usual way, and compute the effective potential in the full theory
to some order in the loop expansion.
The coefficients
of $\phi_0^2/2$ and $\phi_0^4/24$ then give the {\it unresummed}
two and four-point functions, respectively, at zero external momentum.
We then carry out a corresponding calculation of the effective potential
in the effective theory, and the above mentioned coefficients yield the
same correlators in the three dimensional theory.
Now, consider matching at one-loop. The contribution to any correlator
from the $n=0$ mode is canceled against a corresponding contribution
in the effective theory. Hence, the contribution from the
$n\neq 0$ modes to the one-loop effective potential in the full theory
give, up to possible field redefinitions, the one-loop correction to 
any scalar Greens function. 

The determination of the parameters beyond one-loop is more complicated,
since we must be careful with different mode contributions to the correlators. 
We shall make these ideas more precise in the next
section. 

In the previous section we have seen that there are two mass scales
in $\phi^4$ theory, namely $T$ and $gT$. This remark also applies to any
Abelian gauge theory, where the scale $eT$ corresponds to the scale of
electric screening.
In nonabelian gauge theories there is a third scale $g^2T$, which is the
scale of magnetic screening, or the inverse confinement radius.
When the temperature is close to the critical temperature scalar
fields have a mass of order $g^2T$. This implies that we are faced with
three scales in spontaneously broken gauge theories close to a phase 
transition,
even in the Abelian case.
Hence, it is useful to introduce the following definitions due to Kajantie
{\it et al.}~\cite{laine}: 
\begin{itemize}
\item {\it Superheavy modes}. These are modes with masses of order $T$.
The bosonic modes with $n\neq 0$ as well as the fermionic modes
are superheavy. 
\item {\it Heavy modes}. These are modes with masses of order $gT$, where
$g$ is the gauge coupling. The temporal components of the gauge fields acquire
masses proportional to $gT$ and so these fields are heavy. 
For temperatures much bigger than $T_{c}$
the scalar masses is of the order $gT$ and these modes are then heavy.
\item  {\it Light modes}. These are modes with mass of order $g^{2}T$
or less. Near a phase transition the masses of the scalar particles go 
like $g^{2}T$
and these modes are light. The spatial components of the gauge field 
are massless and so these modes are also light.
\end{itemize}
Since we now have three different momentum scales, it is convenient to 
construct a sequence of two effective field theories, which are valid on
successively longer distance scales. This is a fairly straightforward
generalization of the preceding discussion, so we summarize it in the
following recipe:\\ \\
{\bf step 1}\\
Write down the most general Lagrangian consistent with the symmetries
of the system containing light and heavy fields. The parameters are tuned
so that the static correlators of the light and heavy fields
in the full four dimensional field theory
are reproduced by the corresponding correlators in the effective theory,
to some desired accuracy at distance scales $R\gg 1/T$. 
The effective Lagrangian is valid for momenta
$k$ up to order $gT$, and the coefficients encode the physics on the scale $T$,
which is a typical momentum of a particle in the plasma.
The parameters are called {\it short-distance coefficients}.
\\ \\
{\bf step 2}\\
This is an effective field theory where 
${\cal L}_{\mbox{\scriptsize {eff}}}^{\prime}$ 
includes all operators that can be
constructed out of the light fields, and which satisfy the symmetries of the
system. The parameters are determined by demanding that the correlators
of the light fields in the two theories match (to some required accuracy)
at long distance ($R\gg 1/gT$). These coefficients give the contribution
to physical quantities from the scales $T$ and $gT$. Generally, the 
coefficients
of operators in the two effective Lagrangians 
which involve only light terms differ. This renormalization of the parameters
is of course due to the fact that we have integrated out the heavy fields 
and this difference encode the physics on the scale $gT$.
This effective Lagrangian is valid for momenta up to order $g^2T$, and
the parameters are termed {\it middle-distance coefficients}.\\  
\section{The Two-loop Effective Potential}
\heading{Effective Field Theory Approach I}{The Two-loop Effective Potential}
Now, let us apply the ideas of the previous section to 
a model which consists of $N$ charged scalars coupled to an Abelian 
gauge field. $N=1$ corresponds to the Abelian Higgs model, that has
previously been studied by several authors [38,39,89,90].
So the results presented here are a generalization of results that
already appear in the literature. This generalization is fairly 
straightforward, but nevertheless very interesting. The point is
that previous work on this model
by Arnold and Yaffe~\cite{epsilon},
and by Lawrie~\cite{law},  using the epsilon expansion, 
indicate that the nature of the the phase 
transition depends on $N$. More precisely, for $N$ larger than some critical
$N_c\sim 365.9$, the RG-equations have a 
nontrivial fixed point in coupling constant
space in $d=4-\epsilon$ dimensions. 
Such fixed points are taken as evidence for a
second order phase transition~\cite{gins}. Thus from general considerations,
one expects a first order phase transition for $N<N_c$ and a second
order phase transition for $N>N_c$. However, 
perturbation theory normally breaks down for temperatures close to the
critical temperature. Thus lattice simulations may be
the only reliable tool in the determination of the order of the phase
transition and our effective three dimensional field
theory is the starting point of a nonperturbative study
of the $N$-dependence of the phase transition. 

\begin{figure}[htb]
\begin{center}
\mbox{\psfig{figure=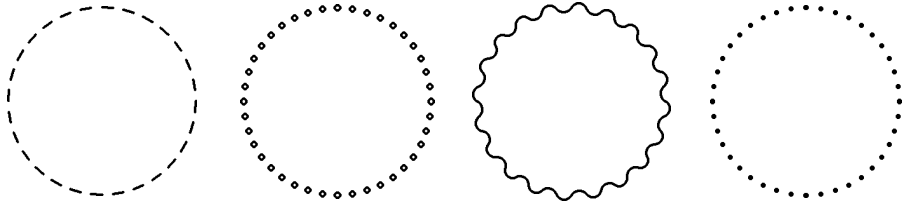}}
\end{center}
\caption[One-loop graphs contributing to the effective potential.]{\protect One-loop graphs contributing to the effective potential.}
\label{kaj1}
\end{figure}

In the Feynman graphs
a dashed line denotes the Higgs field. The Goldstone fields are indicated
by heavy dots, the wiggly line corresponds to the photon, and the 
ghost is denoted by ordinary dotted lines. 
The Euclidean Lagrangian is
\beq
{\cal L}=\frac{1}{4}F_{\mu\nu}F_{\mu\nu}+({\cal D_{\mu}}\Phi )
^{\dagger}({\cal D_{\mu}}\Phi )-\nu^{2}\Phi^{\dagger}\Phi
+\frac{\lambda}{6} (\Phi^{\dagger}\Phi)^{2}
+{\cal L}_{\footnotesize \mbox{gf}}
+{\cal L}_{\footnotesize \mbox{gh}}.
\eeq
Here $D_{\mu}=\partial_{\mu}+ieA_{\mu}$ is the covariant derivative and
$\Phi^{\dagger}=(\Phi_1^{\dagger},\Phi_2^{\dagger},...,\Phi_N^{\dagger})$.
$\Phi$ is the corresponding column vector.\\ \\
We perform the calculations in Landau gauge. The 
propagators and the gauge fixing term are, respectively
\beq
\Delta_{\mu\nu}(k_0,{\bf k})=
\frac{\delta_{\mu\nu}-k_{\mu}k_{\nu}/K^{2}}{K^{2}+m_V^2}
,\hspace{1cm}
\Delta_{H}(k_0,{\bf k})=\frac{1}{K^{2}+m_1^2},\hspace{1cm}
\eeq
\beq
\Delta_{GS}(k_0,{\bf k})=\frac{1}{K^{2}+m_2^2},\hspace{1cm}
{\cal L}_{\footnotesize \mbox{gf}}=\frac{1}{2\alpha}(\partial_{\mu}A_{\mu})^{2}
,\hspace{1cm}\alpha\rightarrow 0.
\eeq
After the shift in the Higgs field, the tree-level masses are
\beq
m_1^2=-\nu^2+\frac{\lambda}{2}\phi_0^2,\hspace{1cm}
m_2^2=-\nu^2+\frac{\lambda}{6}\phi_0^2,\hspace{1cm}
m_V^2=e^2\phi_0^2.
\eeq
The effective theory of the zero modes
consists of $N$ charged scalars coupled to an Abelian 
gauge field in three dimensions, in analogy with the full theory. 
The timelike component of the gauge field
acquires a thermal mass and behaves as real scalar field. This field
also couples to itself. We can then
write
\bqa\nonumber
{\cal L}_{\mbox{\scriptsize eff}}&=&
\frac{1}{4}F_{ij}F_{ij}+({\cal D}_{i}\phi )
^{\dagger}({\cal D}_{i}\phi )+m^2(\Lambda)\phi^{\dagger}\phi
+\frac{\lambda_3(\Lambda)}{6} (\phi^{\dagger}\phi)^{2}
+\frac{1}{2}(\partial_i A_0)^2+\frac{1}{2}m^2_E(\Lambda)A_0^2
\\ &&
+\frac{1}{2}h_E^2(\Lambda)\phi^{\dagger}\phi A_0^2
+
\frac{\lambda_A(\Lambda)}{24}A_0^4
+{\cal L}_{\footnotesize \mbox{gf}}
+{\cal L}_{\footnotesize \mbox{gh}}
+\delta{\cal L}.
\eqa
Here, $\delta {\cal L}$ represent all higher order operators
consistent with the symmetries.
The three dimensional gauge coupling is denoted by $e_E(\Lambda)$.\\ \\
The one-loop diagrams are shown Fig~\ref{kaj1} and they read:
\beq
V_1=\frac{1}{2}C(m_1)+\frac{1}{2}(2N-1)C(m_2)
+\frac{1}{2}(d-1)C(m_V)-\frac{1}{2}C(0).
\eeq
Here, we have defined
\beq
C(m)\equiv\hbox{$\sum$}\!\!\!\!\!\!\int_{P}\ln (P^2+m^2)
=\hbox{$\sum$}\!\!\!\!\!\!\int_{P}\ln P^2-\frac{m^3T}{6\pi}
+m^2\hbox{$\sum$}\!\!\!\!\!\!\int_{P}^{\prime}\frac{1}{P^2}
-\frac{1}{2}
m^4\hbox{$\sum$}\!\!\!\!\!\!\int_{P}^{\prime}\frac{1}{P^4}
+{\cal O}(m^6/T^2).
\eeq

\begin{figure}[htb]
\begin{center}
\mbox{\psfig{figure=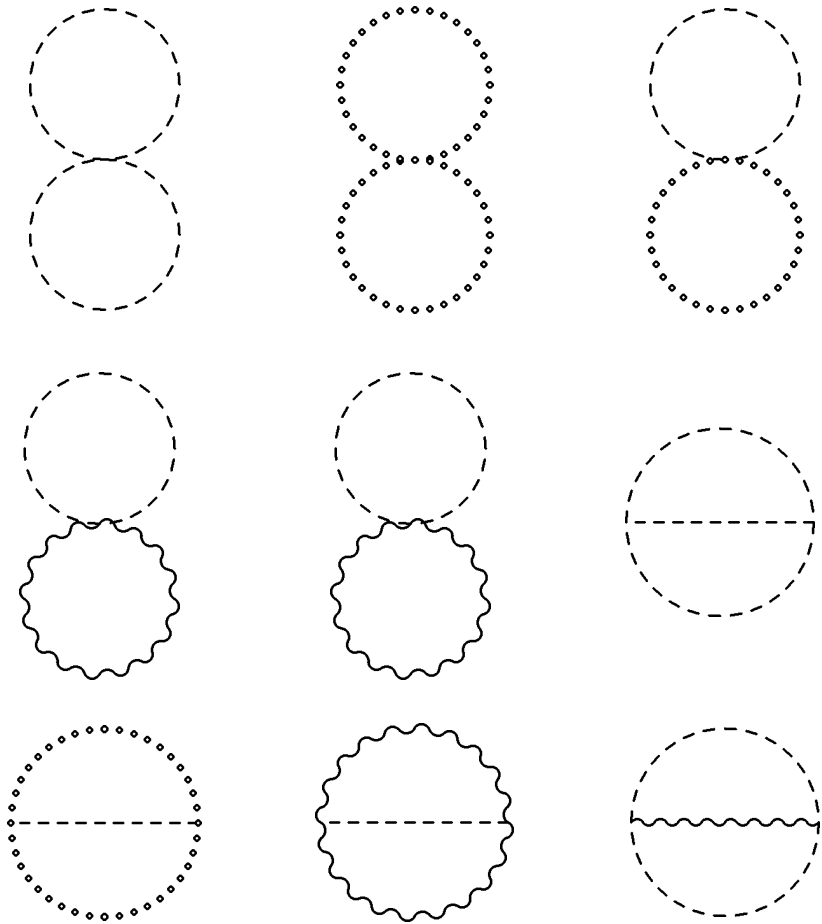}}
\end{center}
\caption[Two-loop graphs contributing to the effective potential.]{\protect Two-loop graphs contributing to the effective potential.}
\label{kaj2}
\end{figure}

The first term is field independent 
and hence irrelevant for the present calculation.
The second term comes from the $n=0$ mode, while the remaining terms arise
from the $n\neq 0$ modes. These terms then give directly 
(up to the renormalization of the fields)
the one-loop corrections to the mass parameter and the quartic coupling.
The two-loop graphs are displayed in Fig~\ref{kaj2}.
Let us focus on the figure-eight diagrams which are in the form
\bqa\nonumber
D_{SS}(m_1,m_2)&\equiv&\hbox{$\sum$}\!\!\!\!\!\!\int_{PQ}\frac{1}{(P^2+m_1^2)(Q^2+m_2^2)} \\ \nonumber
&=&
T^2\int_{pq}\frac{1}{(p^2+m^2_1)(q^2+m_2^2)}+
T\int_p\frac{1}{p^2+m_1^2}\hbox{$\sum$}\!\!\!\!\!\!\int_{Q}^{\prime}
\frac{1}{Q^2+m_2^2}+\\ \nonumber
&&
T\int_p\frac{1}{p^2+m_2^2}\hbox{$\sum$}\!\!\!\!\!\!\int_{Q}^{\prime}
\frac{1}{Q^2+m_1^2}+\hbox{$\sum$}\!\!\!\!\!\!\int_{PQ}^{\prime}
\frac{1}{(P^2+m_1^2)(Q^2+m_2^2)}\\\nonumber
&=&T^2\int_{pq}\frac{1}{(p^2+m^2_1)(q^2+m_2^2)}+
T\int_p\frac{1}{p^2+m_1^2}\hbox{$\sum$}\!\!\!\!\!\!\int_{Q}^{\prime}
\frac{1}{Q^2}\\
&&
T\int_p\frac{1}{p^2+m_2^2}\hbox{$\sum$}\!\!\!\!\!\!\int_{Q}^{\prime}
\frac{1}{Q^2}-(m_1^2+m_2^2)\hbox{$\sum$}\!\!\!\!\!\!\int_{PQ}^{\prime}
\frac{1}{P^2Q^4}+...
\eqa
where the ellipsis 
indicates field independent terms as well as terms of higher order
in $m_1^2$ and $m_2^2$.
The first term which arise when both Matsubara frequencies vanish, is 
reproduced by the effective theory, and hence it is canceled in 
the matching procedure. Furthermore, the second and third terms are also
canceled. There are (at least) two ways of seeing this. In order to obtain the
effective two-loop potential resummation is required. The above-mentioned terms
will then be canceled by the thermal counterterms~\cite{laine}. 
Using the Feynman graph
approach directly. such terms are canceled by one-loop graphs with a mass
insertion in the effective theory. This was explicitly demonstrated in 
in the simpler scalar theory in section~\ref{fin}. 
The last term above, where both zero-frequency modes have been removed,
is then the only part of $D_{SS}(m_1,m_2)$ which contributes to the mass
parameter.\\ \\
Consider first the theta diagram with
three Higgs particles: 
\beq
H(m_1,m_1,m_1)\equiv
\hbox{$\sum$}\!\!\!\!\!\!\int_{PQ}
\frac{1}{(P^2+m_1^2)(Q^2+m_1^2)[(P+Q)^2+m_1^2]}.
\eeq
This diagram is IR-divergent in the limit $m\rightarrow 0$
when both Matsubara frequencies vanish. Moreover, to leading order
in the masses one can safely put $m$ to zero, when at least one Matsubara
frequency is different from zero (the diagram is IR-safe and the 
$m$ is not a relevant scale at leading order. This is completely analogous
to the the discussion in the previous chapter on resummation).
Thus, we may write
\beq
H(m_1,m_1,m_1)=T^2\int_{pq}
\frac{1}{(p^2+m_1^2)(p^2+m_1^2)[({\bf p}+{\bf q})^2+m_1^2]}
+\hbox{$\sum$}\!\!\!\!\!\!\int_{PQ}\frac{1-\delta_{p_0,0}\delta_{p_0,0}}
{P^2Q^2(P+Q)^2}+{\cal O}(m_1^2).
\eeq
As previously noted, the second term above vanishes in dimensional
regularization, and so we are left with
\beq
H(m_1,m_1,m_1)=T^2\int_{pq}
\frac{1}{(p^2+m_1^2)(p^2+m_1^2)[({\bf p}+{\bf q})^2+m_1^2]}+{\cal O}(m_1^2).
\eeq
It is clear that this term is canceled by a corresponding term in the effective
theory, when we match the two-point function. Hence, $H(m_1,m_1,m_1)$
does not contribute
to the mass parameter $m^2(\Lambda)$.\\ \\
The next theta diagram is the graph with two vector propagators and one
scalar propagator. By purely algebraic manipulations it can be written
in terms of $H(m_1,m_2,m_3)$ and double-bubbles. It reads
\bqa\nonumber
D_{SVV}(m_V,m_V,m_1)&\equiv&
\hbox{$\sum$}\!\!\!\!\!\!\int_{PQ}\frac{(\delta_{\mu\nu}-p_{\mu}p_{\nu}/P^2
)(\delta_{\mu\nu}-q_{\mu}q_{\nu}/Q^2)}{(P^2+m_V^2)(Q^2+m_V^2)[(P+Q)^2+m_1^2]}
\\ \nonumber
&=&(d-2)H(m_V,m_V,m_1) \\ \nonumber
&&+\frac{m_1^4}{4m_V^4}\Big[H(m_1,0,0)+H(m_V,m_V,m_1)
-2H(m_V,m_1,0)\Big]\\ \nonumber
&&+\frac{m_1^2}{m_V^2}\Big[H(m_V,m_1,0)-H(m_V,m_V,m_1)\Big]\\ \nonumber
&&+2H(m_V,m_V,m_1)-\frac{1}{2}H(m_V,m_1,0)\\ \nonumber
&&-\frac{m_1^2}{4m_V^4}\Big[D_{SS}(m_V,m_V)
-2D_{SS}(m_V,0)\Big]
\\ \nonumber
&&+\frac{1}{2m_V^2}\Big[D_{SS}(m_V,m_V)+D_{SS}(m_1,0)
-D_{SS}(m_V,m_1)\\ 
&&
-D_{SS}(m_V,0)\Big].
\eqa
From the above expression, we infer that the field dependent contributions 
from the figure-eight
terms cancel. Hence, this diagram does not contribute to the mass parameter
at this stage. \\ \\
The final two-loop diagram can also be written in terms of $D_{SS}(m_1,m_2)$
and $H(m_1,m_2,m_3)$, and it is found to be:
\bqa\nonumber
D_{SSV}(m_V,m_1,m_2)&\equiv&
\hbox{$\sum$}\!\!\!\!\!\!\int_{PQ}\frac{(2P+Q)_{\nu}(2P+Q)_{\nu}
(\delta_{\mu\nu}-q_{\mu}q_{\nu}/Q^2)
}{(P^2+m_1^2)(Q^2+m_V^2)[(P+Q)^2+m_2^2]}\\ \nonumber
&=&D_{SS}(m_V,m_1)+D_{SS}(m_V,m_2)-D_{SS}(m_1,m_2) \\ \nonumber
&&+(m_V^2-2m_1^2-2m_2^2)H(m_V,m_1,m_2)\\ \nonumber
&&+\frac{(m_1^2-m_2^2)^2}{m_V^2}\Big[H(m_V,m_1,m_2)
-H(m_1,m_2,0)\Big]\\ \nonumber
&&+\frac{m_1^2}{m_V^2}\Big[D_{SS}(m_2,0)+D_{SS}(m_V,m_1)-D
_{SS}(m_V,m_2)-D_{SS}(m_1,0)\Big] \\ \nonumber
&&+\frac{m_2^2}{m_V^2}\Big[D_{SS}(m_1,0)+D_{SS}(m_V,m_2)-D
_{SS}(m_V,m_1)-D_{SS}(m_2,0)\Big].\\
\hspace{1cm}&&
\eqa
In contrast with the preceding diagram, this graph contributes to the mass
parameter. More precisely, the second line above yields a contribution
proportional to the mass of the vector particle.\\ \\ 
In terms of the sum-integrals defined above, 
the two-loop effective potential is
\bqa\nonumber
V_2&=&\frac{\lambda}{24}\Big[3D_{SS}(m_1,m_1)+(4N^2-1)D_{SS}(m_2,m_2)+
(4N-2)D_{SS}(m_1,m_2)
\Big]\\ \nonumber
&&+\frac{1}{2}(d-1)e^2\Big[D_{SS}(m_1,m_V)+(2N-1)
D_{SS}(m_2,m_V)\Big]\\ \nonumber
&&-\frac{1}{2}\lambda^2\phi_0^2\Big[
\frac{1}{6}H(m_1,m_1,m_1)+\frac{2N-1}{18}H(m_1,m_2,m_2)
\Big]\\ 
&&+e^4\phi_0^2D_{SVV}(m_V,m_V,m_1)-\frac{1}{2}e^2D_{SSV}(m_1,m_2,m_V)].
\eqa

\begin{figure}[htb]
\begin{center}
\mbox{\psfig{figure=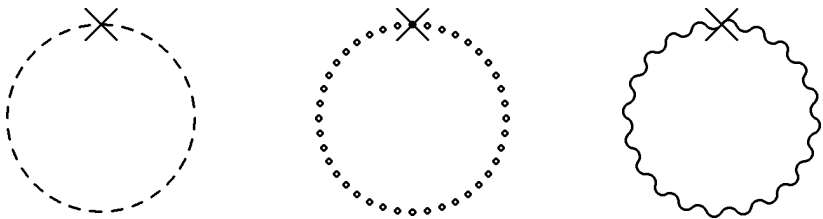}}
\end{center}
\caption[One-loop counterterm diagrams for the effective
potential.]{\protect One-loop counterterm diagrams for the effective potential.}   
\label{kcon}
\end{figure}

Finally, we must consider the counterterm diagrams.
These are shown in Fig~\ref{kcon}. At the one-loop level, there is either
a mass counterterm insertion or an insertion of a wave function 
counterterm. The corresponding integrals are
\bqa\nonumber
V_{\mbox{\scriptsize ct}}=
\frac{1}{2}\hbox{$\sum$}\!\!\!\!\!\!\int_{P}
\frac{\delta m_1^2+\delta Z_{\Phi}P^2}{P^2+m_1^2}
+\frac{1}{2}(2N-1)\hbox{$\sum$}\!\!\!\!\!\!\int_{P}
\frac{\delta m_2^2+\delta Z_{\Phi}P^2}{P^2+m_2^2}+
\frac{1}{2}(d-1)\hbox{$\sum$}\!\!\!\!\!\!\int_{P}
\frac{\delta m_V^2+\delta Z_{A}P^2}{P^2+m_V^2}.\\
&&
\eqa
The counterterms above are those of the four dimensional theory, and they
include those generated by the shift in the Higgs field. The mass counterterms
read
\bqa
\delta m_1^2=-\frac{\nu^2\lambda}{32\pi^2\epsilon}+
\frac{2(N+4)\lambda^2+108e^4}{96\pi^2\epsilon}\phi_0^2,\hspace{0.8cm}
\delta m_2^2=\frac{\delta m_1^2}{3},\hspace{0.8cm}
\delta m_V^2=\frac{3e^4}{16\pi^2\epsilon}\phi_0^2.
\eqa
The wave function renormalization counterterms will be listed in 
subsection~\ref{feltet}.

\clearpage
\newpage
\pagebreak
\section{The Short-distance Coefficients}
\heading{Effective Field Theory Approach I}{The Short-distance Coefficients}
In this section we summarize our results of the discussion in the previous
paragraph. We first calculate the field renormalization constants.
We then use these and the effective potential to obtain the scalar mass
parameter to two-loop order and the scalar self-coupling at the one-loop
level. The other parameters in the effective
theory are also determined to one loop order, and are
found by computing Feynman graphs.
\subsection{The Field Normalization Constants}\label{feltet}
\heading{Effective Field Theory Approach I}{The Field Normalization Constants}
In this subsection we determine the short-distance coefficients
which multiply the fields in the effective theory. At leading order we
have the simple relations:
\beq
\phi(\Lambda)=\frac{1}{\sqrt{T}}\Phi,\hspace{1cm}A_{i}^{3d}(\Lambda)
=\frac{1}{\sqrt{T}}A_{i}
\hspace{1cm} A_{0}^{3d}(\Lambda)=\frac{1}{\sqrt{T}}A_{0}.
\eeq
Beyond leading order this relation breaks down. At next-to-leading order
one may read off the correction from the momentum
dependent part of the two-point correlator,
which is proportional to $k^{2}$ for $\Phi$ and $A_{0}$, and 
$k^{2}\delta_{ij}-k_{i}k_{j}$ for $A_{i}$. The coefficients in front
are denoted by $\Sigma^{\prime}(0)$, $\Pi_{00}^{\prime}(0)$ and
$\Pi^{\prime}(0)$, respectively.
Thus
\beq
\phi (\Lambda)=\frac{\Phi}{\sqrt{T}}\Big[1+\Sigma^{\prime}(0)\Big]^{1/2},
\hspace{0.5cm}A_{0}(\Lambda)=\frac{A_0}{\sqrt{T}}
\Big[1+\Pi_{00}^{\prime}(0)\Big]^{1/2},
\hspace{0.5cm}A_{i}(\Lambda)=\frac{A_{i}}{\sqrt{T}}
\Big[1+\Pi^{\prime}(0)\Big]^{1/2}.
\eeq
For the gauge field the expression is
\beq
\Pi_{\mu\nu}(0,{\bf k})=2N
e^{2}\delta_{\mu\nu}\hbox{$\sum$}\!\!\!\!\!\!\int_P^{\prime}
-Ne^{2}\hbox{$\sum$}\!\!\!\!\!\!\int_P^{\prime}
\frac{(2p+k)_{\mu}
(2p+k)_{\nu}}{P^{2}(P+K)^{2}}.
\eeq
Expanding to order $k^{2}$ and integrating by parts, we find
\bqa
\label{amasse}
\Pi_{00}(0,{\bf k})&=&2Ne^{2}\hbox{$\sum$}\!\!\!\!\!\!\int_P^{\prime}\frac{1}{P^{2}}
-4Ne^{2}\hbox{$\sum$}\!\!\!\!\!\!\int_P^{\prime}\frac{p_0^2}{P^{4}}
+\frac{4Ne^{2}k^{2}}{3}
\hbox{$\sum$}\!\!\!\!\!\!\int_P^{\prime}\frac{p_{0}^{2}}{P^{6}},\\
\Pi_{ij}(0,{\bf k})&=&\frac{Ne^2k^{2}}{3}(\delta_{ij}-k_{i}k_{j}/k^{2})
\hbox{$\sum$}\!\!\!\!\!\!\int_P^{\prime}\frac{1}{P^{4}}.
\eqa
Correspondingly, one finds for the scalar field
\beq
\Sigma^{\prime}(0)=-3e^2\hbox{$\sum$}\!\!\!\!\!\!\int_P^{\prime}
\frac{1}{P^{4}}.
\eeq
After wave function renormalization, using
\beq
Z_{A}=1-\frac{Ne^{2}}{48\pi^{2}\epsilon},\hspace{1cm}
Z_{\Phi}=1+\frac{3e^{2}}{16\pi^2\epsilon},
\eeq
we find $\Sigma^{\prime}(0)$, $\Pi_{00}^{\prime}(0)$ and
$\Pi^{\prime}(0)$. The relations 
between the fields in $3d$ and $4d$ are
\bqa
\label{gdep}
\phi(\Lambda)&=&\frac{1}{\sqrt{T}}\Phi
\Big[1-\frac{3e^{2}}{(4\pi )^{2}}(\ln\frac{\Lambda}{4\pi T}+\gamma_E)
\Big],\\
A_{0}^{3d}(\Lambda)&=&\frac{1}{\sqrt{T}}A_{0}
\Big[1+\frac{Ne^{2}}{3(4\pi )^{2}}(\ln\frac{\Lambda}{4\pi T}+\gamma_E+1)\Big],\\
A_{i}^{3d}(\Lambda)&=&\frac{1}{\sqrt{T}}A_{i}\Big[1+\frac{Ne^{2}}{3(4\pi )^{2}}(\ln\frac{\Lambda}{4\pi T}+\gamma_E)\Big].
\eqa
We would like to emphasize that the first of the
above relations is gauge fixing dependent, 
but the parameters of the effective theory are gauge independent.
Secondly, taking into account the running of the fields in the full theory,
we find that the three dimensional fields are independent of the 
renormalization scale $\Lambda$. This is related to the fact that there is no
wave function renormalization in $3d$.

\begin{figure}[htb]
\begin{center}
\mbox{\psfig{figure=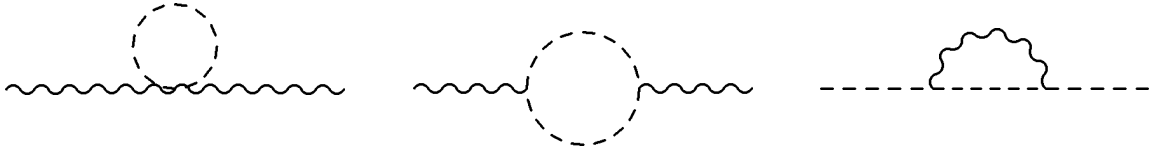}}
\end{center}
\caption[One-loop diagrams relevant for field strength
normalization.]{\protect One-loop diagrams relevant for field strength
normalization.}
\label{vaag}
\end{figure}

\subsection{The Coupling Constants}
\begin{figure}[htb]
\begin{center}
\mbox{\psfig{figure=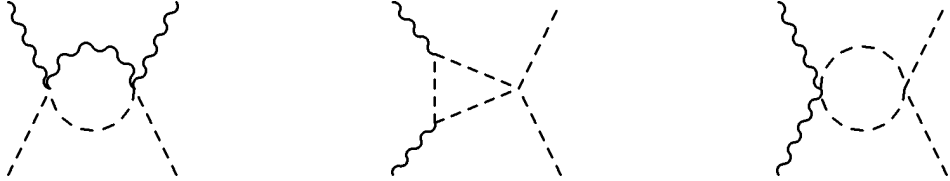}}
\end{center}
\caption[One-loop diagrams needed for the calculating the couplings
$e^2_E(\Lambda)$ and $h^2_E(\Lambda)$.]{\protect One-loop diagrams needed for the calculating the couplings
$e^2_E(\Lambda)$ and $h^2_E(\Lambda)$.}
\label{h3e3}
\end{figure}

\heading{Effective Field Theory Approach I}{The Coupling Constants}
The matching requirement for the quartic coupling yields the equation
\beq
\lambda_3(\Lambda)=\lambda T[1-\Sigma^{\prime}(0)]^2+
T\Gamma^{(4)\prime}_{\phi_1\phi_1\phi_1\phi_1}(0). 
\eeq
The first term takes into account the 
different normalization of the fields, and the second is the one-loop
correction, which may be directly read off from the effective potential
Furthermore, the prime indicates as usual 
that we neglect the zero-frequency mode
in the one-loop graph, and $\phi_1$ is the Higgs field.
This yields 
\bqa
\lambda_{3}(\Lambda)=T\Big[\lambda -
\frac{(N+4)\lambda^2-18\lambda e^2+54e^4}{24\pi^2}(
\ln\frac{\Lambda}{4\pi T}+\gamma_E)+\frac{3e^4}{4\pi^2}\Big].
\eqa
The couplings $e^{2}_{E}(\Lambda)$ 
and $h_{E}^2(\Lambda)$ are computed by 
considering the relevant correlators. The corresponding Feynman graphs are
shown in Fig.~\ref{h3e3}.
The matching relations read
\bqa
\label{e3}
e_{E}^{2}(\Lambda)&=&e^2T\Big[1-\Pi^{\prime}(0)\Big]
\Big[1-\Sigma^{\prime}(0)\Big]
+\frac{T}{2}\Gamma^{(4)\prime}_{\Phi_1^{\dagger}\Phi_1 A_iA_j}(0),\\
\label{h3}
h_{E}^{2}(\Lambda)&=&e^2T\Big[1-\Pi^{\prime}_{00}(0)\Big]
\Big[1-\Sigma^{\prime}(0)\Big]+\frac{T}{2}
\Gamma^{(4)\prime}_{\Phi^{\dagger}_1\Phi_1 A_{0}A_{0}}(0).
\eqa
Again, the first term on the right hand side takes care of the different 
normalization of the fields, while the second term is the one-loop correction.
These are given by 
\bqa
\label{rom}
\Gamma^{(4)}_{\Phi^{\dagger}_1\Phi_1 A_iA_j}(0)&=&
8e^4\hbox{$\sum$}\!\!\!\!\!\!\int_{P}^{\prime}\Big[\frac{p_ip_j}{P^{6}}
-\frac{1}{P^{4}}\Big]+
\frac{(N+3)\lambda e^2}{3}\hbox{$\sum$}\!\!\!\!\!\!\int_{P}^{\prime}\Big[4\frac{p_ip_j}{P^{6}}
-\frac{1}{P^{4}}\Big],\\
\label{tid}
\Gamma^{(4)}_{\Phi^{\dagger}_1\Phi_1 A_{0}A_{0}}(0)&=&
8e^4\hbox{$\sum$}\!\!\!\!\!\!\int_{P}^{\prime}\Big[\frac{p_0^2}{P^{6}}
-\frac{1}{P^{4}}\Big]+
\frac{(N+3)\lambda e^2}{3}\hbox{$\sum$}\!\!\!\!\!\!\int_{P}^{\prime}\Big[4\frac{p_0^2}{P^{6}}
-\frac{1}{P^{4}}\Big].
\eqa
We notice that the term involving $\Sigma^{\prime}(0)$ in Eq.~(\ref{e3})
cancels against the first term in Eq.~(\ref{rom}), after we have integrated
by parts. This reflects
the Ward identity. In the corresponding expression for $h_E(\Lambda)$, there
is not a complete cancelation, but we are left with a finite contribution. 
The net results are
\bqa
e_{E}^{2}(\Lambda)&=&e^{2}T\Big[1-\frac{Ne^2}{24\pi^2}(
\ln\frac{\Lambda}{4\pi T}+\gamma_E)\Big],\\
h_E^2(\Lambda)&=&
e^{2}T\Big[1-\frac{Ne^2}{24\pi^2}(
\ln\frac{\Lambda}{4\pi T}+\gamma_E+1)+\frac{(N+3)\lambda}{48\pi^2}+
\frac{e^2}{8\pi^2}\Big].
\eqa
Here, we have renormalized the electric coupling in the usual way to 
render the expressions finite. The charge renormalization constant is
\beq
Z_{e^2}=1+\frac{Ne^{2}}{48\pi^{2}\epsilon}.
\eeq

We shall also compute the coefficient in front of the quartic self-interacting
term $A_0^{4}$. At leading this can be found by considering the one-loop
contribution to the four-point function for timelike photons at zero external
momenta. The matching condition reads
\beq
\lambda_{A}(\Lambda)=T\Gamma^{(4)\prime}_{A_{0}A_{0}A_{0}A_{0}}(0).
\eeq
Note that the short-distance coefficient multiplying the $3d$ fields does not
affect the coupling at this order in $e$. We obtain
\beq
\Gamma^{(4)\prime}_{A_{0}A_{0}A_{0}A_{0}}(0)=
-12Ne^{4}
\hbox{$\sum$}\!\!\!\!\!\!\int_{P}^{\prime}\frac{1}{P^{4}}
+96Ne^{4}\hbox{$\sum$}\!\!\!\!\!\!\int_{P}^{\prime}\frac{p_{0}^{2}}{P^{6}}-
96Ne^{4}\hbox{$\sum$}\!\!\!\!\!\!\int_{P}^{\prime}\frac{p_{0}^{4}}{P^{8}}.
\eeq
This particular combination is finite, just as the corresponding 
one in QED~\cite{lands}:
\beq
\lambda_{A}(\Lambda)=\frac{Ne^{4}T}{\pi^{2}}.
\eeq
Furthermore, $\lambda_A(\Lambda)$
first runs at order $e^{6}$.

\begin{figure}[htb]
\begin{center}
\mbox{\psfig{figure=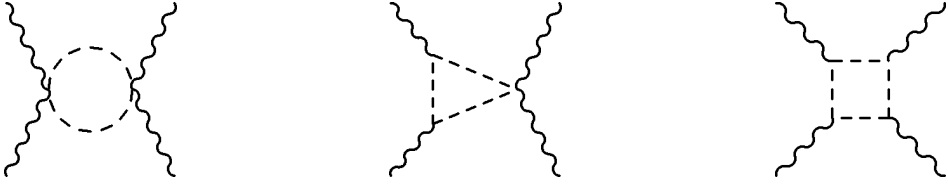}}
\end{center}
\caption[One-loop four-point function with external timelike 
photon lines.]{\protect One-loop, four-point function with external timelike 
photon lines.}
\label{sqquad}
\end{figure}
The coupling constants 
$\lambda_{3}(\Lambda)$, $e^{2}_{E}(\Lambda)$, $h_{3}^{2}(\Lambda)$ and 
$\lambda_{A}(\Lambda)$ are renormalization
group invariant. This can easily be seen by using the renormalization group
equations for the scalar and gauge couplings at leading 
order~\cite{epsilon}:
\bqa
\mu\frac{de^{2}}{d\mu}&=&\frac{Ne^{4}}{24\pi^{2}},\\
\mu\frac{d\lambda}{d\mu}&=&\frac{1}{24\pi^{2}}[(N+4)\lambda^{2}-18e^{2}\lambda
+54e^{4}].
\eqa
Thus, we can trade the scale $\Lambda$ for an arbitrary renormalization 
scale $\mu$.
\subsection{The Mass Parameters}
\heading{Effective Field Theory Approach I}{The Mass parameters}
The mass of the timelike component of the gauge field to one-loop order follows
directly from Eq.~(\ref{amasse})
\beq
m^2_E(\Lambda)=\frac{Ne^2T^2}{3}.
\eeq
Normally, it is sufficient to know $m_E^2$ at leading order in the 
couplings, as above. In the next chapter we determine it at next-to-leading
order (for $N=1$), 
since we are specifically interested in the electric screening mass.
Its generalization to arbitrary $N$ is not to difficult.\\ \\
The scalar mass parameter is determined from the two-lop effective potential.
Schematically we write
\beq
m^2(\Lambda)=-\nu^2 [1-\Sigma^{\prime}(0))]+\Sigma_1(0)
[1-\Sigma^{\prime}(0)]+\tilde{\Sigma}_2(0).
\eeq
Here, $\Sigma_1(0)$ is the one-loop contribution from the nonstatic modes,
and $\tilde{\Sigma}_2(0)$ is the term that survives from the two-loop
graphs. After renormalization,
we are still left with a pole in $\epsilon$. This divergence 
is again a reflection of the logarithmic divergence of the two-point
function in three dimensions.
We cancel it 
by adding a mass counterterm, which thereby is determined to be
\beq
\delta m^{2}=\frac{(N+1)\lambda^{2}T^{2}}{36(4\pi)^{2}\epsilon}
-\frac{(N+1)\lambda e^{2}T^{2}}{6(4\pi)^{2}\epsilon}
+\frac{(N+5)e^{4}T^{2}}{4(4\pi)^{2}\epsilon}.
\eeq
This gives the mass parameter to two-loop order
\bqa\nonumber
m^{2}(\Lambda)&=&
-\nu^{2}\Big[1-\frac{1}{(4\pi )^2}
\Big(\frac{2(N+1)\lambda}{3} -6e^{2}\Big)
\Big(\ln\frac{\Lambda}{4\pi T}+\gamma_E\Big)\Big]
+\frac{(N+1)\lambda T^{2}}{36} +\\ \nonumber
&&\frac{e^2T^{2}}{4}+\frac{\lambda^2}{16\pi^2}\frac{(N+1)T^{2}}{108}
\Big[(4-2N)\ln\frac{\Lambda}{4\pi T}-2(N+1)\gamma_E
+6+
6\frac{\zeta^{\prime}(-1)}{\zeta (-1)}\Big] \\ \nonumber
&& 
-\frac{\lambda e^2}{16\pi^2}\frac{(N+1)T^2}{12}\Big[4\ln\frac{\Lambda}{4\pi T}
+\frac{10}{3}+4\frac{\zeta^{\prime}(-1)}{\zeta (-1)}\Big]\\ \nonumber 
&&
+\frac{e^4}{16\pi^2}\frac{T^2}{36}\Big[(144-6N)\ln\frac{\Lambda}{4\pi T}
+(54-24N)\gamma_E+72+28N+
\\ 
&&
(90+18N)\frac{\zeta^{\prime}(-1)}{\zeta (-1)}
\Big].
\eqa
Note that the scalar mass parameter is explicitly dependent on the scale
$\Lambda$. This dependence is necessary to cancel the logarithmic
ultraviolet divergences in the $3d$ theory. 
\section{The Middle-distance Coefficients}
\heading{Effective Field Theory Approach I}{The Middle-distance Coefficients}
We can also apply the effective potential to the three dimensional effective
theory to integrate out the temporal component of the gauge field.
The only relevant graphs are those with at least one $A_0$ on an internal line.
Diagrams which contain only light particles yield the same contribution
on both side of the matching equation for any $n$-point function.
The corrections to the parameters, found in this section are independent
of $N$. \\ \\
The second effective field theory is identical to the previous one, 
except that the temporal component of the gauge field is left out:
\bqa\nonumber
{\cal L}_{\mbox{\scriptsize eff}}^{\prime}=&=&
\frac{1}{4}\bar{F}_{ij}\bar{F_{ij}}+({\cal D}_{i}\bar{\phi} )^{\dagger}
({\cal D}_{i}\bar{\phi} )+\bar{m}^{2}(\Lambda )
\bar{\phi}^{\dagger}\bar{\phi}
+\frac{\bar{\lambda_3}(\Lambda)}{6}(\bar{\phi}^{\dagger}\bar{\phi})^2
+{\cal L}_{\mbox{\footnotesize gf}}
+{\cal L}_{\mbox{\footnotesize gh}}
+\delta{\cal L}^{\prime}.
\eqa
The gauge coupling is now denoted by $e_M(\Lambda)$, while the other
parameters as well as the fields are barred in order to distinguish them
from those of the previous section.
The dashed line
denotes Higgs field, while the real scalar field is indicated by a solid line
and the Goldstone particles are denoted by dotted lines.
\begin{figure}[htb]
\begin{center}
\mbox{\psfig{figure=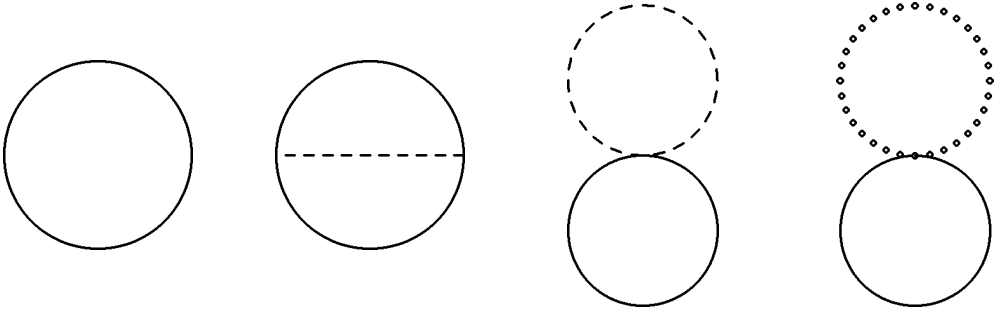}}
\end{center}
\caption[Relevant graphs for integrating over the real scalar 
field $A_0$.]{\protect Relevant graphs for integrating over the real scalar 
field $A_0$.}
\label{toloop}
\end{figure}
After a shifting the Higgs field in the usual way, the masses are
\bqa
m_1^2(\Lambda)&=&
m^2(\Lambda)+\frac{\lambda_3(\Lambda)\phi_0^2}{2},\\
m_2^2(\Lambda)&=&m^2(\Lambda)+\frac{\lambda_3(\Lambda)\phi_0^2}{6},\\
m_{A_0}^2(\Lambda)&=&m^2_E(\Lambda)+e^2_E(\Lambda)\phi_0^2.
\eqa
The relevant diagrams are found in Fig.~\ref{toloop} and the 
contributions are
\bqa\nonumber\!\!\!\!\!\!\!\!\!\!\!\!\!\!
\label{v2}
\bar{V}_2
&=&\frac{1}{2}T\int_p\ln(p^2+m_{A_0}^2)-e_E^4T\phi_0^2
\int_{pq}\frac{1}{(p^2+m_{A_0}^2)(q^2+m_{A_0}^2)[({\bf p}+{\bf q})^2+m_1^2]}
\\
\!\!\!\!\!\!\!\!\!\!\!\!\!\!&&
+\frac{1}{2}e_E^2T\int_{pq}\frac{1}{(p^2+m_{A_{0}}^2)(q^2+m_1^2)}
+(N-\frac{1}{2})e_E^2T\int_{pq}\frac{1}{(p^2+m_{A_0}^2)(q^2+m_2^2)}.
\eqa
The bar on $V_2$ is a reminder that contributions from 
the light fields to the two-loop effective
potential have been omitted. 
We have also multiplied by a factor $T$
so that the effective potential has dimension four.\\ \\
Expanding in powers of $e^2_E(\Lambda)\phi_0^2/m^2_E(\Lambda)$ gives the 
one-loop contribution to the scalar mass parameter as well as to the
quartic coupling constant.
Close to the phase transition scalar mass goes like $e^4T^2$, and hence 
power counting implies that one can ignore the figure-eight graphs.
\subsection{The Field Normalization Constants}
\heading{Effective Field Theory Approach I}{The Field Normalization Constants}
At leading order in the coupling constants the fields in the two effective
theories are related as 
\beq
\label{tree}
\overline{\phi}(\Lambda) =\phi(\Lambda)
,\hspace{1cm}\overline{A}_{i}(\Lambda)=A_{i}(\Lambda).
\eeq
Now, the only one-loop diagram contributing to the scalar two-point
function, where $A_0$ circulates in the loop, is independent of the
external momentum (this is the tadpole). This implies that there is
no renormalization of the field $\overline{\phi}(\Lambda)$ at leading
order in the couplings, and the relation Eq.~(\ref{tree}) still holds.
Similarly, since $A_0$ and $A_i$ do not interact at leading order
in $e^2$, there is no renormalization of $A_i$ either.
This result
is in contrast with the nonabelian case~\cite{laine}, since
$A_0$ and $A_i$ couples directly through the covariant derivative.
\subsection{The Coupling Constants}
\heading{Effective Field Theory Approach I}{The Coupling Constants}
The calculation of the gauge coupling $e_{M}^{2}(\Lambda)$ turns out to be 
particularly simple, since $A_{i}$ and $A_0$ do not interact at leading
order in $e_{E}^{2}(\Lambda)$. 
Hence, the matching condition becomes trivial and we have
\beq
e_{M}^{2}(\Lambda)=e_{E}^{2}(\Lambda).
\eeq
This result is in contrast with nonabelian theories where 
$A_{i}$ and $A_0$ interact directly via
the covariant derivative acting on $A_0$, and
gives rise to a renormalization of the gauge coupling.
Furthermore, $e^2_M(\Lambda)$ is obviously independent of $\Lambda$.\\ \\
According to the discussion above, the quartic coupling gets a 
contribution at the one-loop level:
\bqa
\bar{\lambda}_{3}(\Lambda)&=&\lambda_{3}(\Lambda)
-\frac{3e^4_{E}(\Lambda)}{4\pi m_E}.
\eqa
The quartic coupling is also independent of the scale $\Lambda$.
\subsection{The Mass parameter}
\heading{Effective Field Theory Approach I}{The Mass parameter}
Since the scalar fields does not get renormalized at the one-loop level,
the contribution to the scalar mass parameter is directly given by the
coefficients of $\phi^2/2$ from the sum of the one and two-loop graphs
discussed above:
\beq
\bar{m}^{2}(\Lambda)=m^{2}(\Lambda)-\frac{e^{2}_{E}(\Lambda)m_E}{4\pi}
-\frac{e_{E}^{4}(\Lambda)}{8\pi^{2}}\Big[
\ln\frac{\Lambda}{2m_E}+\frac{1}{2}\Big].
\eeq
Here, we have canceled the pole term in the second integral in Eq.~(\ref{v2})
by a
mass counterterm,
\beq
\delta m^{2}=\frac{e^4_{E}(\Lambda)}{32\pi^2\epsilon},
\eeq
and neglected $m_1$ in comparison with $m_V$, using power counting.
Notice again, that the mass parameter depends on the scale $\Lambda$
due to the logarithmic dependence of the propagator at two loops
in the effective theory.

This concludes our calculations of the parameters of 
${\cal L}^{\prime}_{\mbox{\scriptsize eff}}$ in terms of the temperature,
$T$, $\lambda$, $e^2$ and the renormalization scale $\Lambda$.
Finally, we mention, just for the record, that we have also calculated the
parameters $\bar{m}^2(\Lambda)$ and $\bar{\lambda}_3(\Lambda)$
by evaluating the appropriate Feynman diagrams. We obtain the
same results, and it is a valuable check of our computations.

\chapter{Effective Field Theory Approach II}
\heading{Effective Field Theory Approach II}{Effective Field Theory Approach II}
\section{Introduction}
\heading{Effective Field Theory Approach II}{Introduction}
In the previous chapter we studied in great detail the construction
of effective field theories for the $n=0$ bosonic mode by matching 
Greens functions.   
Instead of explicitly dividing the loop corrections
to static correlators
into contributions from light and heavy modes, 
there exists a convenient alternative
due to Braaten and Nieto~\cite{braaten}. 
This approach is perhaps the most clean and
transparent way of doing effective field theory, and calculations
are greatly simplified compared to resummation methods, since one treats
a single scale at a time. This is, of course, a common property
of every effective field theory approach, but the simplification 
here is that
one does not distinguish between static and nonstatic modes.
The procedure of constructing the effective Lagrangian proceeds, though,
essentially along the lines presented in the previous chapter.

The starting point is the identification of the fields  and the symmetries
in the three dimensional Lagrangian. The parameters are tuned as functions
of the couplings in the full theory, so as to reproduce the static 
correlators at long distances $R\gg 1/T$ in the usual way.
If there are two scales $gT$ and $g^2T$ in the three dimensional
theory as in e.g. QCD, one proceeds to construct a second effective field
theory. We shall discuss this approach in detail in the present
chapter.  

This method was first applied to $\lambda\phi^4$-theory by Braaten and Nieto in
Ref.~\cite{braaten}. 
They computed the free energy to order $\lambda^{5/2}$, which was first 
obtained by Parwani using resummation methods~\cite{parw3}.
They also calculated the screening mass to order $\lambda^2$.
Combining
their results with renormalization group methods, leading logarithms
of the coupling were summed. 
The latter result was new, and an 
improvement of the classic $\lambda^{3/2}$-result of Dolan and 
Jackiw~\cite{dolan}. 
Later, Braaten and Nieto computed the free energy in QCD, through
order $g^5$~\cite{braaten2}, 
and confirmed the resummation results of Zhai and 
Kastening~\cite{kast} 
(which in turn was an extension  of the $g^4$-result of Arnold
and Zhai~\cite{arnold1}).
Moreover, this approach provides a solution to the long-standing 
infrared problem in nonabelian gauge theories at high temperature to
which we shall return in section~\ref{versus}~\cite{braaten2}.

We shall apply these ideas to spinor as well as scalar electrodynamics.
In QED, we obtain both the free energy and the electric screening mass squared
to order $e^5$.
Our calculations reproduce previous results obtained by 
Zhai and Kastening~\cite{kast}, and
independently by Corian\`o and Parwani~\cite{parw2} for the free energy, 
as well
as the screening mass first computed by Blaizot {\it et al}~\cite{parw1}. 
These results were obtained using resummation.

In SQED we also compute the same quantities. In this case
the screening mass squared is calculated to order $e^4$ and $\lambda e^2$, 
and the free energy is derived
to order $\lambda^2$, $\lambda e^2$ and $e^4$.
The former result 
confirms the calculations of Blaizot {\it et al.}
who applied resummation~\cite{parw1}, although they did not include a quartic
self-interaction term for the scalar field. The latter result is 
new and thus represents the present calculational frontier.
\section{QED at High Temperature}
\heading{Effective Field Theory Approach II}{QED at High Temperature}
In this section we discuss QED at high temperature $(T\gg m)$ and the 
construction of the three dimensional effective field theory.
The Euclidean Lagrangian of massless QED reads
\begin{equation}
{\cal L}_{\mbox{\scriptsize QED}}
=\frac{1}{4}F_{\mu\nu}F_{\mu\nu}+\overline{\psi}
\gamma^{\mu}\Big (\partial_{\mu}-ieA_{\mu} \Big )\psi 
+{\cal L}_{\mbox{\footnotesize gf}}+{\cal L}_{\mbox{\footnotesize gh}}.
\end{equation}
In this chapter
all calculations are carried out in the Feynman gauge, but
we emphasize that the parameters in the effective theory are
gauge fixing independent.
The gauge fixing term is then 
\beq
{\cal L}_{\mbox{\footnotesize gf}}=
\frac{1}{2}(\partial_{\mu}A_{\mu})^{2},
\eeq
and the ghost field decouples from the rest of the Lagrangian. \\ \\
We call the corresponding 
effective three dimensional field theory electrostatic 
electrodynamics (EQED), in analogy with the definitions introduced
by Braaten and Nieto in the case of QCD~\cite{braaten2}. 
The first task is to identify
the appropriate fields and the symmetries in EQED. It consists of a
real scalar field coupled to an Abelian gauge field in three dimensions.
The fields can, as usual, be identified (up to normalizations) with the 
zero-frequency modes of the original fields. In particular, the real
massive field is identified with the $n=0$ mode of the timelike component
of the gauge field in the full theory. Note also that there are no fermionic
fields in EQED, since the fermions decouple for reasons that should be clear
at this stage.

Now, ${\cal L}_{\mbox{\scriptsize EQED}}$ must be a gauge invariant function
of the spatial fields $A_i$, up to the usual gauge fixing terms. 
This symmetry
follows from the corresponding symmetry in the full theory
and the Ward-Takahashi identity in the high temperature limit~\cite{lands}.
Since QED is an Abelian gauge theory, there will be no 
magnetic mass~\cite{fradkin}.
Moreover,
the timelike component of the gauge field, $A_0$, behaves as a real
massive self interacting scalar field. 
The fact that $A_{0}$ may develop a thermal mass
is a simple consequence of the lack of Lorentz invariance at 
finite temperature. 
Moreover, there is a rotational symmetry and a discrete symmetry
$A_{0}\rightarrow -A_{0}$. 
The effective Lagrangian then has the general form
\begin{equation}
\label{defleff}
{\cal L}_{\mbox{\scriptsize EQED}}=\frac{1}{4}F_{ij}F_{ij}
+\frac{1}{2}
(\partial_{i}A_0)^{2}
+\frac{1}{2}m^{2}_{E}(\Lambda )A_0^{2}+\frac{\lambda_{E}(\Lambda)}{24}A_0^{4}
+{\cal L}_{\mbox{\footnotesize gf}}
+{\cal L}_{\mbox{\footnotesize gh}}
+\delta {\cal L}.
\end{equation}
Here, $\Lambda$ is the scale introduced in dimensional regularization.
Furthermore, $\delta {\cal L}$ represents all local 
terms that can be constructed out of $A_{i}$ and $A_0$, which 
respect
the symmetries of the theory. This 
includes 
renormalizable terms, such as $g_{E}(\Lambda )A_0^{6}$, as well 
as 
non-renormalizable ones like $h_{E}(\Lambda )(F_{ij}F_{ij})^{2}$.

In Eq.~(\ref{defleff}), we did not include the unit operator. The coefficient
of the unit operator, which we denote $f_{E}(\Lambda)$, 
gives the contribution
to the free energy from the momentum scale $T$. So if we are interested
in calculating the pressure we must determine it, as we determine other
coefficients in EQED. If we are not, it is left out. Generally, this
coefficient (as well the other parameters in EQED) 
depends on the renormalization scale $\Lambda$, in order to
cancel the $\Lambda$-dependence which will arise from the calculations 
in the effective field theory. By including $f_{E}(\Lambda )$ in 
${\cal L}_{\mbox{\scriptsize EQED}}$, we have two equivalent ways of
writing the partition function in QED in terms of its path integral 
representation. In the full theory we have
\begin{equation}
\label{z1}
{\cal Z}=\int {\cal D}\overline{\eta}\,{\cal D}
\eta\,{\cal D}A_{\mu}\,{\cal D}\overline{\psi}
\,{\cal D}
\psi \exp\Big[-\int_{0}^{\beta}d\tau\int d^{3}x\,
{\cal L}\Big ],
\end{equation}
where $\eta$ denotes the ghost field.
The result using the effective three dimensional theory is
\begin{equation}
\label{z2}
{\cal Z}=e^{-f_{E}(\Lambda )V}\int {\cal D}\overline{\eta}
\,{\cal D}\eta\,{\cal D}A_{i}\,{\cal D}
A_{0}\exp
\Big[-\int d^{3}x\,{\cal L}_{\mbox{\scriptsize EQED}}\Big].
\end{equation}
There is another physical quantity for a hot plasma, in addition
to the free energy, which is of great interest, and this is the
electric screening mass. This quantity gives information about the
screening of static electric fields at long distances.
The potential between two static
charges in the plasma
is normally derived in linear response theory~\cite{kapusta},
and reads
\beq
V(R)=Q_{1}Q_{2}\int\frac{d^{3}k}{(2\pi )^{3}}
e^{i{\bf kR}}\frac{1}{k^{2}+\Pi_{00}(0,{\bf k})}.
\eeq
Here, $\Pi_{\mu\nu}(k_{0},{\bf k})$ is the photon polarization tensor.
In the limit $R\rightarrow \infty$, the potential is dominated by the pole in
photon propagator. At leading order this pole is given by the infrared limit
of $\Pi_{00}(0,{\bf k})$, and the potential is thus a modified Coulomb
potential with an inverse screening length 
or electric screening mass $\Pi_{00}(0,{\bf k}\rightarrow 0)$.
This has led
one to {\it define} the electric screening mass as the infrared
limit of the polarization tensor~\cite{kapusta}:
\beq
\label{IR}
m_{s}^{2}=\Pi_{00}(0,{\bf k}\rightarrow 0).
\eeq
This definition cannot be the correct one, since, beyond leading order in
the coupling, it 
is gauge-fixing dependent in nonabelian theories [51,94]. Although 
$\Pi_{\mu\nu}(k_{0},{\bf k})$ is a manifestly gauge-fixing independent
quantity in Abelian theories, the infrared limit is not renormalization 
group invariant, and is so a useless definition even here~\cite{anton1}. 

The electric screening mass is correctly defined as the 
the position of the pole of the propagator at spacelike 
momentum~\cite{anton1}:
\beq
\label{propdef2}
k^{2}+\Pi_{00}(0,{\bf k})=0,\hspace{1cm}k^{2}=-m_{s}^{2}.
\eeq
This definition is gauge fixing independent order by order in perturbation
theory, which can be proved on an algebraic 
level\cite{kobes}\footnote{The pole position
is also independent
of field redefinitions. Since the relation between the fields in the 
underlying theory and the effective theory can be viewed as a field 
redefinition, and since the screening mass is a long-distance quantity, 
one can use the effective theory to compute it.}.
We also note that the two definitions normally coincide at leading order
in the coupling constant.
The above definition can be extended to other theories, 
e.g. $\phi^{4}$-theory.
The polarization tensor is then replaced by the self-energy function 
of the
scalar field, and the screening mass then reflects the screening of
static scalar fields in the plasma.

However, it turns out that one cannot calculate perturbatively the 
screening
mass beyond leading order in nonabelian gauge theories using 
Eq.~(\ref{propdef2})~\cite{anton1}. The problem is a linear mass-shell singularity.
This signals the breakdown of perturbation theory, and calls for a 
gauge-fixing independent
and nonperturbative definition of the electric screening mass~\cite{polyakov}.

In Abelian gauge theories the above definition is equivalent to defining the 
Debye mass as
the correlation length of equal-time electric field correlation function
\cite{polyakov}
\beq
\label{corr}
\langle {\bf E}({\bf x})\cdot{\bf E}(0)\rangle\sim
e^{-m_{s}x}/x^{3},\hspace{1cm}x\rightarrow\infty.
\eeq
Here $x=|{\bf x}|$. 
Unfortunately, the
definition Eq.~(\ref{corr}) is a poor one in nonabelian theories,
since ${\bf E}$ is no longer a gauge invariant quantity.
The above considerations have
led Arnold and Yaffe to define the electric screening
in terms of Polyakov loops~\cite{polyakov}. We shall not pursue this any further, but stick
to the definition based on the pole of the propagator.

\section{The Short-distance Coefficients}
\heading{Effective Field Theory Approach II}{The Short-distance Coefficients}
In this section we determine the parameters of EQED. 
In the Feynman graphs a solid line denotes fermions, the 
photon is a wiggly line, while the ghost is indicated by a dotted line.
In the effective theory the same conventions apply to the gauge field
and the ghost, while the real scalar is denoted by a dashed line.

As we have discussed in some detail in the previous chapter,
strict perturbation theory is ordinary perturbation theory in $e^2$.
In full QED this corresponds to the usual partition of the Lagrangian
into a free and an interacting part
\bqa
({\cal L}_{\mbox{\scriptsize QED}})_{0}&=&\frac{1}{4}
F_{ij}F_{ij}+\bar{\psi}\partial\!\!\!/\psi
+{\cal L}_{\mbox{\footnotesize gf}}
+{\cal L}_{\mbox{\footnotesize gh}},\\
({\cal L}_{\mbox{\scriptsize QED}})_{\mbox{\footnotesize int}}
&=&
-ie\bar{\psi}A\!\!\!/\psi
\eqa
The effective Lagrangian is split the following way:
\bqa
({\cal L}_{\mbox{\scriptsize EQED}})_{0}&=&\frac{1}{4}
F_{ij}F_{ij}+\frac{1}{2}
(\partial_{i}A_0)^{2}+{\cal L}_{\mbox{\footnotesize gf}}
+{\cal L}_{\mbox{\footnotesize gh}},\\
({\cal L}_{\mbox{\scriptsize EQED}})_{\mbox{\footnotesize int}}
&=&\frac{1}{2}m^{2}_{E}(\Lambda )A_0^{2}+\frac{\lambda_{E}(\Lambda)}{24}A_0^{4}
+\delta {\cal L}.
\eqa
\subsection{The Coupling Constant}
\begin{figure}[htb]
\begin{center}
\mbox{\psfig{figure=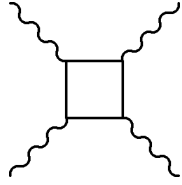}}
\end{center}
\caption[One-loop, four point function with external timelike 
photon lines.]{\protect One-loop four point function with external timelike 
photon lines.}
\label{four}
\end{figure}

\heading{Effective Field Theory Approach II}{The Coupling Constant}
We need the coefficient in front of the quartic
term $A_0^{4}$. 
This coefficient can be found 
by  considering the one-loop contribution to the four-point
function for timelike photons at zero external momenta, and was
first obtained by Landsman in Ref.~\cite{lands}.
Hence, the matching condition is
\beq
\lambda_{E}(\Lambda)=T\Gamma^{(4)}_{\tiny A_{0}A_{0}A_{0}A_{0}}(0).
\eeq
The corresponding Feynman graph is displayed in Fig.~\ref{four}.
It requires the calculation of the following sum-integral, which is
infrared safe since only fermionic propagators are involved:
\bqa \nonumber
\label{fourcoup}
\Gamma^{(4)}_{\tiny A_{0}A_{0}A_{0}A_{0}}(0)&=&
6e^{4}\hbox{$\sum$}\!\!\!\!\!\!\int_{\{P\}}
\mbox{Tr}
\Big[\gamma_{0}\frac{P\!\!\!\!/}{P}\gamma_{0}
\frac{P\!\!\!\!/}{P^2}\gamma_{0}
\frac{P\!\!\!\!/}{P^2}\gamma_{0}\frac{P\!\!\!\!/}{P^2}\Big] \\
&=&6e^{4}\hbox{$\sum$}\!\!\!\!\!\!\int_{\{P\}}
\Big[\frac{32p_{0}^{4}}{P^{8}}-\frac{32p_{0}^{2}}{P^{6}}
+\frac{4}{P^{4}}\Big].
\eqa
The sum of these three integrals is finite in dimensional 
regularization and so the net result turns out to be
\beq
\lambda_{E}(\Lambda )=-\frac{2e^{4}T}{\pi^{2}}.
\eeq
This coefficient is independent of the renormalization
scale $\Lambda$ to order $e^{4}$. Notice the sign in
front of it, which is the opposite as in SQED, as we saw in the
previous chapter. In $SU(N)$ coupled to  fermions, the sign depends
on the ratio $N/N_f$, where $N_f$ is the number of flavours~\cite{lands}.
However, it is not large enough to shift the minimum
of the effective action to a non-zero value of $A_0$.

\subsection{The Mass Parameter}
\heading{Effective Field Theory Approach II}{The Mass Parameter}
In the previous chapter we determined the mass parameters by matching
the propagators for the zero-frequency modes in the two theories.
An equivalent way of determining the mass parameter is by demanding
that the screening mass in the two theories match. 
Generally, the mass parameter (which is the 
unphysical screening mass obtained in strict perturbation theory)
differs 
from the screening mass obtained in resummed perturbation
theory, which
correctly incorporates the effects of electrostatic
screening. 
We shall need the mass parameter $m^2_E(\Lambda)$ at next-to-leading
order in $e$.

In the previous section we saw that the 
screening mass of the particles is defined as the location 
of the 
pole of the propagator for spacelike momentum:
\begin{equation}
\label{propdef1}
k^{2}+\Pi_{00} (0,{\bf k})=0,\hspace{1cm}k^{2}=-m^{2}_{s}.
\end{equation}
The requirement above implies that
\begin{equation}
\label{defeff}
k^{2}+m^{2}(\Lambda )+\Pi_{E}(k,\Lambda )=0,
\hspace{1cm}k^{2}=-m^{2}_{s},
\end{equation}
where $\Pi_{E}\,(k,\Lambda )$ 
is the self-energy of $A_0$
in the effective
theory. One can expand $\Pi (k^{2})\equiv
\Pi_{00}(0,k)$
in a Taylor series around $k^{2}=0$.
To determine the
screening mass squared to 
order $e^{4}$, we must calculate 
$\Pi^{\prime}(0)$ to one loop order and $\Pi (0)$ 
to two loop order, and the screening mass squared is 
then given by \cite{braaten}
\begin{equation}
\label{scrm1}
m_{s}^{2}\approx\Pi_{1}(0)+\Pi_{2}(0)-\Pi_{1}(0)\Pi_{1}^{\prime}(0),
\end{equation}
where $\Pi_{n}(k)$ denotes the $n$'th order contribution
to $\Pi (k)$ in the loop expansion.
The symbol $\approx$ indicates that Eq.~(\ref{scrm1}) 
only holds in strict perturbation theory.
The self-energy to one-loop order in the full theory reads
\begin{eqnarray}\nonumber
\label{scr1}
\Pi_{1}(k^{2}) &=&e^{2}\hbox{$\sum$}\!\!\!\!\!\!\int_{\{P\}}
\mbox{Tr}\Big [\frac{\gamma_{0}
P\!\!\!\!/\gamma_{0}(P\!\!\!\!/+K\!\!\!\!/)}{P^2(P+K)^2}\Big]\\
&=&-4(d-2)e^{2}\hbox{$\sum$}\!\!\!\!\!\!\int_{\{P\}}
\frac{1}{P^{2}}+\frac{2}{3}(d-2)e^{2}k^{2}
\hbox{$\sum$}\!\!\!\!\!\!\int_{\{P\}}
\frac{1}{P^{4}}+{\cal O}(k^{4}/T^2). 
\end{eqnarray}
The corresponding Feynman diagram is shown in 
Fig.~\ref{1qedm}. 
The sum-integrals in Eq.~(\ref{scr1}) are standard and are listed
in appendix A.
The second sum-integral is ultraviolet divergent and this divergence
may be sidestepped by renormalizing the wave function in the usual way.
The field strength renormalization constant to the order required is 
\beq
\label{felt}
Z_{\scriptsize A}=1-\frac{e^{2}}{12\pi^2\epsilon}.
\eeq
This yields
\bqa
\label{1loop}
\Pi_{1}(k^{2})&=&\frac{e^{2}T^{2}}{3}+\frac{e^{2}k^{2}}{12\pi^{2}}
(2\ln \frac{\Lambda }{4\pi T}+2\gamma_{\scriptsize E}-1+4\ln 2)
+{\cal O}(k^{4}/T^2),\\
\Pi_{1}^{\prime}(0)&=&\frac{e^{2}}{12\pi^{2}}
(2\ln \frac{\Lambda }{4\pi T}+2\gamma_{\scriptsize E} -1+4\ln 2).
\eqa

\begin{figure}[htb]
\begin{center}
\mbox{\psfig{figure=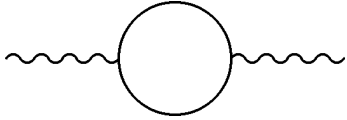}}
\end{center}
\caption[One-loop self-energy graph in full QED.]{\protect One-loop self-energy graph in full QED.}
\label{1qedm}
\end{figure}

\begin{figure}[htb]
\begin{center}
\mbox{\psfig{figure=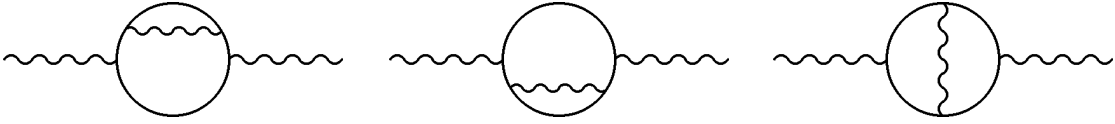}}
\end{center}
\caption[Two-loop self-energy graphs for $\Pi_{00}$
at zero external momentum.]{\protect Two-loop self-energy graphs for $\Pi_{00}$
at zero external momentum.}    
\label{2qedm}
\end{figure}

The two-loop expression for the self energy at zero external
momentum can be found 
either by a direct computation of the two-loop graphs (See Fig~\ref{2qedm})
or by 
applying the
formula (see Ref.~\cite{kapusta})
\begin{equation}
\Pi (0)=-e^{2}\frac{\partial^{2}P}{\partial\mu^{2}},
\end{equation}
where $P$ is the pressure and $\mu$ is the chemical 
potential. 
This requires
the calculation of the free energy to two-loop order 
including 
the chemical
potential. This has been done using contour integration 
in Ref.~\cite{kapusta}, and the two-loop part is
\beq
P_2=
-\frac{e^2}{288}\Big[5T^4+\frac{18}{\pi^2}\mu^2 T^2+\frac{9}{\pi^4}\mu^4\Big].
\eeq
We then find
\begin{equation}
\Pi_{2}(0)=
-\frac{e^{4}T^{2}}{8\pi^{2}}.
\end{equation}

Let us just for the record also list the two-loop graphs which are
shown in Fig.~\ref{2qedm}.
It is interesting to note that diagrams are individually quite
complicated, but the sum is surprisingly simple:
\beq
\Pi_2(0)=
4(d-2)e^4\hbox{$\sum$}\!\!\!\!\!\!\int_{\{PQ\}}
\Big[4\frac{q_0^2}{P^2Q^6}-\frac{1}{P^2Q^4}\Big]-
4(d-2)e^4\hbox{$\sum$}\!\!\!\!\!\!\int_{P\{Q\}}
\Big[4\frac{q_0^2}{P^2Q^6}-\frac{1}{P^2Q^4}\Big].
\eeq

\begin{figure}[htb]
\begin{center}
\mbox{\psfig{figure=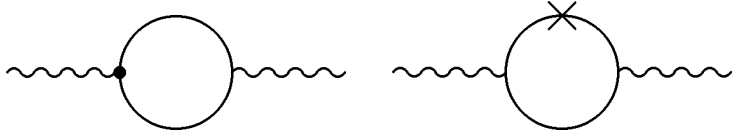}}
\end{center}
\caption[One-loop counterterm diagrams which cancel due to the
Ward identity.]{\protect One-loop counterterm diagrams which cancel due to the
Ward identity.}
\label{1cqedm}
\end{figure}

The result is finite before renormalization, and the reason is that the
two counterterm diagrams cancel as a consequence of the Ward identity 
(See Fig.~\ref{1cqedm}).

\begin{figure}[htb]
\begin{center}
\mbox{\psfig{figure=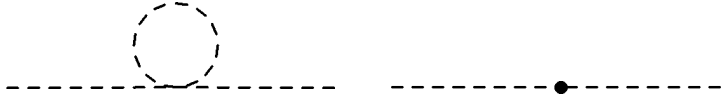}}
\end{center}
\caption[One-loop self-energy correction in the effective theory.]{\protect One-loop self-energy correction in the effective theory.}
\label{1mqed}
\end{figure}

Now, strict perturbation theory in the effective theory 
means 
that the mass term should be treated as a perturbation. 
The corresponding contributions to 
$\Pi_{E}\,(p,\Lambda )$ are 
shown in 
Fig.~\ref{1mqed}, where the blob indicates a mass insertion.
The one-loop diagram vanishes in dimensional
regularization for massless fields, since
the external momentum provides the only mass scale in the
integral.
The matching relation then simply becomes 
$m^{2}_{E}(\Lambda )\approx m_{s}^{2}$.
The mass parameter squared to order $e^{4}$ reads:
\begin{equation}
\label{mass}
m^{2}_{E}(\Lambda)=T^{2}\Big [\frac{e^{2}}{3}
-\frac{e^{4}}{36\pi^{2}}(2
\ln \frac{\Lambda}{4\pi T}+2\gamma_{\scriptsize E} -1+4\ln 2) -\frac{
e^{4}}{8\pi^{2}}\Big ].
\end{equation}
At this point some comments are in order. Firstly, one could obtain
the mass parameter without renormalizing the wave function as an
intermediate step. Instead one uses Eq.~(\ref{scrm1}) directly 
and the divergence there 
is then cancelled by the charge renormalization 
counterterm. 
Secondly, by using the renormalization
group equation for the coupling constant,
\beq
\mu\frac{de^2}{d\mu}=\frac{e^4}{6\pi^2},
\eeq
one can easily demonstrate that Eq.~(\ref{mass}) is 
independent of $\Lambda$.
Thus, up to corrections of order $e^{6}$,
we can replace $\Lambda$ by an arbitrary 
renormalization scale $\mu$. It is interesting to note that the mass parameter
is equal to the physical screening mass at order $e^4$. This is in contrast
with SQED (see subsection~\ref{masse}).

\subsection{The Coefficient of the Unit Operator}
\begin{figure}[htb]
\begin{center}
\mbox{\psfig{figure=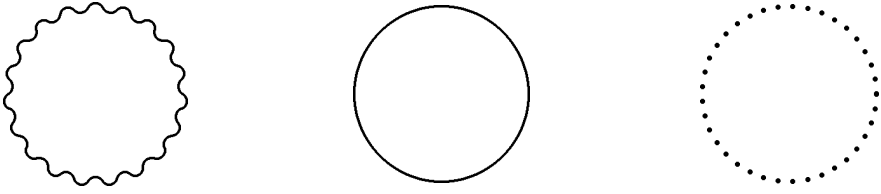}}
\end{center}
\caption[One-loop vacuum diagrams in QED.]{\protect One-loop vacuum diagrams in QED.}
\label{1lqed}
\end{figure}

\heading{Effective Field Theory Approach II}{The Coefficient of the Unit Operator}
We shall now compute $f_{E}(\Lambda )$ to order $e^{4}$ in strict 
perturbation 
theory and we
shall do so by matching calculations of $\ln {\cal Z}$ in the 
full theory and in the effective theory. 
From Eqs. (\ref{z1}) 
and (\ref{z2}),
we see that the matching condition reads
\begin{equation}
\label{mat1}
\ln {\cal Z}=-f_{E}(\Lambda )V+\ln 
{\cal Z}_{\mbox{\scriptsize EQED}}.
\end{equation}
Here, ${\cal Z}_{\mbox{\scriptsize EQED}}$ is the partition function of
the effective theory, which is given by the path integral in 
Eq.~(\ref{z2})
The calculation of $\ln {\cal Z}$ in the full theory
involves one-loop, two-loop and three-loop diagrams, and we shall discuss
them separately in the following. The one-loop contribution is displayed
in Fig.~\ref{1lqed}, and reads
\bqa
-\frac{1}{2}(d-2)\hbox{$\sum$}\!\!\!\!\!\!\int_P\ln P^2
+2\hbox{$\sum$}\!\!\!\!\!\!\int_{\{P\}}\ln P^2=\frac{11\pi^2T^4}{180}.
\eqa
Note that the contribution from the ghost field cancels the contribution
from two of the four polarization states of the photon, and we are left
with the contribution from the two transverse (or physical)
polarization states. The one-loop result is the standard one for a gas
of noninteracting photons and fermions at temperature $T$. 

\begin{figure}[htb]
\begin{center}
\mbox{\psfig{figure=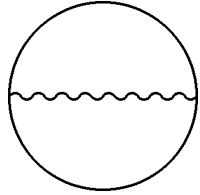}}
\end{center}
\caption[Two-loop vacuum diagrams in QED.]{\protect Two-loop vacuum diagrams in QED.}
\label{2lqed}
\end{figure}

At the two-loop level there is only one vacuum graph (Fig.~\ref{2lqed}), 
which yields
\beq
\label{toqed}
\frac{1}{2}\hbox{$\sum$}\!\!\!\!\!\!\int_{\{P\}}\mbox{Tr}
\Big[\frac{P\!\!\!\!/}{P^2}\Sigma_{f}(P)\Big]
=(d-2)e^2\Big[2\hbox{$\sum$}\!\!\!\!\!\!\int_{P\{Q\}}\frac{1}{P^2Q^2}
-\hbox{$\sum$}\!\!\!\!\!\!\int_{\{PQ\}}\frac{1}{P^2Q^2}\Big].
\eeq
Here, $\Sigma_{f}(P)$ is the fermion self-energy function:
\beq
\Sigma_{f}(P)=e^2\hbox{$\sum$}\!\!\!\!\!\!\int_{\{Q\}}
\frac{\gamma_{\alpha}Q\!\!\!\!/\gamma_{\alpha}}{Q^2(P+Q)^2}.
\eeq
It is interesting to note that this is $(d-2)$ times the contribution of
the corresponding diagram in Yukawa theory. The reason is that the fermion
self-energy function in QED is $(d-2)$ times
the fermion self-energy function in
Yukawa theory.

\begin{figure}[htb]
\begin{center}
\mbox{\psfig{figure=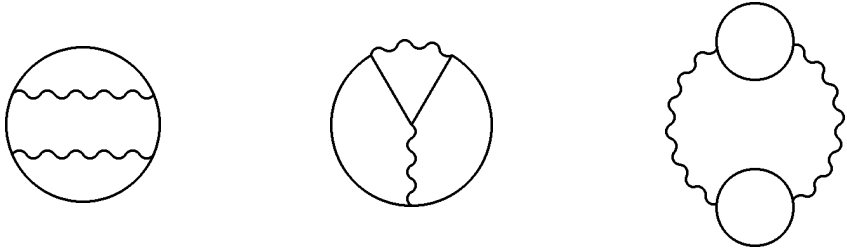}}
\end{center}
\caption[Three-loop vacuum diagrams contributing to three free energy in QED.]{\protect Three-loop vacuum diagrams contributing to three free energy in QED.}
\label{3lqed}
\end{figure}

The three-loop diagrams are displayed in Fig~\ref{3lqed}, and there are several
comments that we wish to make;
The sum of
the first two diagrams in Fig.~\ref{3lqed} is ultraviolet finite. 
The third diagram
in Fig.~\ref{3lqed} has a linear infrared divergence, which is set to zero 
in dimensional regularization. This diagram is the first in the infinite
series of infrared divergent
diagrams (ring diagrams) that are summed to give the first non-analytic
($e^{3}$) contribution to the free energy.

\begin{figure}[htb]
\begin{center}
\mbox{\psfig{figure=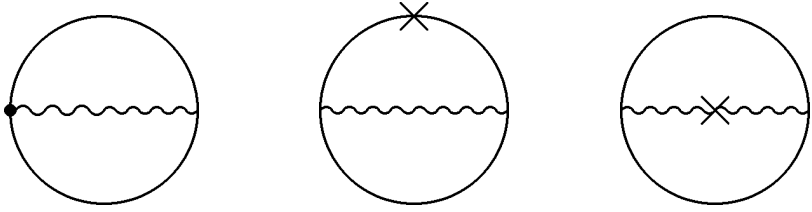}}
\end{center}
\caption[Two-loop counterterm diagrams in QED. The first two cancel since
$Z_1=Z_2$.]{\protect Two-loop counterterm diagrams in QED. The first two cancel since
$Z_1=Z_2$.}
\label{2lcqed}
\end{figure}

The first diagram
can also be written in terms of the fermion self-energy function,
exactly as in the Yukawa case. Recalling the relation between the 
two fermion self-energy functions, the Yukawa result translates into:
\bqa\nonumber
-\frac{1}{2}\hbox{$\sum$}\!\!\!\!\!\!\int_{\{P\}}\mbox{Tr}
\Big[\frac{P\!\!\!\!/}{P^2}\Sigma_{f}(P)
\Big]^2
&=&
-(d-2)^2e^4\hbox{$\sum$}\!\!\!\!\!\!\int_{\{P\}}\frac{1}{P^4}\Big[
\hbox{$\sum$}\!\!\!\!\!\!\int_{\{Q\}}\frac{1}{Q^2}
-\hbox{$\sum$}\!\!\!\!\!\!\int_{Q}\frac{1}{Q^2}\Big]^2\\ \nonumber
&&
-(d-2)^2e^4\hbox{$\sum$}\!\!\!\!\!\!\int_{PQ\{K\}}
\frac{1}{P^2Q^2K^2(P+Q+K)^2}\\ 
&&+2(d-2)^2e^4\hbox{$\sum$}\!\!\!\!\!\!\int_{\{P\}QK}
\frac{QK}{P^2Q^2K^2(P+Q)^2(P+K)^2}.
\eqa
The second diagram cannot be written in any simple way
\bqa\nonumber
-\frac{1}{4}e^4\hbox{$\sum$}\!\!\!\!\!\!\int_{P\{QK\}}\mbox{Tr}
\Big[\frac{\gamma_{\alpha}Q\!\!\!\!/\gamma_{\beta}(P\!\!\!\!/-Q\!\!\!\!/)\gamma_{\alpha}(P\!\!\!\!/-K\!\!\!\!/)\gamma_{\beta}K\!\!\!\!/}{P^2Q^2K^2(P-Q)^2(Q-K)^2(P-K)^2}\Big]&=&\\ \nonumber
\frac{(d-2)(6-d)e^4}{2}\hbox{$\sum$}\!\!\!\!\!\!\int_{\{PQK\}}
\frac{1}{P^2Q^2K^2(P+Q+K)^2}-\\ 
+(d-2)(d-4)e^4\hbox{$\sum$}\!\!\!\!\!\!\int_{PQ\{K\}}
\frac{1}{P^2Q^2K^2(P+Q+K)^2}.
\eqa
The third diagram can also be written in a compact way, involving the
polarization tensor:
\bqa\nonumber
\frac{1}{4}\hbox{$\sum$}\!\!\!\!\!\!\int_{P}\frac{1}{P^4}
\Big[\Pi_{\mu\nu}(P)\Big]^2
&=&4(d-4)e^4\hbox{$\sum$}\!\!\!\!\!\!\int_{P\{QK\}}\frac{1}{P^4Q^2K^2}\\ 
\nonumber
&&+(d-4)e^4\hbox{$\sum$}\!\!\!\!\!\!\int_{\{PQK\}}
\frac{1}{P^2Q^2K^2(P+Q+K)^2}\\ 
&&
+16e^4\hbox{$\sum$}\!\!\!\!\!\!\int_{P\{QK\}}\frac{(QK)^2}{P^4Q^2K^2(P+Q)^2(P+K)^2}.
\eqa
Now, let us consider the two-loop counterterm diagrams. 
These appear in Fig.~\ref{2lcqed}. Gauge invariance
implies the Ward identity which ensures that $Z_1=Z_2$. The corresponding
graphs then cancel, and we are left with the diagram with a photon
wave function counterterm insertion. This diagram equals the two-loop
diagram times $(1-Z_{A})$. The divergence here cancels
divergence from the three-loop diagram.
Alternatively, one can
carry out charge renormalization by the substitution
$e^2\rightarrow Z_{e^2}e^2$ in the two-loop graph, where 
\beq
Z_{e^2}=1+\frac{e^{2}}{12\pi^2\epsilon}.
\eeq

Collecting our results, we find 
\begin{eqnarray}
\label{unit1}\nonumber
\frac{T\ln {\cal Z}}{V}&\approx&\frac{11\pi^{2}T^{4}}{180}
-\frac{5e^{2}T^4}{288}-
\frac{e^{4}}{16\pi^2}\Big(\frac{T^2}{12}\Big)^2\Big[
-\frac{20}{3}
\ln (\frac{\Lambda}{4\pi T}) 
-4\gamma_{\scriptsize E} -\frac{319}{12} +\frac{208}{5}
\ln 2
\Big.  \\
&&+\frac{8}{3}
\frac{\zeta '(-3)}{\zeta (-3)}
-\frac{16}{3}\frac{\zeta '(-1)}{\zeta (-1)}
\Big.\Big.\Big ].
\end{eqnarray}
The scale $\Lambda$ may be traded for an arbitrary scale $\mu$ by using the
renormalization group equation for the running gauge coupling.

\begin{figure}[htb]
\begin{center}
\mbox{\psfig{figure=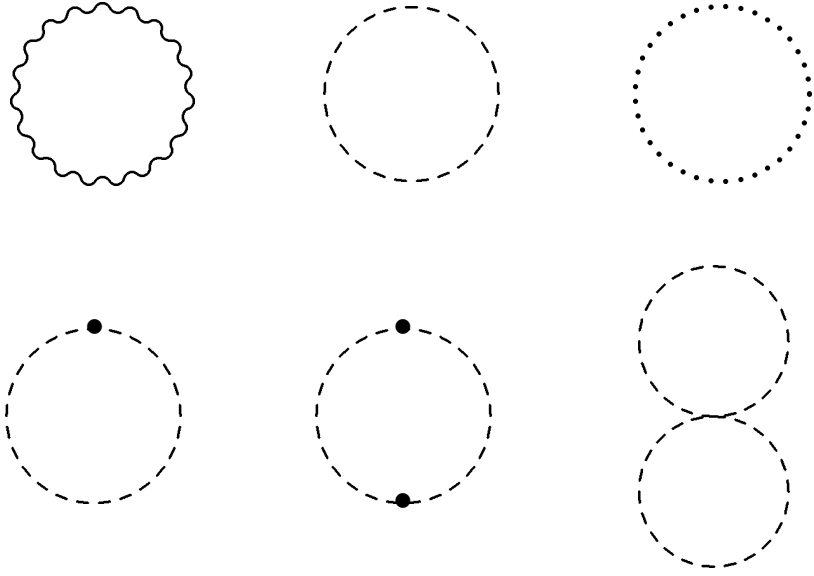}}
\put(-153.5,91){\makebox(0,0){\large$\bullet$} }
\put(-270,91){\makebox(0,0){\large$\bullet$} }
\put(-153.5,25){\makebox(0,0){\large$\bullet$} }
\end{center}
\caption[Loop diagrams in the effective theory.]{\protect Loop diagrams in the effective theory.}    
\label{1eff}
\end{figure}

We now turn to the effective theory.
The mass parameter is viewed as a perturbation in the 
effective theory, as explained above. This implies that
$\ln {\cal Z}_{\mbox{\scriptsize EQED}}$ 
is given
by ordinary one and two-loop diagrams as well as one-loop diagrams
with mass insertions 
(which is indicated by a blob in Fig.~\ref{1eff}).
The computation is
rather simple since loop diagrams involving massless 
fields vanish
identically in dimensional regularization. 
Therefore, 
$\ln {\cal Z}_{\mbox{\scriptsize EQED}}$
vanishes in {\it strict perturbation theory} and
the matching condition turns out to be
\begin{equation}
\frac{T\ln {\cal Z}}{V}\approx-f_{E}(\Lambda )T.
\end{equation}
$f_E(\Lambda)$ is then given by minus the right hand side of Eq.~(\ref{unit1}) 
divided by $T$.

With the comments after Eq.~(\ref{unit1}) in mind, it is 
clear that
$f_E(\Lambda )$ has no dependence on $\Lambda$
at the order we are calculating. The function $F=f_E(\Lambda) T$ 
can be viewed as the 
contribution to the free
energy from the momentum scale $T$, which is a typical
momentum of a particle in the plasma.

\newpage \vspace{1cm}
\section{The Free Energy and the Electric Screening Mass}\label{discussion}
\heading{Effective Field Theory Approach II}{The Free Energy and the Electric Screening Mass}
Now that we have determined the short distance coefficients 
$\lambda_{E}(\Lambda )$, $m^{2}_{E}(\Lambda )$ and $f_{E}(\Lambda)$ 
in the effective 
theory, we can calculate the screening mass squared as
well as the free energy in QED to order $e^{5}$. 
In order to do so, we must take 
properly into account the effects of electric
screening. This corresponds
to the following decomposition of the Lagrangian:
\bqa
({\cal L}_{\mbox{\scriptsize EQED}})_{0}&=&
\frac{1}{4}F_{ij}F_{ij}+\frac{1}{2}
(\partial_{i}A_0)^{2}+\frac{1}{2}m^{2}_{E}(\Lambda )A_0^{2}
+{\cal L}_{\mbox{\footnotesize gf}}
+{\cal L}_{\mbox{\footnotesize gh}}\\
({\cal L}_{\mbox{\scriptsize EQED}})_{\mbox{\footnotesize int}}
&=&\frac{\lambda_{E}(\Lambda)}{24}A_0^{4}
+\delta {\cal L}.
\eqa
The electric screening mass is given by the location
of the pole in the propagator and at leading order
it is simply $m_{s}=m_{E}(\Lambda )$. To leading
order, the self-energy function $\Pi_E\,
(k,\Lambda)$ is given by the first Feynman diagram 
in figure~\ref{1mqed} and is independent of the 
external momentum. Eq.~({\ref{defeff}) then turns into
\bqa\nonumber
m_s^2&=&m_E^2+\Pi_{E}\,(k,\Lambda )\\
&=&m^2_E+\frac{\lambda_E}
{2}\int_p\frac{1}{p^{2}+m^{2}_E}.
\eqa
Using this result, the expression of $\lambda_E(\Lambda)$,
and expanding the mass parameter in 
powers of $e$, we obtain the electric screening mass squared to order $e^{5}$:
\begin{equation}
\label{pscre}
m_{s}^{2}=T^{2}\Big [\frac{e^{2}}{3}-\frac{e^{4}}
{36\pi^{2}}(2\ln \frac{\Lambda}
{4\pi T}+2\gamma_{\scriptsize E} -1+4\ln 2) 
-\frac{e^{4}}{8\pi^{2}}+\frac{e^{5}}{4\sqrt{3}\pi^{3}}\Big ].
\end{equation}
It is easily checked that the result is RG-invariant, 
as required. Furthermore,
our result agrees with the calculation of Blaizot 
{\it et al} \cite{parw1}. Note also that there is no 
$e^{3}$ term in the expression
for the screening mass squared in contrast with both 
$\phi^{4}$-theory and SQED.
The reason is that there are no bosonic propagators
in the one-loop self-energy graph in QED and fermions
need no resummation, since their Matsubara frequencies
are never zero.
Finally, the $e^5$ term is non-analytic in $e^2$ and so corresponds to
the summation of an infinite number of diagrams in terms of bare perturbation
theory. The diagrams are the two-loop graphs in Fig.~\ref{2qedm} 
with insertions of
any number of $\Pi_{00}(k_0,k)$ on the internal photon lines. 
This is in complete analogy with the infinite string of diagrams which we
discussed in connection with the evaluation of the screening mass in 
Yukawa theory in chapter two. 
This is a good example of the efficiency of the effective field theory
approach. Instead of summing an infinite number of diagrams, we simply
perform a one-loop computation in three dimensions.

The calculation of the free energy in the effective 
theory is 
straightforward. Bearing in mind the fact that the self interaction
term contributes to the free energy first at order $e^{6}$,
we only need to perform a one-loop computation,
and so we obtain
\beq
\frac{T\ln {\cal Z}_{\mbox{\scriptsize EQED}}}{V}=-
\frac{1}{2}T\int_p\ln (p^{2}
+m^{2}_E)-\frac{1}{2}(d-3)T\int_p\ln p^{2}.
\eeq
The relevant Feynman graphs are displayed in 
Fig.~\ref{1eff}, except that
the diagrams with a mass insertion, as well as the two-loop graph are
not included.
The contributions from the gauge field and ghost vanish. Using 
the expression for the mass of the scalar field and 
expanding it 
in powers of $e$ yields the following contribution to the 
free energy:
\begin{equation}
\label{logeff}
\frac{T\ln {\cal Z}_{\mbox{\scriptsize EQED}}}{V}=
\frac{e^3T^{4}}{36\sqrt{3}\pi}
-\frac{e^{5}}{576\sqrt{3}\pi^{3}}\Big(4\ln\frac{\Lambda}
{4\pi T}+4\gamma_{\scriptsize E}+7+8\ln 2\Big).
\end{equation}
This term takes into account the effects from long distance 
scales of order
$1/(eT)$, which can be associated with the 
scale of electric screening. 
Using Eqs.~(\ref{mat1}), (\ref{unit}) and 
(\ref{logeff}), one finally obtains
\begin{eqnarray}\nonumber
\label{res}
\frac{T\ln{\cal Z}}{V}&=&\frac{11\pi^{2}T^{4}}{180}
-\frac{5e^{2}T^4}{288}
+\frac{e^{3}T^4}{36\sqrt{3}\pi}
-\frac{e^{4}}{16\pi^{2}}
\Big(\frac{T^2}{12}\Big)^2
\Big[-\frac{20}{3}\ln (\frac{\Lambda}{4\pi T}) 
-4\gamma_E \\ \nonumber
&& 
-\frac{319}{12} +\frac{208}{5}
\ln 2+\frac{8}{3}\frac{\zeta '(-3)}{\zeta (-3)}
-\frac{16}{3}\frac{\zeta '(-1)}{\zeta (-1)}
\Big]-\frac{e^{5}T^4}{576\sqrt{3}\pi^{3}}
\Big[4\ln\frac{\Lambda}{4\pi T}\Big. \\
&&\Big.+4\gamma_{\scriptsize E}+7+8\ln 2\Big ].
\end{eqnarray} 
This result is 
renormalization group invariant as required for a physical quantity.
This can 
easily be checked by using
the one-loop $\beta $-function in QED.
Moreover, it is in 
agreement with the computation of Parwani~\cite{parw2}, and Zhai 
and Kastening \cite{kast},
who use resummed perturbation theory.
The advantage of the effective field theory approach should now be clear;
In order to to extract the $e^3$ and $e^5$ contributions 
to the free energy using
resummation, one must use the resummed propagator in every diagram
and find the subleading pieces by subtracting the leading ones, as we 
indicated
in the previous chapter. This is at least a rather tedious task. Here, we 
obtain the non-analytic terms in the free energy
from a straightforward one-loop calculation.

It is interesting to note that there are no terms in the expressions for the
screening mass or the free energy which involve logarithms of the
coupling constant. This is in contrast with QCD and SQED, where the 
free energy contains a term proportional to $g^4\ln g$~\cite{braaten2}.
This can be understood in terms of the renormalization group
and the renormalization of the parameters in the effective theory.

The short-distance coefficients in ${\cal L}_{\mbox{\scriptsize EQED}}$
are obtained by integrating out the non-zero Matsubara frequencies, and
they are polynomials in $\ln (\Lambda/4\pi T)$. So in order to avoid large
logs, we must choose the cutoff of order $T$ or $2\pi T$. The latter
is a more physical choice in the sense that the heavy modes have masses
$2\pi T$ or more. 

Moreover, in the effective theory one generally
encounters logs of $\Lambda /m_{E}$ in the perturbative expansion. This 
implies that we must choose the scale $\Lambda$ in the effective theory
of order $m_E$ to control the perturbative expansion.
Thus, one must take the parameters in EQED
from the scale $2\pi T$ to the 
scale $m_E$, using the equations which govern their evolution
with the scale. These evolution equations follow from the requirement
that physical quantities be independent of the cutoff, and are in the form
\beq
\Lambda\frac{dC_n(\Lambda)}{d\Lambda}=\beta_n (C(\Lambda)).
\eeq
The beta-functions can be written as power series in the coupling constants
of the effective theory. Using dimensional arguments, we can infer the 
general structure. Consider first $m_{E}^2(\Lambda)$. Its expansion
must be a quadratic polynomial in $e^{2}_{E}(\Lambda)$ and 
$\lambda_{E}(\Lambda)$, and other coefficients, so that the 
dimension of every term
is two. The only coefficient that contribute to order
$e^4$ is $e^{4}_{E}(\Lambda)$ (Recall that $\lambda_{E}(\Lambda)\sim e^4$
and that other coefficients are of even higher powers of $e$).
Reasoning along the same lines, we conclude that the beta-function
of $f_{E}(\Lambda)$ involves the terms 
$e^{2}_{E}(\Lambda)m^{2}_{E}(\Lambda)$,
$\lambda_{E}(\Lambda)m^{2}_{E}(\Lambda)$, as well as 
cubic polynomials in $e^{2}_{E}(\Lambda)$ and 
$\lambda_{E}(\Lambda)$ (and other terms which have dimension three).
The only relevant term at the order $e^4$ is
$e^{2}_{E}(\Lambda)m^{2}_{E}(\Lambda)$. 
Finally, we mention that the beta-functions for $e_E(\Lambda)$
and $\lambda_E(\Lambda)$
vanishes for superrenormalizable interactions~\cite{braaten2}. This implies
that the beta-functions are highly suppressed by powers of the coupling,
since these receive contributions only from higher order operators.

However, we have already noted that $m_{E}^2(\Lambda)$,
$f_{E}(\Lambda)$ and $\lambda_E(\Lambda)$ 
are independent of $\Lambda$ at this order so we have
\beq
\Lambda\frac{dm_{E}^2(\Lambda)}{d\Lambda}={\cal O}(e^6),
\hspace{1cm}\Lambda\frac{df_{E}(\Lambda)}{d\Lambda}={\cal O}(e^6),
\hspace{1cm}\Lambda\frac{d\lambda_{E}(\Lambda)}{d\Lambda}={\cal O}(e^6).
\eeq
The dependence of $\Lambda$ in the parameters in 
${\cal L}_{\mbox{\scriptsize EQED}}$ are canceled by the cutoff
dependence in the effective theory, and so the vanishing of $\beta$-functions
for these parameters explains that no logs of $e$ occur.
More specifically, the fact that $m_{E}^2(\Lambda)$ does not
run implies the non-existence of a term $e^4\ln e$.
Moreover, since $m_E^2(\Lambda)$ has an expansion in $e^2$ and 
$\lambda_E(\Lambda)$ does not run, 
there can be no term $e^5\ln e$ term either. 
Similarly, the vanishing of $\beta$-function for the unit operator
is responsible for the fact that a term $e^4\ln e$ is absent in the
expression for the free energy.
Since $f_E(\Lambda)$ has an expansion in even powers of $e$, an
$e^5\ln e$ can only arise from the mass parameter. However, since
$m_E^2(\Lambda)$ does not run at next-to-leading order in $e$, this
explain the absence of such a term.
\section{QED Versus QCD}\label{versus}
\heading{Effective Field Theory Approach II}{QED Versus QCD}
Finally, we would like to discuss a computational 
as well a principal difference between
QED and QCD, when calculating the free energy beyond the fifth
order in the coupling constant.
In QCD the computation of the free energy 
involves 
the construction of {\it two} effective field theories, which 
reflects the fact that 
there are contributions from three different momentum scales 
($T$, $gT$ and 
$g^{2}T$, where $g$ is the gauge coupling) \cite{braaten2}. 
The first effective field
theory, called electrostatic QCD (EQCD), consists of the 
magnetostatic $A_{i}^{a}$ field and the electrostatic field
$A_{0}^{a}$. The unit operator 
$f_E(\Lambda)$ as well as the 
other parameters in EQCD
are then determined by the usual matching procedure
and $f_E(\Lambda)$ gives the contribution 
to the free energy 
from the short distance scale $1/T$.

The second effective field theory is called magnetostatic 
QCD (MQCD) and consists
simply of the self-interacting magnetostatic gauge field 
$A_{i}^{a}$.
Again, the unit operator $f_M(\Lambda)$ and the coupling constants
of this effective theory 
can be determined by matching calculations,
and $f_M(\Lambda)$ 
yields the contribution to the free energy from the 
distance scale 
$1/(gT)$. Now, the
perturbative expansion in MQCD is plagued with infrared 
divergences, implying
that the functional integral can only be calculated  
non-perturbatively, e.g.
by putting MQCD on a lattice. Using lattice simulations 
the path integral
may be computed, so that one obtains the contribution 
to the free energy
from the scale $1/(g^{2}T)$. It can be written as a
power series in $g$ starting at order $g^{6}$.
The leading contribution, ${\cal O}(g^6)$, has very recently been determined
by Karsch {\it et al.}~\cite{karsk2}. 

One can, of course, construct a second effective field theory, 
which naturally is termed magnetostatic QED 
(MQED), but it is completely unnecessary. Although this is obvious from a 
physical point of view (there are only two scales in QED), it is
instructive to see this in practice.
Now, MQED contains all operators which can be constructed
out of the fields $A_i$, which satisfy the symmetries, such as gauge 
invariance and rotational symmetry
. We can then schematically write
\beq
{\cal L}_{\mbox{\scriptsize MQED}}=\frac{1}{4}F_{ij}F_{ij}
+g_{M}(\Lambda)F_{ij}\nabla^2F_{ij}
+h_{M}(\Lambda)(F_{ij}F_{ij})^2+...
\eeq 
The second operator is the analog to the Uehling term which is well-known
from non-relativistic atomic physics~\cite{ueh}. The coefficient has be 
determined by Landsman~\cite{lands} and is $7\zeta (3)e^2/960\pi^4T^2$.
The third term corresponds to one of the operators in the
famous Euler-Heisenberg Lagrangian~\cite{euler}. Its coefficient has been
worked out in Ref.~\cite{worked}, but it is in an extremely
complicated form. Nevertheless, we shall consider it as an example. 
The two-loop graph arising from this interaction is shown in Fig.~\ref{vanish}.
Using power counting arguments it is
easy to verify that only power ultraviolet divergences occur.
The canonical dimensionality of $F_{ij}$ is $3/2$, and so 
$h_{M}(\Lambda)$ must be proportional to $e^4/T^3$ 
(at leading order in $e$).
The contribution then goes like
\beq
\frac{e^4}{T^3}\int_{pq}\frac{f(p,q)}{p^2q^2}.
\eeq
Here, $f(p,q)$ has dimension four, since the dimension for the free
energy is three.
The power ultraviolet divergences
are artifacts of the regulator, and in dimensional regularization 
they are set to zero~\cite{priv}. Thus, the contribution vanishes.
One can use similar arguments to conclude that every operator,
except for the unit operator, gives zero contribution 
to the free energy in magnetostatic QED.
The point is now that we are actually computing the unit operator, 
$f_M(\Lambda)$ when we do perturbative calculations in EQED and
include the mass term $m_E(\Lambda)$ in the unperturbed part of the
Lagrangian. Thus, there is no need for determining the other coefficients
in MQED.
We then have ${\cal F}=f_E(\Lambda)T+f_M(\Lambda)T$.

\begin{figure}[htb]
\begin{center}
\mbox{\psfig{figure=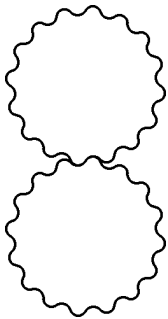}}
\end{center}
\caption[Example of a vanishing two-loop diagrams in magnetostatic QED.]{\protect Example of a vanishing two-loop diagrams in magnetostatic QED.}
\label{vanish}
\end{figure}

Let us close this section by some comments on the infrared catastrophe
in QCD and its solution by the present method~\cite{braaten2}.
It is a well-known fact that 
the free energy of 
nonabelian gauge theories may
be calculated to fifth order in the coupling using resummed 
perturbation theory. 
However, the method 
breaks down at order $g^{6}$, due to infrared divergences,
as first pointed out by Linde~\cite{linde}. 
These divergences arise from regions where all internal energies
vanish, and so the singularities are the same as in
three-dimensional pure QCD. Thus, the breakdown of perturbation
theory simply reflects the infrared problems 
appearing in a perturbative treatment of any nonabelian
gauge theory in three dimensions, in particular MQCD
(although it is well behaved
nonperturbatively with a mass gap of order $g^{2}T$).
In the present approach one can compute order by order the contributions
to the free energy, although some coefficients must be evaluated numerically.
The infrared problems can naturally be avoided if one uses lattice 
simulations directly in four dimensions. However, this is extremely
time consuming in comparison with MQCD, and the time savings here
arise from
the reduction of the problem from four to three dimensions, and 
also by integrating out the fermions.

\section{Electrostatic Scalar Electrodynamics}
\heading{Effective Field Theory Approach II}{Electrostatic Scalar Electrodynamics}
In this section we continue our study of effective field theories by
investigating scalar electrodynamics.
The Euclidean Lagrangian of massless SQED is
\beq
{\cal L}_{\mbox{\scriptsize SQED}}
=\frac{1}{4}F_{\mu\nu}F_{\mu\nu}+({\cal D_{\mu}}\Phi )
^{\dagger}({\cal D_{\mu}}\Phi )+\frac{\lambda}{6}(\Phi^{\dagger}\Phi)^2
+{\cal L}_{\mbox{\footnotesize gf}}
+{\cal L}_{\mbox{\footnotesize gh}},
\eeq
The effective field theory is called electrostatic scalar electrodynamics
(ESQED), and consists of a real massive scalar field (the 
temporal component of the gauge field) coupled to scalar electrodynamics
in three dimensions.
According to the preceding
discussion we must write down the most general Lagrangian which respects
the symmetries at high $T$. Lorentz invariance is broken at finite temperature,
so we must allow for a mass term and self-interactions 
for the timelike component of the gauge field.
Moreover, the Ward identity at high temperature implies that the effective
Lagrangian is a gauge invariant function of the fields $A_{i}$. 
Finally, there is a $Z_2$-symmetry for the fields $\phi$ and $A_0$. This
symmetry follows from the corresponding symmetries in the full theory.
The effective Lagrangian then has the general form
\bqa \nonumber
{\cal L}_{\mbox{\scriptsize ESQED}}&=&
\frac{1}{4}F_{ij}F_{ij}+({\cal D}_{i}\phi )^{\dagger}
({\cal D}_{i}\phi )+M^{2}(\Lambda )
\phi^{\dagger}\phi
+\frac{1}{2}(\partial_{i}A_{0})
(\partial_{i}A_{0})+\\
&&
\frac{1}{2}m^{2}_{E}(\Lambda )A_{0}^{2}
+e^{2}_{E}(\Lambda)\phi^{\dagger}\phi A_{0}^{2}
+\frac{\lambda_{3}(\Lambda)}{6}(\phi^{\dagger}\phi )^2
+{\cal L}_{\mbox{\footnotesize gf}}
+{\cal L}_{\mbox{\footnotesize gh}}
+\delta{\cal L}.
\eqa
Here $\delta {\cal L}$ represents all other terms consistent
with the symmetries.
Examples of such terms
are $\lambda_{E}(\Lambda )A_{0}^{4}$, which is 
superrenormalizable and $h_{E}(\Lambda )(F_{ij}F_{ij})^{2}$, which is 
nonrenormalizable.
\section{The Parameters in ESQED}
\heading{Effective Field Theory Approach II}{The Parameters in ESQED}
The coefficients in ESQED
are again determined
by ordinary perturbation theory in powers
of $e^{2}$ and $\lambda$, 
neglecting resummation. In the full theory we then split the 
Lagrangian into a free part and an interaction part accordingly:
\bqa \nonumber
({\cal L}_{\mbox{\scriptsize SQED}})_{0}&=&\frac{1}{4}
F_{\mu\nu}F_{\mu\nu}+(\partial_{\mu}
\Phi )^{\dagger}(\partial_{\mu}\Phi )
+{\cal L}_{\mbox{\footnotesize gf}}
+{\cal L}_{\mbox{\footnotesize gh}},
\\ \nonumber
({\cal L}_{\mbox{\scriptsize SQED}})
_{\mbox{\footnotesize int}}&=&
e^{2}\Phi^{\dagger}\Phi A_{\mu}^{2}
-ieA_{\mu}
(\Phi^{\dagger}\partial_{\mu}\Phi-\Phi\partial_{\mu}
\Phi^{\dagger})+\frac{\lambda}{6}(\Phi^{\dagger}\Phi )^2.
\eqa
In the effective theory the masses as well as higher order operators
are treated as perturbations. We then write 
${\cal L}_{\mbox{\scriptsize ESQED}}=
({\cal L}_{\mbox{\scriptsize ESQED}})_{0}
+({\cal L}_{\mbox{\scriptsize ESQED}})
_{\mbox{\footnotesize int}}$
and strict perturbation theory
corresponds to the following partition of the 
effective Lagrangian 
\bqa \nonumber
({\cal L}_{\mbox{\scriptsize ESQED}})_{0}&=&\frac{1}{4}
F_{ij}F_{ij}+(\partial_{i}
\phi )^{\dagger}(\partial_{i}\phi )
+\frac{1}{2}(\partial_{i}A_{0})^{2}
+{\cal L}_{\mbox{\footnotesize gf}}
+{\cal L}_{\mbox{\footnotesize gh}},
\\ \nonumber
({\cal L}_{\mbox{\scriptsize ESQED}})
_{\mbox{\footnotesize int}}&=&\frac{1}{2}m^{2}_{E}(\Lambda )A_{0}^{2}+
M^{2}(\Lambda )\phi^{\dagger}
\phi+
e^{2}_{E}(\Lambda)\phi^{\dagger}\phi (A_{i}^{2}+A_{0}^{2})\\
&&-ie_{E}(\Lambda)A_{i}
(\phi^{\dagger}\partial_{i}\phi-\phi\partial_{i}
\phi^{\dagger})+\frac{\lambda_{3}(\Lambda)}{6}(\phi^{\dagger}\phi )^2
+\delta{\cal L}.
\eqa
In SQED the scalar field is denoted
by a dashed line, the photon by a wiggly line, and the ghost
by a dotted line.
In ESQED we have the additional convention that the real scalar field
is indicated by a solid line.
\subsection{The Coupling Constants}
\heading{Effective Field Theory Approach II}{The Coupling Constants}
For the present calculations, we need the gauge coupling 
$e_{E}(\Lambda)$
only to leading order in $e$ and $\lambda$. By using the 
relation between the gauge
fields in the two theories
\beq
A_{i}^{3d}=\frac{1}{\sqrt{T}}A_{i},
\eeq
and comparing ${\cal L}_{\mbox{\scriptsize ESQED}}$ 
with $\int_{0}^{\beta}d\tau{\cal L}_{\mbox{\scriptsize SQED}}$, we find 
\beq
e_{\footnotesize E}^{2}(\Lambda)=e^{2}T.
\eeq
At this order there is no dependence on the renormalization 
scale $\Lambda$. Similarly one finds at leading order 
\beq
\lambda_{3}(\Lambda)=\lambda T.
\eeq
\subsection{The Mass Parameters}\label{masse}
\heading{Effective Field Theory Approach II}{The Mass Parameters}
In this subsection we calculate the parameters $M^{2}(\Lambda)$ and 
$m^{2}_E(\Lambda)$ at leading and next-to-leading order 
in the coupling constants $e$ and $\lambda$, respectively. 
The physical interpretation of a mass parameter 
is that it is the 
contribution to the physical screening mass from 
momenta of order $T$.
The simplest way of determining the
mass parameters is to match the screening masses in SQED and
in ESQED. Denoting the self-energy
for the field $\Phi$ by $\Sigma (k_{0},{\bf k})$, 
the {\it scalar} screening mass is the solution to the 
equation\footnote{We remind the reader that this screening mass has nothing
to do with the screening of electric fields.
As previously noted, this is a quantity which gives information
about the screening of static scalar fields due to rearrangements in the 
plasma.}
\beq
\label{propdef}
k^{2}+\Sigma(0,{\bf k})=0,\hspace{1cm}k^{2}=-m^{2}_{s}.
\end{equation}
The matching requirement implies that 
\beq
\label{sscr}
k^{2}+M^{2}(\Lambda )+\bar{\Sigma}(k,\Lambda )=0,
\hspace{1cm}k^{2}=-m^{2}_{s},
\eeq
where $\bar{\Sigma}\,(k,\Lambda )$ 
is the self-energy of the field $\phi$ in the effective
theory. 
The self-consistent solution to Eq.~(\ref{propdef}) is to leading order
in the coupling constants
$m_{s}^{2}\approx \tilde{\Sigma}_{1}(0)$. Here $\tilde{\Sigma}_{n}(k^{2})
\equiv\Sigma_{n}(0,{\bf k})$ denotes the
nth order contribution to the self-energy in the loop expansion and
the symbol $\approx$ is a reminder that this unphysical screening mass
is obtained in strict perturbation theory.
The relevant diagrams are depicted in Fig.~\ref{1skalar}
and the one-loop self-energy at zero external momentum is given by 
\beq
\label{skmass}
\tilde{\Sigma}
_{1}(0)=(d-1)e^{2}\hbox{$\sum$}\!\!\!\!\!\!\int_{P}\frac{1}{P^{2}}
+\frac{2\lambda}{3}\hbox{$\sum$}\!\!\!\!\!\!\int_{P}\frac{1}{P^{2}}.
\eeq
The limit $d\rightarrow 4$ is perfectly finite, and 
this immediately gives
\beq
\tilde{\Sigma}_{1}(0)=\frac{e^{2}T^{2}}{4}+\frac{\lambda T^2}{18}.
\eeq

\begin{figure}[htb]
\begin{center}
\mbox{\psfig{figure=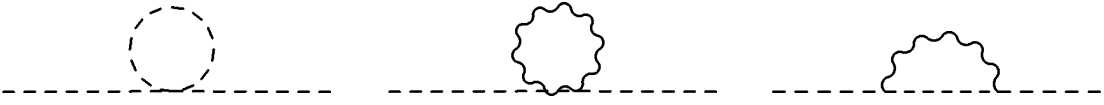}}
\end{center}
\caption[One-loop scalar self-energy diagrams in SQED.]{\protect One-loop scalar self-energy diagrams in SQED.}
\label{1skalar}
\end{figure}

The self-energy function $\bar{\Sigma}(k,\Lambda)$ vanishes in strict 
perturbation theory,
since all the propagators are massless. Hence the matching requirement
gives $m_{s}^{2}\approx M^{2}(\Lambda)$ and the mass parameter is
\beq
M^{2}(\Lambda)=\frac{e^{2}T^{2}}{4}+\frac{\lambda T^2}{18}.
\eeq
At this order $M^{2}(\Lambda)$ is independent of the scale $\Lambda$.

Let us now turn to the mass parameter $m^{2}_E(\Lambda)$. The 
screening mass is again defined as the pole of the propagator at spacelike
momentum
\beq
\label{prop}
k^{2}+\Pi_{00}(0,{\bf k})=0,\hspace{1cm}k^{2}=-m^{2}_{s}.
\end{equation}
The self-energy function is given by a series expansion in $e^{2}$ and
can also be expanded in a Taylor series around $k^{2}=0$. 
The self-consistent
solution to Eq.~(\ref{prop}) at next-to-leading order in the coupling
constant is in analogy with QED
\beq
\label{scrm}
m_{s}^{2}\approx\Big[\Pi_{1}(0)+\Pi_{2}(0)\Big]
\Big [1-\Pi_{1}^{\prime}(0)\Big].
\eeq 
Here, we have again defined
$\Pi(k^{2})\equiv\Pi_{00}(0,{\bf k})$ and $\Pi_{n}(k^{2})$ denotes the 
nth order contribution
to $\Pi (k^{2})$ in the loop expansion. The one-loop self-energy 
is shown in Fig~\ref{1lsqedm}.
and equals
\bqa 
\Pi_{1}(k^{2})&=&2e^{2}\hbox{$\sum$}\!\!\!\!\!\!\int_{P}\frac{1}{P^{2}}
-4e^{2}\hbox{$\sum$}\!\!\!\!\!\!\int_{P}\frac{p^{2}_{0}}{P^{2}(P+K)^{2}} 
\label{divergent}.
\eqa
Expanding in powers of the external momentum and integrating by parts
in $d-1$ dimensions yields
\bqa
\Pi_{1}(k^{2})&=&2e^{2}\hbox{$\sum$}\!\!\!\!\!\!\int_{P}\frac{1}{P^{2}}
-4e^{2}\hbox{$\sum$}\!\!\!\!\!\!\int_{P}\frac{p_{0}^{2}}{P^{4}}
+\frac{4}{3}e^{2}k^{2}\hbox{$\sum$}\!\!\!\!\!\!\int_{P}\frac{p_{0}^{2}}{P^{6}}
+O(k^{4}/T^{2}). 
\eqa
The last sum-integral is ultraviolet divergent and this 
divergence may
be removed by the wave function renormalization counterterm:
\beq
Z_{\mbox{\scriptsize A}}=1-\frac{e^{2}}{3(4\pi)^{2}\epsilon}.
\eeq
One then obtains
\bqa
\label{1qedloop}\nonumber
\Pi_{1}(k^{2})&=&\frac{e^{2}T^{2}}{3}+\frac{2k^{2}}{3(4\pi )^{2}}
(\ln \frac{\Lambda }{4\pi T}+\gamma_{E}+1)+O(k^{4}/T^{2}),\\
\Pi_{1}^{\prime}(0)&=&\frac{2}{3(4\pi )^{2}}
(\ln \frac{\Lambda }{4\pi T}+\gamma_{E}+1).\
\eqa

\begin{figure}[htb]
\begin{center}
\mbox{\psfig{figure=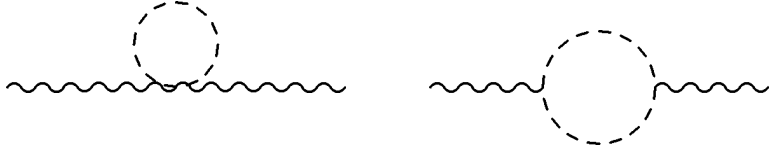}}
\end{center}
\caption[One-loop self-energy diagrams in the full theory.]{\protect One-loop self-energy diagrams in the full theory.}
\label{1lsqedm}
\end{figure}

We also need the self-energy at zero external momentum to 
two loop order.
The contributing diagrams are displayed in Fig.~\ref{2lsqedm}. Many of the
two-loop sum-integrals
vanish in dimensional regularization, while others factorize into products
of one-loop sum-integrals. After some 
calculations we find
\bqa\nonumber
\Pi_{2}(0)&=&8(d-1)e^{4}\hbox{$\sum$}\!\!\!\!\!\!\int_{PQ}
\frac{p_0^2}{P^{6}Q^{2}}
-2(d-1)e^{4}\hbox{$\sum$}\!\!\!\!\!\!\int_{PQ}
\frac{1}{P^{4}Q^{2}}\\
&&
-\frac{4\lambda e^2}{3}\hbox{$\sum$}\!\!\!\!\!\!\int_{PQ}\frac{1}{P^4Q^2}
+\frac{16\lambda e^2}{3}\hbox{$\sum$}\!\!\!\!\!\!\int_{PQ}\frac{p_0^2}{P^6Q^2}.
\eqa
The ultraviolet divergences in the above 
sum-integrals actually cancel, and so we 
are left with a finite expression for $\Pi_{2}(0)$. This cancelation 
is exactly the same as the one we encountered in QED and it
reflects the Ward identity.
Using the tabulated one-loop
sum-integrals in appendix A, one obtains
\beq
\Pi_{2}(0)=\frac{e^{4}T^{2}}{(4\pi)^{2}}+\frac{\lambda e^2T^2}{72\pi^2}.
\eeq
Using these results, we finally obtain $m_{s}^{2}$ to 
order $e^{4}$:
\beq
m_{s}^{2}\approx\frac{e^{2}T^{2}}{3}\Big[1-\Big(
\frac{2}{3}\ln \frac{\Lambda }{4\pi T}+\frac{2}{3}\gamma_{E}
-\frac{7}{3}\Big)
\frac{e^{2}}{(4\pi )^{2}}\Big]+\frac{\lambda e^2T^2}{72\pi^2}.
\eeq
Note that one could have obtained this result without carrying out
wave function renormalization in Eq.~(\ref{divergent}). Instead one uses 
Eq.~(\ref{scrm}) and the divergence there is canceled by the charge 
renormalization counterterm.

In the effective theory the contributing diagrams are the usual
one - and two-loop graphs plus one-loop graphs with mass insertions.
Denoting the self-energy by $\Pi_{E}(k,\Lambda)$, the
screening mass is given by the solution to the equation
\beq
\label{els}
k^{2}+m^{2}_{\scriptsize E}(\Lambda )+\Pi_{\scriptsize E}(k,\Lambda)=0,
\hspace{1cm}k^{2}=-m^{2}_{s}.
\eeq
We are now familiar with the fact that all loop integrals involve
massless fields and these vanish in dimensional
regularization. Hence $\Pi_{E}(k,\Lambda)=0$, and so the matching 
relation becomes 
$m^{2}(\Lambda )\approx m_{s}^{2}$.
Thus 
\bqa
m^{2}_{\scriptsize E}(\Lambda)&=&\frac{e^{2}T^{2}}{3}\Big[1-\Big(
\frac{2}{3}\ln \frac{\Lambda }{4\pi T}+
\frac{2}{3}\gamma_{E}-\frac{7}{3}\Big)
\frac{e^{2}}{(4\pi )^{2}}\Big]+\frac{\lambda e^2T^2}{72\pi^2}.
\eqa 
One can verify that the apparent $\Lambda$-dependence of 
$m^{2}_{\scriptsize E}(\Lambda)$
is illusory. This implies that, up to correction of order $e^{6}$, one can
trade $\Lambda$ for an arbitrary renormalization scale $\mu$. 
The reason behind this fact is that the physical screening mass does
not receive logarithmic corrections in the effective theory to order $e^{4}$ 
(see section~\ref{esqed}).

\begin{figure}[htb]
\begin{center}
\mbox{\psfig{figure=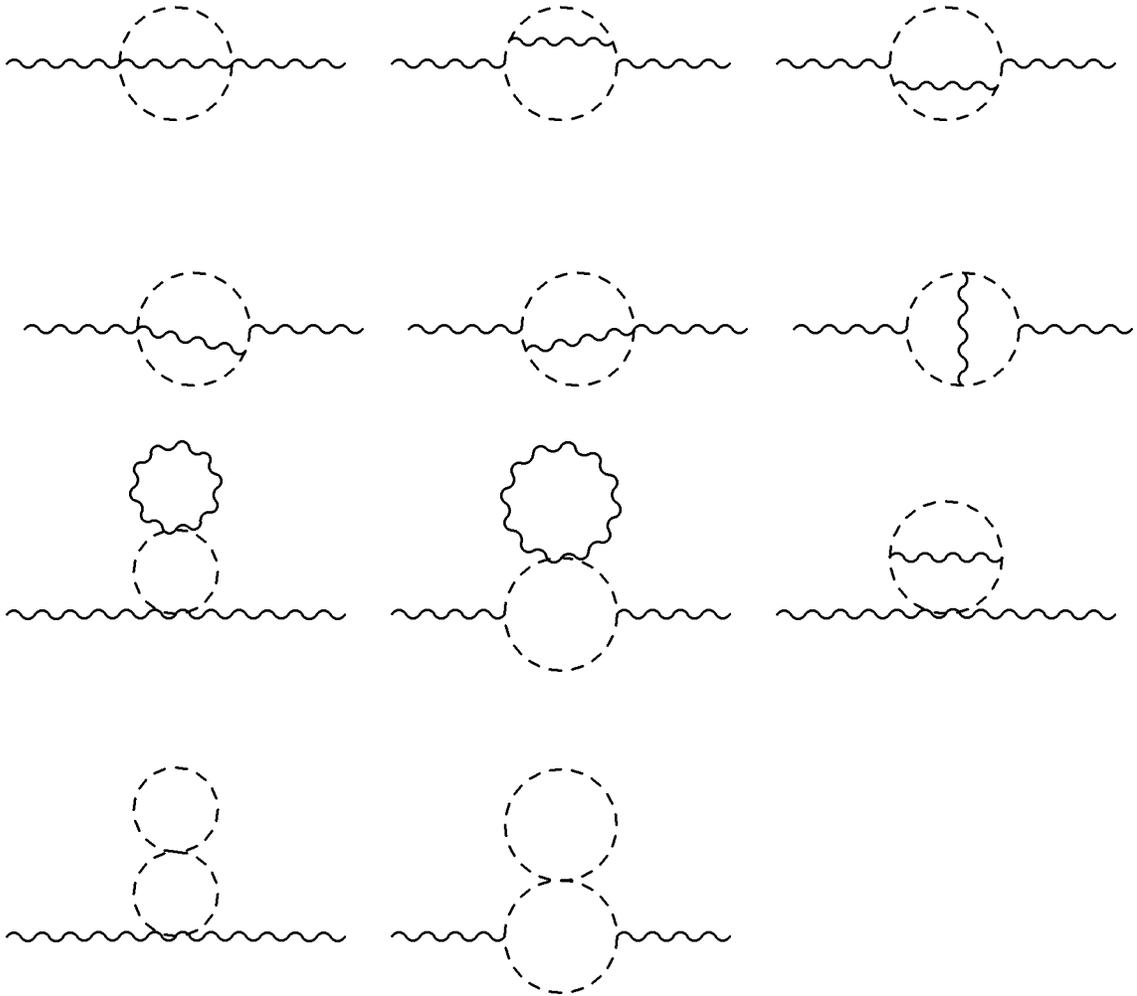}}
\end{center}
\caption[Two-loop self-energy diagrams in the full theory.]{\protect Two-loop self-energy diagrams in the full theory.}
\label{2lsqedm}
\end{figure}

\subsection{The Coefficient of the Unit Operator}
\heading{Effective Field Theory Approach II}{The Coefficient of the Unit Operator}
In this subsection we shall determine the coefficient of the
unit operator. 
We shall consider the one, two and three-loop contributions as well
the contributions from the counterterms diagrams separately. 
The matching condition we use to determine 
$f_E(\Lambda)$ 
follows
from the two path integral representations of the partition function
in complete analogy with QED:
\begin{equation}
\ln {\cal Z}=-f_E(\Lambda )V+\ln 
{\cal Z}_{\mbox{\scriptsize ESQED}}.
\end{equation}
Let us first focus on the SQED. 
The one-loop graphs in the underlying theory are depicted in Fig.~\ref{1lsqed}
and the corresponding contribution
reads
\beq
\frac{d}{2}\hbox{$\sum$}\!\!\!\!\!\!\int_{P}\ln P^2=-\frac{2\pi^2}{45}T^4.
\eeq

\begin{figure}[htb]
\begin{center}
\mbox{\psfig{figure=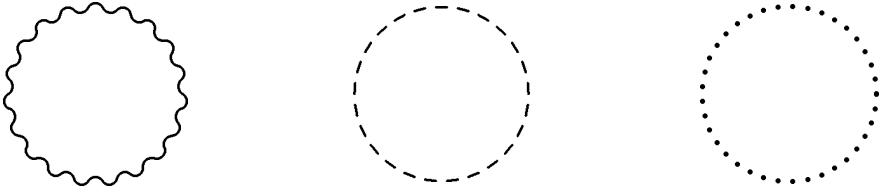}}
\end{center}
\caption[One-loop vacuum diagrams in SQED.]{\protect One-loop vacuum diagrams in SQED.}
\label{1lsqed}
\end{figure}

The two-loop diagrams are shown Fig.~\ref{2lsqed}. 
After some purely algebraic manipulations
they factorize into products of simpler one-loop sum-integrals. The result is
\beq
\frac{\lambda}{3}\hbox{$\sum$}\!\!\!\!\!\!\int_{PQ}\frac{1}{P^2Q^2}
+(d-\frac{3}{2})e^2\hbox{$\sum$}\!\!\!\!\!\!\int_{PQ}\frac{1}{P^2Q^2}
=\Big(\frac{T^2}{12}\Big)^2\Big[\frac{\lambda}{3}+\frac{5e^2}{2}\Big].
\eeq

\begin{figure}[htb]
\begin{center}
\mbox{\psfig{figure=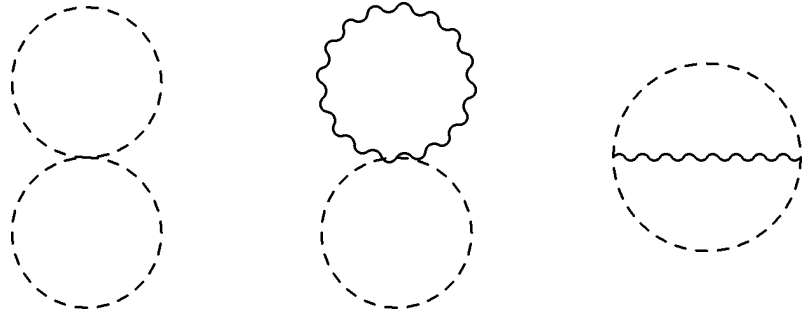}}
\end{center}
\caption[Two-loop diagrams for the free energy.]{\protect Two-loop diagrams for the free energy.}
\label{2lsqed}
\end{figure}

Let us now turn to the three-loop diagrams. These are displayed in 
Fig.~\ref{sqed3}. Here, the shaded blob means insertion of the 
one-loop polarization
tensor $\Pi_{\mu\nu}(k_0,{\bf k})$, while the black blob implies insertion of the
scalar self-energy function $\Sigma (k_0,{\bf k})$, also at one loop.

The first four diagrams can expressed entirely in terms of the bosonic
basketball. After some purely algebraic manipulations one finds
\beq
-\Big[\frac{\lambda^2}{18}+(d-13/4)e^4\Big]
\hbox{$\sum$}\!\!\!\!\!\!\int_{PQK}\frac{1}{P^2Q^2K^2(P+Q+K)^2}.
\eeq
The fifth diagram gives a contribution
\bqa\nonumber
-\frac{1}{4}
\hbox{$\sum$}\!\!\!\!\!\!\int_{P}\frac{1}{P^4}\Big[\Pi_{\mu\nu}(P)
\Big]^2&=&
-e^4\hbox{$\sum$}\!\!\!\!\!\!\int_{PQK}\frac{(K-Q)^4}{P^4Q^2K^2(P-K)^2(P-Q)^2}
-(d-6)e^4\hbox{$\sum$}\!\!\!\!\!\!\int_{PQK}\frac{1}{P^4Q^2K^2}\\
&&
+\frac{e^4}{4}\hbox{$\sum$}\!\!\!\!\!\!\int_{PQK}
\frac{1}{P^2Q^2K^2(P+Q+K)^2}.
\eqa
The last graph reads
\bqa\nonumber
-\frac{1}{2}
\hbox{$\sum$}\!\!\!\!\!\!\int_{P}\frac{1}{P^4}\Big[\Sigma (p_0,{\bf p})\Big]^2&=&
-2e^4\hbox{$\sum$}\!\!\!\!\!\!\int_{PQK}\frac{1}{P^2Q^2K^2(P+Q+K)^2}\\
&&-\Big[\frac{2\lambda^2}{9}+\frac{2(d-1)\lambda e^2}{3}
+\frac{(d-1)^2e^4}{2}\Big]\hbox{$\sum$}\!\!\!\!\!\!\int_{PQK}
\frac{1}{P^4Q^2K^2}.
\eqa

\begin{figure}[htb]
\begin{center}
\mbox{\psfig{figure=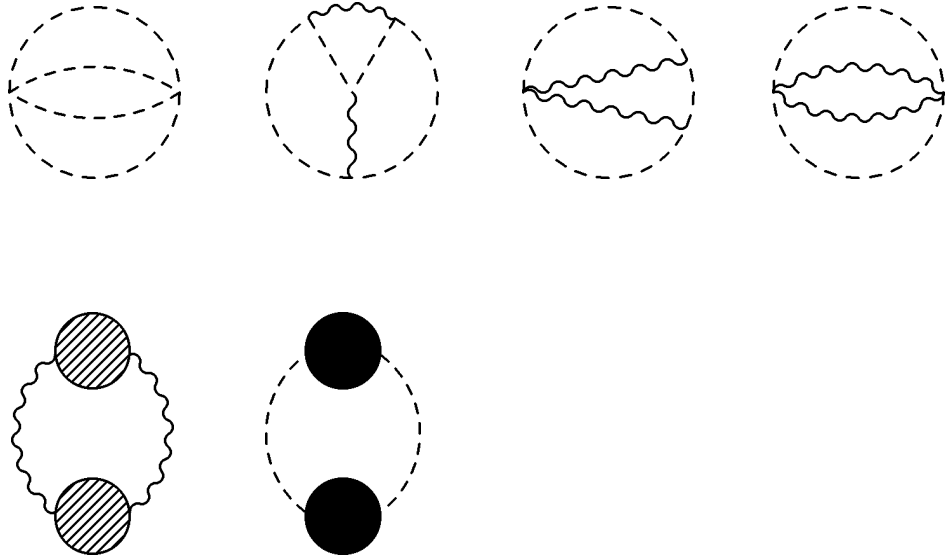}}
\end{center}
\caption[Three-loop diagrams for the free energy.]{\protect Three-loop diagrams for the free energy.}
\label{sqed3}
\end{figure}

With our experience of both Yukawa theory and QED in mind, we carry out
renormalization by the substitutions $\lambda\rightarrow Z_{\lambda}\lambda$
and $e^2\rightarrow Z_{e^2}e^2$. To the order needed,
the renormalization constants
are:
\beq
Z_{\lambda}\lambda=\lambda
+\frac{5\lambda^2-18\lambda e^2+54e^4}{48\pi^2\epsilon},
\hspace{1cm}Z_{e^2}=1+\frac{e^2}{48\pi^2\epsilon}.
\eeq
After we have carried out coupling constant renormalization, we are 
still left with a term proportional to $1/\epsilon$.
This term is canceled by
$\delta f_E(\Lambda)$, 
which is the counterterm for $f_E(\Lambda )$. 
This is the only nonvanishing term in the calculation of
$\ln {\cal Z}_{\mbox{\scriptsize ESQED}}$ 
in strict perturbation theory (we remind the reader
that this involves massless fields, and the loop-integrals are therefore
zero).
According to Ref.~\cite{braaten}, $\delta f_E(\Lambda)$ 
can be computed by considering the 
ultraviolet logarithmic divergences in the effective
theory, when ones uses dimensional 
regularization. 
Generally, $\delta f_E(\Lambda)$, 
is a power series in 
$M^2(\Lambda)$, $m^2_{E}(\Lambda)$, $e^2_{E}(\Lambda)$
and $\lambda_3(\Lambda)$. \\ \\
At leading order it turns out that it is given by
\beq
\label{df}
\delta f_E(\Lambda)=-\frac{e_{E}^2M^2}{2(4\pi )^2}
\frac{1}{\epsilon},
\eeq
which follows from a two-loop calculation in the next section.
Since the mass $M$ is multiplied by $1/\epsilon$ it is necessary
to expand it to first order in $\epsilon$ when expressing
$\delta f_E(\Lambda)$ in terms of $e^2$, $\lambda$ and $T$. 
From Eq. (\ref{skmass}) one finds
\beq
\frac{\partial M^2}{\partial\epsilon}\Bigg|_{\epsilon = 0}=
\frac{e^2T^2}{12}\Big[6\ln\frac{\Lambda}{4\pi T}+4+6\frac{\zeta^{\prime}(-1)}{\zeta (-1)}\Big]+
\frac{\lambda T^2}{18}\Big[2\ln\frac{\Lambda}{4\pi T}+2+2
\frac{\zeta^{\prime}(-1)}{\zeta (-1)}\Big].
\eeq
This implies that
\bqa\nonumber
\delta f_E(\Lambda)T
&=&-\frac{e^4}{(4\pi )^2}\Big(\frac{T^2}{12}\Big)
\Big[\frac{18}{\epsilon}+36
\ln\frac{\Lambda}{4\pi T}+24+36\frac{\zeta^{\prime}(-1)}{\zeta (-1)}\Big]\\
&&-\frac{\lambda e^2}{(4\pi )^2}\Big(\frac{T^2}{12}\Big)
\Big[\frac{4}{\epsilon}+8
\ln\frac{\Lambda}{4\pi T}+8+8\frac{\zeta^{\prime}(-1)}{\zeta (-1)}\Big].
\eqa
Putting our results together, one finally obtains
\bqa\nonumber
\label{unit}
\!\!\!\!\!\!\!\!\hspace{-0.5cm}f_E(\Lambda)T&=&
-\frac{2\pi^2T^4}{45}+\Big(\frac{T^2}{12}\Big)^2\Big[\frac{\lambda}{3}
+\frac{5e^2}{2}\Big]\\ \nonumber
&&-\frac{\lambda^2}{16\pi^2}\Big(\frac{T^2}{12}\Big)^2
\Big[\frac{10}{9}\ln\frac{\Lambda}{4\pi T}+\frac{4}{9}\gamma_{E}
+\frac{31}{45}
-\frac{2}{3}\frac{\zeta^{\prime}(-3)}{\zeta (-3)}
+\frac{4}{3}\frac{\zeta^{\prime}(-1)}{\zeta (-1)}
\Big]\\ \nonumber
&&-\frac{\lambda e^2}{16\pi^2}\Big(\frac{T^2}{12}\Big)^2
\Big[12\ln\frac{\Lambda}{4\pi T}+4\gamma_{E}+\frac{20}{3}
+8\frac{\zeta^{\prime}(-1)}{\zeta (-1)}
\Big]\\
&&-\frac{e^4}{16\pi^2}\Big(\frac{T^2}{12}\Big)^2
\Big[\frac{257}{3}\ln\frac{\Lambda}{4\pi T}+13\gamma_{E}
+\frac{164}{3}-\frac{110}{3}\frac{\zeta^{\prime}(-3)}{\zeta (-3)}
+\frac{328}{3}\frac{\zeta^{\prime}(-1)}{\zeta (-1)}
\Big].
\eqa
In contrast to the corresponding three-loop calculation in QED, 
$f_{\mbox{\scriptsize E}}(\Lambda)$ is not 
renormalization group invariant. This follows easily from the
renormalization group equations, 
and the reason is that there
is a logarithmic ultraviolet divergence at two-loop order in ESQED.
\bqa
\label{rg}
\frac{de^2}{d\mu}&=&\frac{e^4}{24\pi^2},\\
\frac{d\lambda}{d\mu}&=&\frac{5\lambda^2-18\lambda e^2+54e^4}{24\pi^2}.
\eqa
Instead, $f_E(\Lambda)$ satisfies an evolution
equation~\cite{braaten2}, which follows from Eq.~(\ref{df})
\beq
\Lambda\frac{df_E(\Lambda)}{d\Lambda}=-\frac{e_{E}^2M^2}{2(4\pi )^2}.
\eeq

\section{Calculations in ESQED}\label{esqed}
\heading{Effective Field Theory Approach II}{Calculations in ESQED}
Now that we have determined the short-distance coefficients we shall 
use the effective three-dimensional
field theory and calculate the electric screening mass
and the free energy.
We shall do so using perturbation theory
and in order to take the physical effect of
screening into account, 
we must again include the mass parameters in the free part of the 
effective Lagrangian.
This corresponds to the 
following partition of ${\cal L}_{\mbox{\scriptsize ESQED}}$:
\bqa \nonumber
({\cal L}_{\mbox{\scriptsize ESQED}})_{0}&=&\frac{1}{4}
F_{ij}F_{ij}+(\partial_{i}
\phi^{\dagger})(\partial_{i}\phi )+M^{2}(\Lambda )\phi^{\dagger}
\phi+\frac{1}{2}(\partial_{i}A_{0})^{2}\\ 
&&
+\frac{1}{2}
m^{2}_E(\Lambda )A_{0}^{2}
+{\cal L}_{\mbox{\footnotesize gf}}
+{\cal L}_{\mbox{\footnotesize gh}}, \\ \nonumber
({\cal L}_{\mbox{\scriptsize ESQED}})_{\mbox{\footnotesize int}}&=&
e^{2}_{E}(\Lambda)\phi^{\dagger}\phi (
A_{i}^{2}+A_{0}^{2})-ie_{E}(\Lambda )A_{i}
(\phi^{\dagger}\partial_{i}\phi-\phi\partial_{i}
\phi^{\dagger})\\
&&+\frac{\lambda_3(\Lambda)}{6}(\phi^{\dagger}\phi)^2+\delta{\cal L}.
\eqa
The physical screening masses are given by the
self-consistent solutions to Eqs.~(\ref{sscr}) and~(\ref{els}).
The solution to 
Eq.~(\ref{sscr}) to
leading order in coupling is equal to the mass parameter 
$M^{2}(\Lambda)$.
However, recently it has been realized 
that 
this equation has no self-consistent
solution beyond leading order in perturbation theory [52,53].
The problem is the last diagram in Fig.~\ref{1skalar}, which 
has a branch point singularity 
at $k=im_{s}$. The problem is the same as in QCD, namely a scalar field
interacting with a massless gauge field in three dimensions.
In QCD this singularity may be screened by a magnetic mass 
of nonperturbative origin. In SQED the magnetic mass is absent
since it is an Abelian theory~\cite{fradkin}, and so
the problem cannot be solved this way.
We shall not discuss this any further, but refer to Ref.~\cite{nonp}
where a nonperturbative definition of the scalar screening mass is discussed
in detail.

The one and two-loop diagrams that contribute to the
electric
screening mass in ESQED are displayed in Figs.~\ref{12eff}.
We then find
\bqa\nonumber 
\Pi_{E}\,(k,\Lambda )&
=&2e^{2}_{E}\int_{p}\frac{1}{p^{2}+M^{2}}
-2e^{4}_{E}\int_{pq}\frac{\delta_{ii}}{q^{2}(p^{2}+M^{2})}
\\ \nonumber
&&+2
e^{4}_{E}\int_{pq}\frac{({\bf p}+{\bf q})^{2}}{({\bf p}
-{\bf q})^{2}(p^{2}+M^{2})^{2}(q^{2}+M^{2})}
-2e^{4}_{E}\int_{pq}\frac{1}{(q^{2}+M^{2})^{2}(p^{2}+m^{2}_E)}\\\nonumber
&&-4e^{4}_{E}\int_{pq}\frac{1}{(p^{2}+M^{2})
(p^2+q^{2}+m^{2}_E)[({\bf p}+{\bf q}+{\bf k})^{2}+M^{2}]}\\
&&-\frac{4\lambda_{3}e^2_{E}}{3}\int_{pq}\frac{1}{(p^2+M^2)(q^2+M^2)^2}.
\eqa
The integrals may be reduced to known ones by algebraic 
manipulations, which
involve changes of variables.
The integrals needed are tabulated in appendix B.
The second integral above, which corresponds to the the fifth graph in
Fig.~\ref{12eff}, vanishes in dimensional regularization due to the 
masslessness of the photon.
Moreover, the fifth integral is dependent on the external momentum.
This is 
the same as the scalar setting sun diagram we met in the 
chapter on resummation, albeit with different masses.
We recall that the self-consistent solution to Eq.~(\ref{els}) 
is found by evaluating the integral at 
the point $k=im_{s}$.
The calculation of this diagram is carried out in some detail in appendix C.
Notice also that the logarithmic divergence from this integral is exactly
canceled by a corresponding term in the second two-loop integral above.
Adding the different pieces, we obtain the physical screening mass 
squared to order
$e^{4}$ and $\lambda e^2$:
\bqa\nonumber
\label{mainres}
m_{s}^{2}&=&T^{2}\Big[\frac{e^{2}}{3}-\frac{Me^{2}}{2\pi}+
\frac{e^{4}}{(2\pi )^{2}}\Big(-1+\frac{m}{4M}
+(1+\frac{M}{m})
\ln (1+\frac{m}{M})\Big)\\
&&-\frac{2e^{4}}{(12\pi )^{2}}
\Big(1+\gamma_{E}
+\ln\frac{\Lambda}{4\pi T}\Big)\Big]+\frac{\lambda e^2T^2}{18\pi^2}.
\eqa
Settting $\lambda$ to zero, our calculations reproduce the result of
Blaizot {\it et al.}~\cite{parw1},
who used resummation methods. The inclusion of a scalar 
self-interaction term only produces a modification of the
scalar mass parameter $M^2(\Lambda)$ proportional to 
$\lambda$, and an almost trivial term in the
expression for the electric screening mass squared, proportional
to $\lambda e^2$.

Using the renormalization group equation
for $e$, we find that the physical screening mass is independent of the 
renormalization scale $\Lambda$ up to corrections of order $e^{5}$.
We have for completeness also checked that the incorrect definition
$m_{s}^{2}=\Pi_{00}(0,{\bf k}\rightarrow 0)$ does not satisfy the
RG-equation, exactly as in QED, first pointed out by Rebhan~\cite{anton1}.
The lesson we can learn from this, is that gauge fixing independence
is only a {\it necessary}, but not a sufficient criterion for a 
quantity to be physical (recall that $\Pi_{\mu\nu}(k_0,{\bf k})$ is
manifestly gauge fixing independent in Abelian gauge theories).

\begin{figure}[htb]
\begin{center}
\mbox{\psfig{figure=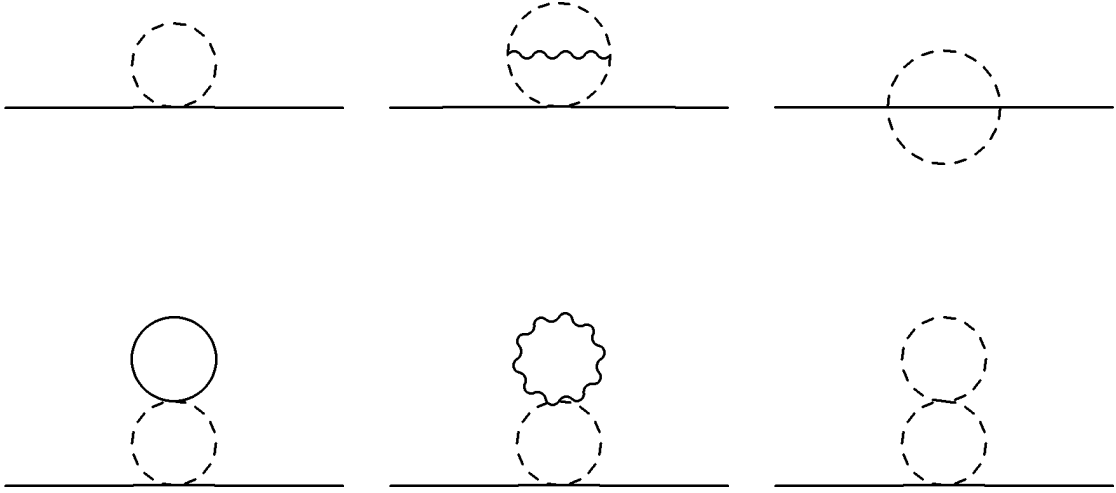}}
\end{center}
\caption[One and two-loop self-energy diagrams in the effective
theory.]{\protect One and two-loop self-energy diagrams in the effective
theory.}
\label{12eff}
\end{figure}

Let us now move on to the calculation of the free energy in ESQED.
The one and two-loop contributions are depicted in Fig.~\ref{2lesqed}
and yield
\bqa\nonumber
\label{frieff}
\frac{T\ln{\cal Z}_{\scriptsize{\mbox{\scriptsize ESQED}}}}{V}&=&
-\frac{1}{2}T\int_p\ln (p^2+m^2_E)-T\int_p\ln (p^2+M^2)
-\frac{1}{2}(d-3)T\int_p\ln p^2\\ \nonumber
&&+\frac{1}{2}
e^2_{\scriptsize{E}}T\int_{pq}\frac{({\bf p}+{\bf q})^2}
{(p^2+M^2)(q^2+M^2)({\bf p}-{\bf q})^2}-d
e^2_{\scriptsize{E}}T\int_{pq}\frac{1}{(p^2+M^2)q^2}\\ \nonumber
&&-e^2_{\scriptsize{E}}\int_{pq}\frac{1}{(p^2+M^2)(q^2+m^2_E)}
-\frac{\lambda_3}{3}\int_{pq}\frac{1}{(p^2+M^2)(q^2+M^2)}\\
&&-\delta f_E(\Lambda).
\eqa
The two-loop contributions may be reduced to products of one-loop
integrals and two-loop integrals which are tabulated in appendix B.
Using this and Eq.~(\ref{unit}) 
we obtain the
free energy through order $\lambda^2$, $\lambda e^2$ and $e^4$:
\bqa\nonumber
-{\cal F}&=&\frac{2\pi^2T^4}{45}
-\Big(\frac{T^2}{12}\Big)^2\Big[\frac{\lambda}{3}+\frac{5e^2}{2}\Big]
+\frac{e^3T^4}{36\pi\sqrt{3}}+\frac{M^3T}{6\pi}
-\frac{e^3MT^3}{16\pi^2\sqrt{3}}
\\ \nonumber
&&+\frac{\lambda^2}{16\pi^2}\Big(\frac{T^2}{12}\Big)^2
\Big[\frac{10}{9}\ln\frac{\Lambda}{4\pi T}+\frac{4}{9}\gamma_{E}
-\frac{89}{45}
-\frac{2}{3}\frac{\zeta^{\prime}(-3)}{\zeta (-3)}]
+\frac{4}{3}\frac{\zeta^{\prime}(-1)}{\zeta (-1)}
\Big]\\ \nonumber
&&+\frac{\lambda e^2}{16\pi^2}\Big(\frac{T^2}{12}\Big)^2
\Big[12\ln\frac{\Lambda}{4\pi T}-16\ln\frac{\Lambda}{2M}
+4\gamma_{E}-\frac{52}{3}
+8\frac{\zeta^{\prime}(-1)}{\zeta (-1)}
\Big]\\ \nonumber
&&+\frac{e^4}{16\pi^2}\Big(\frac{T^2}{12}\Big)^2
\Big[\frac{257}{3}\ln\frac{\Lambda}{4\pi T}-72\ln\frac{\Lambda}{2M}
+13\gamma_{E}+\frac{2}{3}
-\frac{110}{3}\frac{\zeta^{\prime}(-3)}{\zeta (-3)}
+\frac{328}{3}\frac{\zeta^{\prime}(-1)}{\zeta (-1)}
\Big].
\eqa

\begin{figure}[htb]
\begin{center}
\mbox{\psfig{figure=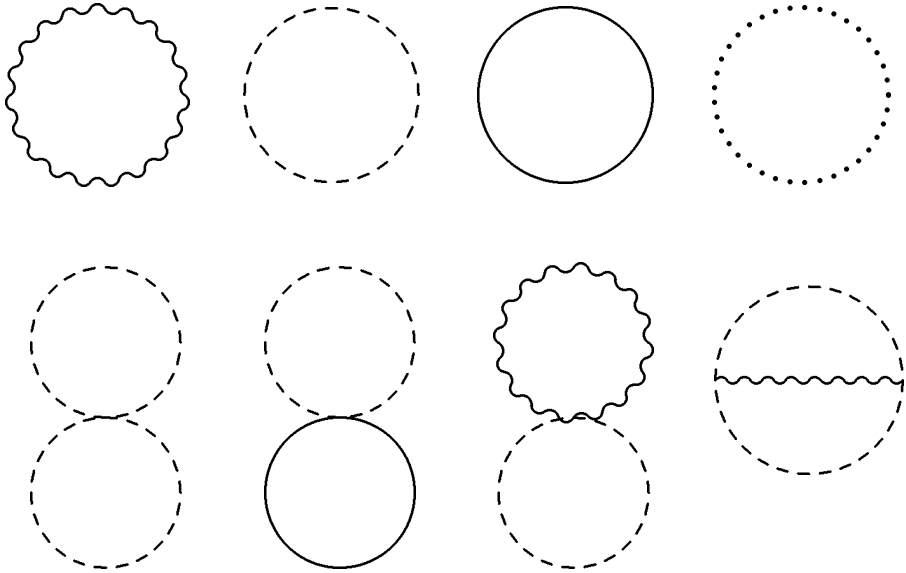}}
\end{center}
\caption[One and two-loop diagrams contributing to
the free energy in ESQED.]{\protect One and two-loop diagrams contributing to
the free energy in ESQED.}
\label{2lesqed}
\end{figure}

Firstly, we note that the two-loop contribution in the gauge sector is
exactly the same as in spinor QED. Secondly, 
our result is renormalization group
invariant up to order $\lambda^2$, $\lambda e^2$ and $e^4$, 
as it must be. This can be easily checked
by using the RG-equations for the running coupling constants.
Thirdly, we notice the appearance of $e^4\ln(\Lambda /M)$ 
and $\lambda e^2\ln(\Lambda /M)$ terms. These are 
necessary in order to cancel the $\Lambda$-dependence in $f_E(\Lambda)$.
The fact that logarithms of the coupling constants occur is then
attributed to the renormalization of $f_{E}(\Lambda )$.
A corresponding term $g^4\ln (\Lambda/m_E)$ was also found by Braaten and 
Nieto in the case of QCD~\cite{braaten2}.
No terms of order $\lambda^2\ln (\Lambda /m)$ arise
in $\lambda\phi^4$-theory, since $f(\Lambda)$ does not
run at next-to-leading order in $\lambda$~\cite{braaten}.

Moreover, from chapter three, we know that the scalar mass parameter
$M^2(\Lambda)$ is not renormalization group invariant, in contrast with
the corresponding mass parameter in QCD. From the general arguments
given in section~\ref{discussion}, this explains the absence of
a $g^5\ln g$ term in expression for the free energy in QCD, and allows
us to predict such terms in the free energy in SQED (just as in pure
$\lambda\phi^4$-theory).
\section{Summary and further Outlook}
\heading{Effective Field Theory Approach II}{Summary and further Outlook}
In this chapter we carried out detailed calculations in gauge 
theories at finite temperature using effective field theory to unravel
the contributions to physical quantities from the scales $T$ and 
$eT\sim\sqrt{\lambda}T$.
This is the advantage of the effective field approach
over the more conventional
resummation procedure; the latter complicates calculations  
unnecessary because the sum-integrals involve both $T$ and $m$.
The simplifications of effective field theory are more transparent
as we go to higher orders in the loop expansion. 

In contrast with QCD, one can, in principle, compute the free energy 
in Abelian gauge theories to any order desired in the coupling 
constant~\cite{braaten2}. 
So let us briefly
outline what it takes to push the calculational frontier
to the next order. Consider first QED. In order to obtain the free energy
to sixth order, we must know $f_{E}(\Lambda)$ to order 
$e^6$. This requires the
calculation of four-loop diagrams in the full theory. The mass parameter
$m_{E}^2(\Lambda)$ 
contribute only at odd powers in $e$ in a one-loop calculation in EQED.
This follows from the facts that one can write its coefficient in powers of
$e^2$, and that the free energy at one-loop  goes like $m_{E}^3(\Lambda)T$.
So the next contribution is first at order $e^7$. Moreover, there is only
one operator in addition to $f_{E}(\Lambda )$ that contributes at order $e^6$,
namely the quartic self-interaction of $A_0$. Since we already know its
coefficient, we are left with a straightforward two-loop calculation in EQED
(the double bubble). Hence, the challenge is the four-loop diagrams in 
full QED.

In the case of SQED, we need the parameters $f_{E}(\Lambda)$, 
$m_{E}^2(\Lambda)$ and $M^2(\Lambda)$ to order $\lambda^2$, $\lambda e^2$
and $e^4$. We already know these coefficients, 
although the latter parameter was 
determined in chapter three, using the effective potential.
There are no new operators which contribute at this order, so all that
remains is calculating the three-loop diagrams in ESQED. 
Moreover, the evolution equations should play a greater role here than
in QED, since the beta-function of the scalar mass parameter
is non-vanishing, exactly as in the somewhat simpler 
$\lambda\phi^4$-theory~\cite{braaten}. In particular, they can be used
to sum up leading logarithms. 
I have not yet performed the three-loop calculation, 
but it should be a manageable task, and will be the subject of 
further investigation.

\cleardoublepage
\cleardoublepage
\chapter{Conclusions}
\heading{Conclusions}{Conclusions}
It is time to summarize and draw some conclusions from the present work.
In the first chapter, we studied a rather old problem, namely that
of charged particles in external magnetic fields, to demonstrate calculations
where the results could be expressed in closed form. 
The fermions in two spatial 
dimensions show nontrivial properties such as de-Haas van Alphen oscillations,
induced vacuum charge and induced gauge-noninvariant terms in the
effective action. The bosons are somewhat more trivial, except for the
fact that they go from a diamagnetic phase to a paramagnetic one.

The second chapter was devoted to resummation, which has been the dominant
way of doing consistent calculations at finite temperature.
We applied the formalism to calculate the screening mass and the 
free energy in Yukawa theory, which are both static quantities.
In many respects, resummation has been beaten by effective field theory,
but it will be with us for years to come. It it still the only way
of solving dynamical problems at finite temperature, since these rely
on an unambiguous analytic continuation of Greens functions
from imaginary time to real time,
or direct calculations in real time.
 
The bulk of the thesis is on effective field theory. I have tried to bring 
about the philosophy and the general ideas about effective field theory by
some examples and discussion. The effective field theory program has
led to an increased understanding of quantum field theory,
conceptually speaking, and a better insight into concrete physical
systems. This approach opens up a new way of attacking difficult
problems in physics. I am sure that the future will provide examples
where effective field theory either solves problems which cannot
be solved by more traditional methods, or is used to push the
calculational frontier to higher orders in perturbation theory.
The message is (at least) two-fold: 

Firstly, there is the modern view on
renormalization which is intimately connected with the fact the every
quantum field theory should be looked upon as an effective description
at some scale. From this perspective renormalizability is no longer
a requirement of a consistent and useful quantum field theory.
This has given meaning to gravity and the nonlinear sigma model
beyond the tree approximation.
The next generation of text books certainly should be rewritten on this
point!

Secondly, I have tried to demonstrate the {\it efficiency} of the
effective field theory program: identify the scales in your problem
and take care of them, one by one, by integrating them out successively.
Thus, effective field theory unravels the contributions to physical
quantities from different momentum scales, and has become
an important tool in practical calculations.
In order for this program to work, the energy scales in the system
must be widely separated.
I have applied these ideas to quantum field theories at finite
temperature, and reproduced known results relatively easily.
I have also obtained new results with the effective field theory approach
which minimizes the calculational efforts. Effective field theories are here
to stay!

\pagebreak
\appendix
\chapter{Sum-integrals in the Full Theory}
\heading{Sum-integrals in the Full Theory}{Sum-integrals in the Full Theory}
In this appendix we summarize our conventions and define the 
sum-integrals used
in the calculations. We use the imaginary time formalism, 
where the four-momentum is $P=(p_{0},{\bf p})$
with $P^{2}=p_{0}^{2}+{\bf p}^{2}$. 
The Euclidean energy takes on discrete values, $p_{0}=2n\pi T$
for bosons and $p_{0}=(2n+1)\pi T$ for fermions.
Dimensional regularization is used to
regularize both infrared and ultraviolet divergences by working
in $d=4-2\epsilon$ dimensions, 
and we apply the $\overline{\mbox{MS}}$ 
renormalization scheme. \\ \\ 
The shorthand notations for the sum-integrals in the
bosonic and fermionic cases are, respectively:
\bqa
\hbox{$\sum$}\!\!\!\!\!\!\int_P f(P)&\equiv& 
\Big( \frac{e^{\gamma_{\tiny E}}\mu^{2}}
{4\pi}\Big )^{\epsilon}\,\,
T\!\!\!\!\!
\sum_{p_{0}=2\pi nT}\int\frac{d^{3-2\epsilon}k}
{(2\pi)^{3-2\epsilon}}f(P),\\
\hbox{$\sum$}\!\!\!\!\!\!\int_{\{P\}} f(P)&\equiv& 
\Big( \frac{e^{\gamma_{\tiny E}}\mu^{2}}
{4\pi}\Big )^{\epsilon}\,\,
T\!\!\!\!\!
\sum_{p_{0}=2\pi (n+1/2)T}\int\frac{d^{3-2\epsilon}k}
{(2\pi)^{3-2\epsilon}}f(P).
\eqa
All of the sum-integrals used in this thesis
have been calculated and tabulated by Arnold and
Zhai~\cite{arnold1}, Zhai and Kastening~\cite{kast}, 
and Braaten and Nieto~\cite{braaten2}. \\ \\ 
The general formulas for the bosonic one-loop
sum-integrals are
\bqa
\label{spe}
\hbox{$\sum$}\!\!\!\!\!\!\int_P\frac{1}{(P^2)^{m}}&=&\Big( \frac{e^{\gamma_{\tiny E}}\mu^{2}}
{4\pi^2T^2}\Big )^{\epsilon}\frac{T^{4-2m}}{2^{2m-1}}\pi^{3/2-2m-\epsilon}
\frac{\Gamma (m-3/2+\epsilon)}{\Gamma (m)}\zeta (2m-3+2\epsilon ),\\
\hbox{$\sum$}\!\!\!\!\!\!\int_P\frac{p_0^2}{(P^2)^{m}}&=&\Big( \frac{e^{\gamma_{\tiny E}}\mu^{2}}
{4\pi^2T^2}\Big )^{\epsilon}\frac{T^{6-2m}}{2^{2m-3}}\pi^{7/2-2m-\epsilon}
\frac{\Gamma (m-3/2+\epsilon)}{\Gamma (m)}\zeta (2m-5+2\epsilon ),\\
\hbox{$\sum$}\!\!\!\!\!\!\int_P\frac{p_0^4}{(P^2)^{m}}&=&\Big( \frac{e^{\gamma_{\tiny E}}\mu^{2}}
{4\pi^2T^2}\Big )^{\epsilon}\frac{T^{8-2m}}{2^{2m-5}}\pi^{11/2-2m-\epsilon}
\frac{\Gamma (m-3/2+\epsilon)}{\Gamma (m)}\zeta (2m-7+2\epsilon ).
\eqa
In appendix C, we give an example of how to calculate the above sum-integrals.
More specifically we need
\bqa
\hbox{$\sum$}\!\!\!\!\!\!\int_P\ln P^{2}&=&-\frac{\pi^{2}T^{4}}{45}\Big[1
+O(\epsilon )\Big], \\ 
\hbox{$\sum$}\!\!\!\!\!\!\int_P\frac{1}{P^{2}}&=&\frac{T^{2}}{12}
\Big[1+\Big(2\ln\frac{\mu}{4\pi T}+
2+2\frac{\zeta^{\prime}(-1)}{\zeta (-1)}\Big)\epsilon
+O(\epsilon^{2})\Big],\\
\hbox{$\sum$}\!\!\!\!\!\!\int_P\frac{1}{(P^{2})^{2}}&=&\frac{1}{(4\pi )^{2}}
\Big[\frac{1}{\epsilon}+2\ln\frac{\mu}{4\pi T}
+2\gamma_{E}+O(\epsilon )\Big],\\
\hbox{$\sum$}\!\!\!\!\!\!\int_P\frac{p_{0}^{2}}{(P^{2})^{2}}
&=&-\frac{T^{2}}{24}
\Big[1+\Big(2\ln\frac{\mu}{4\pi T}
+2\frac{\zeta^{\prime}(-1)}{\zeta (-1)}\Big)\epsilon
+O(\epsilon^{2} )\Big],\\
\hbox{$\sum$}\!\!\!\!\!\!\int_P\frac{p_{0}^{2}}{(P^{2})^{3}}&=&
\frac{1}{4(4\pi )^{2}}
\Big[\frac{1}{\epsilon}+2\ln\frac{\mu}{4\pi T}+2\gamma_{E}+2+O(\epsilon)\Big],\\
\hbox{$\sum$}\!\!\!\!\!\!\int_P\frac{p_{0}^{4}}{(P^{2})^{4}}&=&
\frac{1}{8(4\pi )^{2}}
\Big[\frac{1}{\epsilon}+2\ln\frac{\mu}{4\pi T}+2\gamma_{E}+\frac{8}{3}+O(\epsilon)\Big].
\eqa
We also need some fermionic one-loop sum-integrals. They can be obtained
from the bosonic ones by scaling arguments~\cite{parw1}. The relations are
\bqa
\hbox{$\sum$}\!\!\!\!\!\!\int_{\{P\}}\frac{1}{(P^{2})^{m}}&=&
\Big(2^{2m+1-d}-1\Big)\hbox{$\sum$}\!\!\!\!\!\!\int_{P}\frac{1}{(P^{2})^{m}},\\
\hbox{$\sum$}\!\!\!\!\!\!\int_{\{P\}}\frac{p_{0}^{2}}{(P^{2})^{m}}&=&
\Big(2^{2m-1-d}-1\Big)\hbox{$\sum$}\!\!\!\!\!\!\int_{P}\frac{p_{0}^{2}}{(P^{2})^{m}},\\
\hbox{$\sum$}\!\!\!\!\!\!\int_{\{P\}}\frac{p_{0}^{4}}{(P^{2})^{m}}&=&
\Big(2^{2m-3-d}-1\Big)\hbox{$\sum$}\!\!\!\!\!\!\int_{P}\frac{p_{0}^{4}}{(P^{2})^{m}}.
\eqa
The specific sum-integrals needed are
\begin{eqnarray}
\hbox{$\sum$}\!\!\!\!\!\!\int_{\{P\}}\ln P^2&=&\frac{7\pi^2T^4}{360}\Big[1
+O(\epsilon )\Big], \\ 
\hbox{$\sum$}\!\!\!\!\!\!\int_{\{P\}}\frac{1}{P^{2}}&=&-\frac{T^{2}}{24}\Big 
[1+(2\ln \frac{\mu}{4\pi T}+2-2\ln 2
+2\frac{\zeta '(-1)}{\zeta (-1)})\epsilon \Big]+{\cal  O}(\epsilon^{2}), \\
\hbox{$\sum$}\!\!\!\!\!\!\int_{\{P\}}\frac{1}{P^{4}}&=&\frac{1}{(4\pi)^{2}}
\Big ( \frac{1}{\epsilon}+2
\ln \frac{\mu}{4\pi T}+2\gamma_{E}+4\ln 2\Big)+{\cal  O}(\epsilon ), \\
\hbox{$\sum$}\!\!\!\!\!\!\int_{\{P\}}\frac{p_{0}^{2}}{P^{6}}&=&\frac{1}{4(4\pi)^{2}}\Big[\frac{1}{\epsilon}+2\ln\frac{\mu}{4\pi T}+2\gamma_{E}+2+4\ln2\Big]
+{\cal  O}(\epsilon ),\\
\hbox{$\sum$}\!\!\!\!\!\!\int_{\{P\}}\frac{p_{0}^{4}}{P^{8}}&=&\frac{1}{8(4\pi)^{2}}\Big[\frac{1}{\epsilon}+2\ln\frac{\mu}{4\pi T}+2\gamma_{E}+\frac{8}{3}+4\ln2\Big]+{\cal  O}(\epsilon ).
\end{eqnarray}
The two-loop integrals we need are
\bqa
\hbox{$\sum$}\!\!\!\!\!\!\int_{PQ}\frac{1}{P^{2}Q^{2}(P+Q)^{2}}&=&0,\\
\hbox{$\sum$}\!\!\!\!\!\!\int_{P\{Q\}}\frac{1}{P^{2}Q^{2}(P+Q)^{2}}&=&0,\\
\hbox{$\sum$}\!\!\!\!\!\!\int_{\{PQ\}}\frac{1}{P^{2}Q^{2}(P+Q)^{2}}&=&0.
\eqa  
The last two-loop sum-integral can be obtained from the second by a change
of variables. Moreover, in appendix C, we demonstrate how to 
calculate them.\\ \\
The simplest three-loop sum diagrams are the bosonic, fermionic and mixed
 basketballs. They read 
\bqa\nonumber
\hbox{$\sum$}\!\!\!\!\!\!\int_{PQK}\frac{1}{P^2Q^2K^2(P+K+Q)^2}&=&
\frac{1}{(4\pi )^2}\Big(\frac{T^2}{12}\Big)^{2}\Big[\frac{6}{\epsilon}+36\ln\frac{\mu}{4\pi T}+\frac{182}{5}\\
&&-12\frac{\zeta^{\prime}(-3)}{\zeta (-3)}
+48\frac{\zeta^{\prime}(-1)}{\zeta (-1)}\Big]+O(\epsilon ),\\
\nonumber
\hbox{$\sum$}\!\!\!\!\!\!\int_{\{PQK\}}\frac{1}{P^2Q^2K^2(P+K+Q)^2}&=&
\frac{1}{(4\pi )^2}\Big(\frac{T^2}{12}\Big)^{2}\Big[\frac{3}{2\epsilon}
+9\ln\frac{\mu}{4\pi T}+\frac{173}{120}\\
&&-\frac{63}{5}\ln 2-3\frac{\zeta^{\prime}(-3)}{\zeta (-3)}
+12\frac{\zeta^{\prime}(-1)}{\zeta (-1)}\Big]+O(\epsilon ),\\  \nonumber
\label{bf}\nonumber
\hbox{$\sum$}\!\!\!\!\!\!\int_{PQ\{K\}}\frac{1}{P^2Q^2K^2(P+K+Q)^2}&=&
\frac{1}{(4\pi )^2}\Big(\frac{T^2}{12}\Big)^{2}\Big[-\frac{3}{4\epsilon}
-\frac{9}{2}\ln\frac{\mu}{4\pi T}-\frac{179}{40}\\
&&+\frac{51}{10}\ln 2+\frac{3}{2}\frac{\zeta^{\prime}(-3)}{\zeta (-3)}
-6\frac{\zeta^{\prime}(-1)}{\zeta (-1)}\Big]+O(\epsilon ).
\eqa
In appendix C, we outline the calculation of the fermionic basketball.
Finally, there are some more difficult three-loop sum-integrals:
\bqa\nonumber
\hbox{$\sum$}\!\!\!\!\!\!\int_{\{P\}QK
}\frac{(QK)}{P^2Q^2K^2(P+K)^2(P+Q)^2}&=&
\frac{1}{(4\pi )^2}\Big(\frac{T^2}{12}\Big)^{2}\Big[\frac{3}{8\epsilon}+\frac{9}{4}\ln\frac{\mu}{4\pi T}+\frac{9}{4}\gamma_{E}
+\frac{361}{160}\\
&&-\frac{57}{10}\ln 2
+\frac{3}{2}\frac{\zeta^{\prime}(-3)}{\zeta (-3)} -\frac{3}{2}\frac{\zeta^{\prime}(-1)}{\zeta (-1)}\Big]+O(\epsilon ),\\ \nonumber
\hbox{$\sum$}\!\!\!\!\!\!\int_{P\{QK\}
}\frac{(QK)^2}{P^4Q^2K^2(P+K)^2(P+Q)^2}&=&
\frac{1}{(4\pi )^2}\Big(\frac{T^2}{12}\Big)^{2}\Big[\frac{5}{24\epsilon}+\frac{5}{4}\ln\frac{\mu}{4\pi T}+\frac{1}{4}\gamma_{E}
+\frac{23}{24}\\
&&-\frac{8}{5}\ln 2
-\frac{1}{6}\frac{\zeta^{\prime}(-3)}{\zeta (-3)} 
+\frac{7}{6}\frac{\zeta^{\prime}(-1)}{\zeta (-1)}\Big]+O(\epsilon ),\\ \nonumber
\hbox{$\sum$}\!\!\!\!\!\!\int_{PQK}\frac{(Q-K)^4}{P^4Q^2K^2(P+K)^2(P+Q)^2}&=&
\frac{2}{3(4\pi )^2}\Big(\frac{T^2}{12}\Big)^{2}\Big[\frac{11}{\epsilon}+66\ln\frac{\mu}{4\pi T}
+\frac{73}{2}+
12\gamma_{E}\\
&&
-10\frac{\zeta^{\prime}(-3)}{\zeta (-3)}+64\frac{\zeta^{\prime}(-1)}{\zeta (-1)}\Big]+O(\epsilon ).
\eqa
\cleardoublepage
\chapter{Integrals in the Effective Theory}
\heading{Integrals in the Effective Theory}{Integrals in the Effective Theory}
In this appendix we define the integrals we need in the calculations in the
effective three dimensional Euclidean field theory.
We employ dimensional regularization
in $3-2\epsilon$ dimensions to regularize infrared and ultraviolet 
divergences. 
In analogy with Appendix A, we define
\beq
\int_{p}f(p)\equiv\Big( \frac{e^{\gamma_{\tiny E}}\mu^{2}}
{4\pi}\Big )^{\epsilon}\int\frac{d^{3-2\epsilon}p}
{(2\pi)^{3-2\epsilon}}f(p).
\eeq
Again $\mu$ coincides with the renormalization scale in the 
modified minimal subtraction renormalization
scheme. \\ \\
In the effective theory we need the following one-loop integrals
\bqa
\int_{p}\ln (p^{2}+m^{2})&=&-\frac{m^{3}}{6\pi}
\Big[1+\Big(2\ln\frac{\mu}{2m}+\frac{8}{3}\Big)\epsilon
+O(\epsilon^{2})\Big ], \\
\int_{p} \frac{1}{p^{2}+m^{2}}&=&-\frac{m}{4\pi}
\Big[1+\Big(2\ln\frac{\mu}{2m}+2\Big)\epsilon
+O(\epsilon^{2})\Big],\\
\int_{p} \frac{1}{(p^{2}+m^{2})^{2}}&=&\frac{1}{8\pi m}
\Big[1+\Big(2\ln\frac{\mu}{2m}\Big)\epsilon
+O(\epsilon^{2})\Big].
\eqa
All integrals are
straightforward to evaluate in 
dimensional regularization. Details may be found in Ref.~\cite{ryder}.\\ \\
The specific two-loop integrals needed are
\bqa\nonumber
\int_{pq} \frac{1}{(p^{2}+m_1^{2})(q^{2}+m_2^{2})[({\bf p}+{\bf q})^{2}+m_3^2]}&=&
\frac{1}{(4\pi )^{2}}
\Big[\frac{1}{4\epsilon}
+\frac{1}{2}+\\&&
\ln\frac{\mu}{m_1+m_2+m_3}+O(\epsilon )\Big],\\
\int_{pq} \frac{1}{(p^{2}+m^{2})^{2}(q^{2}+m^{2})({\bf p}-{\bf q})^{2}}&=&
\frac{1}{(4\pi )^{2} m^{2}}
\Big[\frac{1}{4}+
O(\epsilon )\Big],\\  \nonumber
\label{dimjens}
\left.\int_{pq} \frac{1}{(p^{2}+M^{2})(q^{2}
+M^{2})[({\bf p}+{\bf q}+{\bf k})^{2}+m^{2}]}\right|_{k=im}
&=&
\frac{1}{(8\pi )^{2}}\Big[\frac{1}{\epsilon}
+6\\
-
4\ln [\frac{2(M+m)}{\mu}]
+4\frac{M}{m}\ln\frac{M}{M+m}+O(\epsilon )\Big].&&
\eqa
The first two of these integrals can be found in Refs.[36,73].
The integral in Eq.~(\ref{dimjens}) has previously been calculated by
Braaten and Nieto for $m=M$ in Ref.~\cite{braaten}. 
In appendix C, we calculate it for the more general case $m\neq M$.
\chapter{Some Sample Calculations}
\heading{Some Sample Calculations}{Some Sample Calculations}
This appendix is devoted to the explicit calculation of
some sum-integrals in order to
illustrate the methods invented by Arnold and Zhai~\cite{arnold1}. 
We will give some examples of how one computes
one, two and three-loop diagrams and choose to present the calculations
of sum-integrals that have not explicitly worked out in the literature.\\ \\
Let us start with the following one-loop sum-integral:
\bqa\nonumber
\hbox{$\sum$}\!\!\!\!\!\!\int_{P}\frac{p_0^4}{(P^2)^{m}}&=&
\Big( \frac{e^{\gamma_{\tiny E}}\mu^{2}}
{4\pi}\Big )^{\epsilon}\,\,
T\sum_{n}\int\frac{d^{3-2\epsilon}p}
{(2\pi)^{3-2\epsilon}}\frac{(2\pi nT)^4}{[p^2+(2\pi nT)^2]^{m}}\\\nonumber
&=&\Big( \frac{e^{\gamma_{\tiny E}}\mu^{2}}
{4\pi}\Big )^{\epsilon}\,\,
T(2\pi T)^4\int\frac{d^{3-2\epsilon}p}
{(2\pi)^{3-2\epsilon}}\frac{1}{[p^2+(2\pi T)^2]^{m}}\sum_{n}n^{7-2m-2\epsilon}
\\ 
&=&\Big( \frac{e^{\gamma_{\tiny E}}\mu^{2}}
{4\pi^2T^2}\Big )^{\epsilon}\,\,
\frac{T^{8-2m}}{2^{2m-5}}\pi^{11/2-2m-\epsilon}
\frac{\Gamma (m-3/2+\epsilon)}{\Gamma (m)}
\zeta (2m-7+2\epsilon).
\eqa 
In the second line we have changed variables, and in the last line we have
employed the definitions of the $\Gamma$ and $\zeta$-functions, as well as
performing a standard one-integral using dimensional 
regularization~\cite{ryder}. \\ \\
The next sum-integral we consider comes from the 
fermionic setting sun diagram:
\beq
\hbox{$\sum$}\!\!\!\!\!\!\int_{\{PQ\}}\frac{1}{P^2Q^2(P+Q)^2}\equiv
\hbox{$\sum$}\!\!\!\!\!\!\int_{P}\frac{\Pi_{f}(P)}{P^2}.
\eeq 
Here, we have defined the fermionic self-energy
\beq
\Pi_{f}(P)=\hbox{$\sum$}\!\!\!\!\!\!\int_{\{Q\}}\frac{1}{Q^2(P+Q)^2}.
\eeq
There are different ways to compute this sum-integral. One is the contour
method. One rewrites the sum over $p_0$ and $q_0$ as contour integrals.
One then finds terms independent, linear and quadratic in the distribution
functions, which separately can be treated using dimensional 
regularization~\cite{par2}.
We shall use the method of Arnold and Zhai, who have developed a new
machinery to computing difficult multi-loop sum-integrals~\cite{arnold1}.

The idea is as follows. The calculation of an $n$-loop sum-integral
obviously requires the evaluation of $n$ sums and $n$ three-dimensional
integrations, if we work in momentum space.
In coordinate space it only requires one four dimensional integration.
At first sight, it therefore seems that it would be simpler to compute
the sum-integrals in coordinate space. This would certainly be the case
if the expressions were finite, so that we could evaluate the expressions
directly in four dimensions. However, the sum-integrals are UV-divergent,
and we must subtract off these divergences that arise at $T=0$.
This is most easily carried out in momentum space. The remainder
may then be evaluated in four dimensions, and this is then done
using the Fourier transform of the momentum space propagator:
\beq
\tilde{\Delta}(q_0,r)=\frac{e^{-|p_0|r}}{4\pi r}.
\eeq
First, we separate the fermionic self-energy into a $T=0$ piece and
a finite temperature term by writing
\beq
\Pi_{f}(P)=\Pi_{f}^{(0)}(P)+\Pi_{f}^{(T)}(P).
\eeq 
The temperature independent part is, using standard
results from dimensional regularization~\cite{ryder}
\beq
\Pi_{f}^{(0)}=\mu^{2\epsilon}\int\frac{d^{d}q}{(2\pi )^d}\frac{1}{Q^2(P+Q)^2}=
\frac{1}{(4\pi )^2}(\frac{4\pi\mu^2}{P^2})^{\epsilon}
\Big[\frac{1}{\epsilon}-\gamma_{E}+2+{\cal O}(\epsilon)\Big].
\eeq
By using Eq.~(\ref{spe}), one finds
\beq
\label{analog}
\hbox{$\sum$}\!\!\!\!\!\!\int_{P}\frac{\Pi_{f}^{(0)}(P)}{P^2}=
\frac{1}{(4\pi )^2}\frac{T^2}{12}\Big(\frac{1}{\epsilon}+4\ln\frac{\mu}{4\pi T}
+6+4\frac{\zeta^{\prime}(-1)}{\zeta (-1)}\Big).
\eeq
We now need $\Pi_{f}^{(T)}(P)$, and in order to obtain it we shall
compute $\Pi_{f}(P)$
and subtract off its $T=0$ limit. The fermionic self-energy is 
\bqa\nonumber
\Pi_f(P)&=&T\sum_{\{q_0\}}\int d^3r\tilde{\Delta}(q_0,r)
\tilde{\Delta}(q_0+p_0,r)e^{ip\cdot r}\\
\label{pif}
&=&\frac{T}{(4\pi )^2}\sum_{\{q_0\}}\int\frac{d^3r}{r^2}e^{-|q_0|r}
e^{-|p_0+q_0|r}e^{ip\cdot r}.
\eqa
The sum over $q_0$ is given by
\beq
\label{sumf}
\sum_{\{q_0\}}e^{-|q_0|r}
e^{-|p_0+q_0|r}=[\mbox{csch}\bar{r}+|\bar{p}_0|]e^{-|p_0|r}.
\eeq
Here, $\bar{r}=2\pi rT$ and $\bar{p}_0=p_0/2\pi rT$.
This formula can obtained by splitting the sum 
into three parts depending on the
sign of $q_0$ and $(q_0+p_0)$, and then use known results for geometric series.
Substituting Eq.~(\ref{sumf}) into Eq.~(\ref{pif}) and 
letting $T\rightarrow 0$, we obtain $\Pi^{(0)}_f(P)$. Subtracting this
from $\Pi_f(P)$, we find
\beq
\label{pif2}
\Pi_f^{T}(P)=\frac{T}{(4\pi )^2}\int\frac{d^3r}{r^2}e^{ip\cdot r}
[\mbox{csch}\bar{r}-1/\bar{r}]e^{-|p_0|r}+{\cal O}(\epsilon ).
\eeq
Although $\Pi_f^{(T)}(P)$ is finite for $P\rightarrow \infty$, 
$\Pi_f^{T}(P)/P^2$ is logarithmically divergent, because the former goes like 
$1/P^2$ in this limit:
\beq
\label{h1}
\Pi_{f}^{(T)}(P)\rightarrow 
-\frac{2}{P^2}\hbox{$\sum$}\!\!\!\!\!\!\int_{\{Q\}}\frac{1}{Q^2}.
\eeq
This behaviour can be inferred by the contour trick: One rewrites the
sum over Matsubara frequencies as a contour integral and study its high $P$
limit~\cite{arnold1}. 
Moreover, the high momentum behaviour is also given by the small 
$r$ behaviour of the integrand in Eq.~(\ref{pif2}),  and 
using the series expansion of $\mbox{csch}\bar{r}$ one finds
\beq
\label{h2}
\Pi_{f}^{(T)}(P)\rightarrow 
-\frac{1}{6}\frac{T}{(4\pi )^2}\int\frac{d^3r}{r^2}\bar{r}
e^{ip\cdot r}e^{-|p_0|r}.
\eeq
We can now write the finite temperature part of the fermionic
setting-sum diagram as
\beq
\label{setsun}
\frac{T^2}{(4\pi )^2}\hbox{$\sum$}\!\!\!\!\!\!\int_{P}\frac{1}{P^2}
\int\frac{d^3r}{r^2}[\mbox{csch}\bar{r}-1/\bar{r}+\bar{r}/6]
e^{ip\cdot r}e^{-|p_0|r}
+2\hbox{$\sum$}\!\!\!\!\!\!\int_{P\{Q\}}\frac{1}{P^4Q^2}.
\eeq 
Consider the first term above, which we denote by $I$.
Integration over $p$ simply gives the propagator in coordinate space,
while the summation over $p_0$ yields a factor $\mbox{coth}\bar{r}$, since
\beq
\label{bsum}
\sum_{q_0}=e^{-|q_0|r}
e^{-|p_0+q_0|r}=[\coth \bar{r}+|\bar{p}_0|]e^{-|p_0|r}.
\eeq
Furthermore,
integration over the sphere gives the usual $4\pi$, and we find 
\beq
I=\frac{T^2}{(4\pi)^2}\Big[\int\frac{dr}{r}[\mbox{csch}\bar{r}-1/\bar{r}
+\bar{r}/6]\Big]\mbox{coth}\bar{r}.
\eeq 
The integral that has been obtained are convergent.
One can then calculate it numerically.
However, it is divergent term by term, but Arnold and Zhai have developed
a clever way to compute them analytically, using methods similar to
dimensional regularization~\cite{arnold1}. 
Below we shall discuss the derivation of them.\\ \\
The integral can be expressed in terms of $\Gamma$-functions and 
$\zeta$-functions:
\bqa
I=\frac{T^2}{(4\pi)^2}\Big[(2-2^{-z}-2^{-z-1})\Gamma (z)\zeta (z)
+\frac{1}{6}2^{-z+1}\Gamma (z+2)\zeta (z+2)\Big]
\hspace{1cm}z\rightarrow -1.
\eqa
We proceed by expanding the $\Gamma$ and 
$\zeta$-functions around the pole $z=-1$. This produces:
\beq
\frac{1}{(4\pi )^2}\frac{T^2}{12}\Big[2\gamma_{E}-4-2\ln 2
-2\frac{\zeta^{\prime}(-1)}{\zeta (-1)}\Big].
\eeq
Finally, we have the second term in Eq.~(\ref{setsun}), which is easily
evaluated using appendix A:
\beq
2\hbox{$\sum$}\!\!\!\!\!\!\int_{P\{Q\}}\frac{1}{P^4Q^2}=
-\frac{1}{(4\pi)^2}\frac{T^2}{12}\Big[
\frac{1}{\epsilon}+4\ln\frac{\mu}{4\pi T}+2\gamma_{E}+2-2\ln 2
+2\frac{\zeta^{\prime}(-1)}{\zeta (-1)}\Big].
\eeq
Adding up the different pieces, we conclude that the fermionic setting sun 
graph vanishes:
\beq
\hbox{$\sum$}\!\!\!\!\!\!\int_{\{PQ\}}\frac{1}{P^2Q^2(P+Q)^2}=0.
\eeq 
Let us now move on and
discuss the evaluation of the mixed boson-fermion basketball 
diagram. We shall be slightly more sketchy this time.
The sum-integral reads
\beq
J=\hbox{$\sum$}\!\!\!\!\!\!\int_{PQ\{K\}}\frac{1}{P^2Q^2K^2(P+Q+K)^2}.
\eeq
By changing variables this may be rewritten as
\beq
\hbox{$\sum$}\!\!\!\!\!\!\int_{P}\Pi_{f}(P)\Pi_{b}(P).
\eeq
Here, $\Pi_{b}(P)$ is the bosonic self-energy
\beq
\hbox{$\sum$}\!\!\!\!\!\!\int_{Q}\frac{1}{Q^2(P+Q)^2}.
\eeq
In complete analogy with the fermionic case, we write
\beq
\Pi_{b}(P)=\Pi^{(0)}_{b}(P)+\Pi^{(T)}_{b}(P).
\eeq
Note that $\Pi_b^{(0)}(P)=\Pi_f^{(0)}(P)$. The mixed basketball then reads
\bqa\nonumber
\label{idef}
J&=&\hbox{$\sum$}\!\!\!\!\!\!\int_{P}\Pi_{f}^{(0)}(P)\Pi^{(0)}_{b}(P)
+\hbox{$\sum$}\!\!\!\!\!\!\int_{P}\Pi_{f}^{(T)}(P)\Pi^{(0)}_{b}(P)
+\hbox{$\sum$}\!\!\!\!\!\!\int_{P}\Pi_{f}^{(0)}(P)\Pi^{(T)}_{b}(P)\\
&&
+\hbox{$\sum$}\!\!\!\!\!\!\int_{P}\Pi_{f}^{(T)}(P)\Pi^{(T)}_{b}(P).
\eqa
By using Eq~(\ref{spe}) once more we derive
the result
\beq
\hbox{$\sum$}\!\!\!\!\!\!\int_{P}\Pi_{f}^{(0)}(P)\Pi_b^{(0)} (P)=
\frac{1}{(4\pi )^2}\Big(\frac{T^2}{12}\Big)^2\Big[\frac{2}{5\epsilon}
+\frac{12}{5}\ln\frac{\mu}{4\pi T}
+\frac{24}{5}+\frac{12}{5}\frac{\zeta^{\prime}(-3)}{\zeta (-3)}\Big].
\eeq

The next two terms are calculated using similar methods~\cite{arnold1}. 
In order not to get overwhelmed by calculational details, we 
simply state the result~\cite{arnold1}:
\bqa\nonumber
\hbox{$\sum$}\!\!\!\!\!\!\int_{P}\Pi_{f}^{(T)}(P)\Pi_b^{(0)}(P)&=&
\frac{1}{(4\pi )^2}\Big(\frac{T^2}{12}\Big)^2\Big[\frac{1}{20\epsilon}
+\frac{3}{10}\ln\frac{\mu}{4\pi T}-\frac{301}{120}\\
&&
-\frac{37}{10}\ln 2
-\frac{37}{10}\frac{\zeta^{\prime}(-3)}{\zeta (-3)}
+4\frac{\zeta^{\prime}(-1)}{\zeta (-1)}\Big],\\ \nonumber
\hbox{$\sum$}\!\!\!\!\!\!\int_{P}\Pi_{f}^{0}(P)\Pi^{(T)}_b(P)&=&
\frac{1}{(4\pi )^2}\Big(\frac{T^2}{12}\Big)^2\Big[\frac{4}{5\epsilon}
+\frac{24}{5}\ln\frac{\mu}{4\pi T}+\frac{103}{15}\\
&&
+\frac{4}{5}\frac{\zeta^{\prime}(-3)}{\zeta (-3)}
+4\frac{\zeta^{\prime}(-1)}{\zeta (-1)}\Big].
\eqa
The bosonic self-energy may be written as
\beq
\Pi_{b}(P)=\frac{T}{(4\pi )^2}\sum_{q_0}\int\tilde{\Delta}(q_0,r)
\tilde{\Delta}(q_0+p_0,r)e^{ip\cdot r}.
\eeq
As in the previous calculation, we find $\Pi_b^{(T)}(P)$ by 
subtracting off its $T=0$ part
\beq
\Pi_b^{(T)}=\frac{T}{(4\pi )^2}\int\frac{d^3r}{r^2}e^{ip\cdot r}
[\coth\bar{r}-1/\bar{r}]e^{-|p_0|r}.
\eeq
The high momentum behaviour of the bosonic self-energy is
obtained in analogy with the Eqs.~(\ref{h1}) and~(\ref{h2}) in the fermionic case:
\bqa
\Pi_{b}^{(T)}(P)&\rightarrow &\frac{1}{3}\frac{T}{(4\pi)^2}
\int\frac{d^3r}{r^2}\bar{r}e^{ip\cdot r}e^{-|p_0|r}.\\
\Pi_{b}^{(T)}(P)&\rightarrow & -\frac{2}{P^2}
\hbox{$\sum$}\!\!\!\!\!\!\int_{Q}\frac{1}{Q^2}.
\eqa
The last term in Eq.~(\ref{idef}) can then be rewritten as
\beq
\frac{T^4}{32\pi^3}\int\frac{d\bar{r}}{\bar{r}^2}\Big[(\mbox{csch}\bar{r}-1/\bar{r}
)(\mbox{coth}\bar{r}-1/\bar{r})(\mbox{coth}\bar{r}-1)+\frac{\bar{r}^2}{18}
(\mbox{coth}\bar{r}-1)\Big]+4\hbox{$\sum$}\!\!\!\!\!\!\int_{PQ\{K\}}
\frac{1}{P^4Q^2K^2}.
\eeq
Here we have performed the sum over Matsubara frequencies using 
Eq.~(\ref{bsum}). This first integral above is again finite, but infinite
term by term. Using the regularization techniques of Arnold and Zhai
the first term above yields
\beq
\frac{1}{(4\pi )^2}\Big(\frac{T^2}{12}\Big)^2\Big[
4\gamma_E-\frac{169}{30}
+\frac{24}{5}\ln 2+
2\frac{\zeta^{\prime}(-3)}{\zeta (-3)}
-6\frac{\zeta^{\prime}(-1)}{\zeta (-1)}\Big],
\eeq
while the second reads
\beq
-\frac{1}{(4\pi )^2}\Big(\frac{T^2}{12}\Big)^2\Big[\frac{2}{\epsilon}+
12\ln\frac{\Lambda}{4\pi T}+4\gamma_E +8
-4\ln 2+
8\frac{\zeta^{\prime}(-1)}{\zeta (-1)}\Big].
\eeq
Adding all the terms, we obtain Eq.~(\ref{bf}). \\

We would also like to mention that it has been noted that 
one can obtain this diagram from the bosonic and fermionic basketball diagrams
by scaling arguments~\cite{par2}. One finds
\bqa\nonumber
\hbox{$\sum$}\!\!\!\!\!\!\int_{PQ\{K\}}\frac{1}{P^2Q^2K^2(P+K+Q)^2}&=&
-\frac{1}{6}[1-2^{11-3d}]\hbox{$\sum$}\!\!\!\!\!\!\int_{PKQ}\frac{1}{P^2Q^2K^2(P+K+Q)^2}\\
&&
-\frac{1}{6}
\hbox{$\sum$}\!\!\!\!\!\!\int_{\{PQK\}}\frac{1}{P^2Q^2K^2(P+K+Q)^2}.
\eqa
We have checked that our results agree.\\ \\
Finally, we must consider the calculation of the divergent integrals
that appeared above. \\ \\
The first needed is
\beq
\label{van}
\int_0^{\infty}drr^z.
\eeq
Depending on the value of $z$, the contribution to the integral from one of the
limits vanishes (and the other blows up). Thus, if one analytically continue
$z$ independently to regulate the behaviour at $r=0$ and $r=\infty$, the
integral in Eq.~(\ref{van}) vanishes. \\ \\
We also have the result
\beq
\int_0^{\infty}drr^ze^{-ar}=a^{-1-z}\Gamma (1+z),
\eeq
for the values of $z$ for which the integral is well defined. The integral
is then defined for all values of $z$ by analytic continuation. 
This makes it possible to attack integrals of hyperbolic
functions times powers of $r$:
\bqa\nonumber
\int_0^{\infty}drr^z\mbox{csch}r&=&\int_0^{\infty}r^z\Big[2e^{-r}\sum_{n=0}^{\infty}e^{-2nr}
\Big]\\ \nonumber
&=&2^{-z}\Gamma (1+z)\sum_{n=0}^{\infty}\frac{1}{(n+\frac{1}{2})^{z+1}}\\
&=&(2-2^{-z})\Gamma (1+z)\zeta (1+z).
\eqa
In the last line, we have used the functional relation between the
Riemann Zeta-function and the Hurwitz Zeta function~\cite{tab}:
\beq
\zeta (z)=\frac{1}{2^z-1}\zeta (z,\frac{1}{2}).
\eeq
In a similar fashion one can obtain
\beq
\int_0^{\infty}drr^z\coth r=
2^{-z}\Gamma (z+1)\zeta (z+1).
\eeq
More complicated integrals, such as products between $\mbox{csch} \bar{r}$
and $\coth \bar{r}$ can be computed from the above formulas using 
integration by parts. We then have all regulated integrals needed 
to evaluate the expressions above.\\ \\
Let us now turn to the three-dimensional integrals.
All, except for one integral in the effective have been computed by Braaten
and Nieto in Ref.~\cite{braaten}. In order to illustrate their methods. 
we shall calculate this integral, which is
\beq
\left.\int_{pq} \frac{1}{(p^{2}+M^{2})(q^{2}
+M^{2})[({\bf p}+{\bf q}+{\bf k})^{2}+m^{2}]}\right|_{k=im}
\eeq
The integral has been computed in the less general case $m=M$~\cite{braaten}. 
It can best be computed by going to coordinate space. The 
Fourier
transform of the propagator is
\beq
V_{m}(R)=\int_p e^{i{\bf p}\cdot {\bf R}}\,\frac{1}{p^{2}+m^{2}}.
\eeq
It can be expressed in terms of a modified Bessel function
\beq
V_{m}(R)=\Big(\frac{e^{\gamma_{E}}\mu^{2}}{4\pi}\Big)
^{\epsilon}\frac{1}{(2\pi )^{3/2-\epsilon}}
\Big(\frac{m}{R}\Big)^{1/2-\epsilon}K_{1/2-\epsilon}(mR).
\eeq
In three dimensions ($\epsilon =0$) this is the Yukawa potential:
\beq
\tilde{V}_{m}(R)=\frac{e^{-mR}}{4\pi R}.
\eeq
For small $R$ it can be written as a sum of two Laurent series in $R^{2}$.
One of these is singular beginning with an $R^{-1+2\epsilon}$ 
term and the other
is regular which begins with an $R^{0}$ term:
\bqa
V_{m}(R)&=&\Big(\frac{e^{\gamma_{E}}\mu^{2}}{4}\Big)
^{\epsilon}\frac{\Gamma (\frac{1}{2}-\epsilon)}
{\Gamma (\frac{1}{2})}\frac{1}{4\pi}R^{-1+2\epsilon}
\Big[1+\frac{m^{2}R^{2}}{2(1+2\epsilon)}+O(m^{4}R^{4})\Big]\\
&&-(e^{\gamma_{E}}\mu^{2})^{\epsilon}
\frac{\Gamma (-\frac{1}{2}+\epsilon)}
{\Gamma (-\frac{1}{2})}\frac{1}{4\pi}m^{1-2\epsilon}
\Big[1+\frac{m^{2}R^{2}}{2(3-2\epsilon)}+O(m^{4}R^{4})\Big].
\eqa
The integral can be written
\beq
\int_{pq} \frac{1}{(p^{2}+M^{2})(q^{2}+M^{2})
[({\bf p}+{\bf q}+{\bf k})^{2}+m^{2}]}
=\int_{R} e^{i{\bf kR}}V^{2}_{M}(R)V_{m}(R).
\eeq
The radial integration is now split into two regions, 
$0<R<r$ and $r<R<\infty$.
The ultraviolet divergences arise from the region $R\rightarrow 0$. 
This implies that we can set $\epsilon =0$ in the region 
where $r<R<\infty$.
Hence, one can write the integral as
\bqa\nonumber
\int e^{i{\bf kR}}V^{2}_{M}(R)V_{m}(R)&=&
\Big(\frac{e^{\gamma_{E}}\mu^{2}}{2k}\Big)^
{-\epsilon}\frac{(2\pi)^{3/2}}{\sqrt{k}}
\int_{0}^{r}dRR^{3/2-\epsilon}J_{1/2-\epsilon}(kR)V^{2}_{M}(R)
V_{m}(R)\\
&&+\frac{4\pi}{k}\int_{r}^{\infty}dRR\sin (kR)\tilde{V}^{2}_{M}(R)
\tilde{V}_{m}(R).
\eqa
Here, $J_{\nu}(x)$ is an ordinary Bessel function. The Bessel function has 
the following expansion for small $R$:
\beq
J_{1/2-\epsilon}(kR)=\frac{1}{\Gamma (\frac{3}{2}-\epsilon)}
\Big(\frac{1}{2}kR\Big)^{1/2-\epsilon}[1+O(k^{2}R^{2})].
\eeq
Using this expansion and the small $R$ expansion of the 
potential, the first
integral is, after dropping terms that vanish in the limit 
$r\rightarrow 0$
\bqa \nonumber
\Big(\frac{e^{\gamma_{E}}\mu^{2}}{2k}\Big)
^{-\epsilon}\frac{(2\pi)^{3/2}}{\sqrt{k}}
\int_{0}^{r}dRR^{3/2-\epsilon}J_{1/2-\epsilon}
(kR)V^{2}_{M}(R)V_{m}(R)&=&
\frac{1}{(8\pi )^{2}}\Big[\frac{1}{\epsilon}+
4\ln \mu r\\
&&+2+4\gamma_{E}\Big]+O(\epsilon ).
\eqa
The second integral can be found in e.g Ref.~\cite{tab} and equals
\beq
\frac{i}{2k(4\pi)^{2}}\Big[(2M+m+ik)\Gamma [-1,(2M+m+ik)r]
-(2M+m-ik)\Gamma [(2M+m-ik)r]\Big].
\eeq
Evaluating this at $k=im$ yields
\bqa \nonumber
\left.\frac{4\pi}{k}\int_{r}^{\infty}dRR\sin (kR)\tilde{V}^{2}_{M}(R)
\tilde{V}_{m}(R)\right|_{k=im}&=&
\frac{1}{(4\pi )^{2}}\Big[\frac{M}{m}\ln\frac{M}{M+m}-\gamma_{E}+1
\\
&&-\ln [2(M+m)r]\Big]+O(\epsilon),
\eqa
where we have used the series expansion of the incomplete gamma 
function
\beq
\Gamma [-1,x]=\frac{1}{x}+\gamma_{E}-1+\ln x +O(x^{2}),
\eeq
and dropped terms that vanish as $r\rightarrow 0$.
Collecting our results we obtain Eq.~(\ref{dimjens}).
The logarithms of $r$ cancel and our result
reduces to the one found in Ref.~\cite{braaten} 
in the case $m=M$, as it should.

\end{document}